\documentclass[iop]{emulateapj}
\usepackage{natbib,float,longtable,subfigure}
\usepackage{amsmath,amssymb}
\usepackage{mathptmx}
\usepackage{graphicx}
\usepackage{multirow}
\usepackage{color}

\newcommand{\todo}{\ifmmode \text{\Huge{\(\bullet\)}} \else {\Huge$\bullet$}\fi}
\newcommand{\tido}{\ifmmode {\bullet} \else $\bullet$\fi}

\newcommand{\E        }[1]{\ifmmode 10^{#1} \else $10^{#1}$\fi}
\newcommand{\tE        }[1]{\ifmmode \times10^{#1} \else $\times10^{#1}$\fi}
\newcommand{\til}{\ifmmode \sim \else $\sim$\fi}
\renewcommand{\~} {\ifmmode \sim \else $\sim$\fi}

\newcommand{\pc}	{\ifmmode {\rm pc} \else pc\fi}
\newcommand{\ld}	{\ifmmode {\rm l.d.} \else l.d.\fi}
\newcommand{\kms}	{\ifmmode {\rm km\,s}^{-1} \else km\,s$^{-1}$\fi}
\newcommand{\cc}	{\ifmmode {\rm cm}^{-3}    \else cm$^{-3}$\fi}
\newcommand{\cmii}	{\ifmmode {\rm cm}^{-2}    \else cm$^{-2}$\fi}
\newcommand{\ergs}	{\ifmmode {\rm erg\,s}^{-1} \else erg s$^{-1}$\fi}
\newcommand{\ergcms}	{\ifmmode {\rm erg\,cm}^{-2}\,{\rm s}^{-1} \else erg\,cm$^{-2}$\,s$^{-1}$\fi}
\newcommand{\ergcmsA}	{\ifmmode {\rm erg\,cm}^{-2}\,{\rm s}^{-1}\,{\rm\AA}^{-1}
\else erg\,cm$^{-2}$\,s$^{-1}$\,\AA$^{-1}$\fi}
\newcommand{  \ergcmsHz  }{\ifmmode{\rm erg\,cm}^{-2}\,{\rm s}^{-1}\,{\rm Hz}^{-1}
                       \else ergs\,cm$^{-2}$\,s$^{-1}$\,Hz$^{-1}$\fi}
\newcommand{\kev}	{\ifmmode {\rm keV} \else keV\fi}

\newcommand{\mic}	{\ifmmode {\rm \mu m} \else $\mu$m\fi}
\newcommand{\vFWHM}	{\ifmmode v_{\mbox{\tiny FWHM}} \else $v_{\mbox{\tiny FWHM}}$\fi}
\newcommand{\vBLR}	{\ifmmode v_{\mbox{\tiny BLR}} \else $v_{\mbox{\tiny BLR}}$\fi}
\newcommand{\sigBLR}	{\ifmmode \sigma_{\mbox{\tiny BLR}} \else $\sigma_{\mbox{\tiny BLR}}$\fi}
\newcommand{\vNLR}	{\ifmmode v_{\mbox{\tiny NLR}} \else $v_{\mbox{\tiny NLR}}$\fi}
\newcommand{\tauBLR}	{\ifmmode \tau_{\mbox{\tiny BLR}} \else $\tau_{\mbox{\tiny BLR}}$\fi}

\newcommand{\Hubble}	{\ifmmode {\rm km\,s}^{-1}\,{\rm Mpc}^{-1} \else km\,s$^{-1}$\,Mpc$^{-1}$\fi}
\newcommand{\NDunit}	{\ifmmode {\rm Mpc}^{-3} \else Mpc$^{-3}$\fi}
\newcommand{\LFunit}	{\ifmmode {\rm Mpc}^{-3}\,{\rm mag}^{-1} \else Mpc$^{-3}$\,mag$^{-1}$\fi}
\newcommand{\MFunit}	{\ifmmode {\rm Mpc}^{-3}\,{\rm dex}^{-1} \else Mpc$^{-3}$\,dex$^{-1}$\fi}

\newcommand{\Msun}{\ifmmode M_{\odot} \else $M_{\odot}$\fi}
\newcommand{\Lsun}{\ifmmode L_{\odot} \else $L_{\odot}$\fi}
\newcommand{\Zsun}{\ifmmode Z_{\odot} \else $Z_{\odot}$\fi}
\newcommand{\mpyr}{\ifmmode \Msun\,{\rm yr}^{-1} \else $\Msun\,{\rm yr}^{-1}$\fi}

\newcommand{\Msol}{\Msun}

\newcommand{\qnote}{\ifmmode q_{0} \else $q_{0}$\fi}
\newcommand{\Hnote}{\ifmmode H_{0} \else $H_{0}$\fi}
\newcommand{\hnote}{\ifmmode h_{0} \else $h_{0}$\fi}
\newcommand{\anote}{\ifmmode a_{0} \else $a_{0}$\fi}
\newcommand{\tnote}{\ifmmode t_{0} \else $t_{0}$\fi}

\newcommand{  \Halpha   }{\ifmmode {\rm H}\alpha \else H$\alpha$\fi}
\newcommand{  \ha   	}{\ifmmode {\rm H}\alpha \else H$\alpha$\fi}
\newcommand{  \Hbeta    }{\ifmmode {\rm H}\beta \else H$\beta$\fi}
\newcommand{  \hb    	}{\ifmmode {\rm H}\beta \else H$\beta$\fi}
\newcommand{  \Hgamma   }{\ifmmode {\rm H}\gamma \else H$\gamma$\fi}
\newcommand{  \Hdelta   }{\ifmmode {\rm H}\delta \else H$\delta$\fi}
\newcommand{  \Lya      }{\ifmmode {\rm Ly}\alpha \else Ly$\alpha$\fi}
\newcommand{  \Lyb      }{\ifmmode {\rm Ly}\beta \else Ly$\beta$\fi}
\newcommand{  \Pa       }{\ifmmode {\rm P}\alpha \else P$\alpha$\fi}
\newcommand{  \Pb       }{\ifmmode {\rm P}\beta \else P$\beta$\fi}
\newcommand{  \Bra      }{\ifmmode {\rm Br}\alpha \else Br$\alpha$\fi}
\newcommand{  \Brg      }{\ifmmode {\rm Br}\gamma \else Br$\gamma$\fi}
\newcommand{  \hii      }{\ifmmode {\rm H}\,\textsc{ii} \else H\,\textsc{ii}\fi}
\newcommand{  \hei      }{\ifmmode {\rm He}\,\textsc{i} \else He\,\textsc{i}\fi}
\newcommand{  \heii     }{\ifmmode {\rm He}\,\textsc{ii} \else He\,\textsc{ii}\fi}
\newcommand{  \HeIIuv   }{\ifmmode {\rm He}\,\textsc{ii}\,\lambda1640 \else He\,\textsc{ii}\,$\lambda1640$\fi}
\newcommand{  \HeIIop   }{\ifmmode {\rm He}\,\textsc{ii}\,\lambda4686 \else He\,\textsc{ii}\,$\lambda4686$\fi}
\newcommand{  \cii      }{\ifmmode {\rm C}\,\textsc{ii}  \else C\,\textsc{ii}\fi}
\newcommand{  \ciii     }{\ifmmode {\rm C}\,\textsc{iii}\right] \else C\,\textsc{iii}]\fi}
\newcommand{  \CIII     }{\ifmmode {\rm C}\,\textsc{iii}\right]\,\lambda1909 \else C\,\textsc{iii}]\,$\lambda1909$\fi}
\newcommand{  \civ      }{\ifmmode {\rm C}\,\textsc{iv}  \else C\,\textsc{iv}\fi}
\newcommand{  \CIV      }{\ifmmode {\rm C}\,\textsc{iv}\,\lambda1549 \else C\,\textsc{iv}\,$\lambda1549$\fi}
\newcommand{  \nii      }{\ifmmode [{\rm N}\,\textsc{ii}]  \else [N\,\textsc{ii}]\fi}
\newcommand{  \niii     }{\ifmmode {\rm N}\,\textsc{iii} \else N\,\textsc{iii}\fi}
\newcommand{  \niv      }{\ifmmode {\rm N}\,\textsc{iv}  \else N\,\textsc{iv}\fi}
\newcommand{  \NIVuv    }{\ifmmode {\rm N}\,\textsc{iv}\,\lambda1486 \else N\,\textsc{iv}\,$\lambda1486$\fi}
\newcommand{  \nv       }{\ifmmode {\rm N}\,\textsc{v}   \else N\,\textsc{v}\fi}
\newcommand{\oi}{\ifmmode \left[{\rm O}\,\textsc{i}\right] \else [O\,{\sc i}]\fi}
\newcommand{\OI}{\ifmmode \left[{\rm O}\,\textsc{i}\right]\,\lambda6300 \else [O\,{\sc i}]$\,\lambda6300$\fi}

\newcommand{\oii}{\ifmmode \left[{\rm O}\,\textsc{ii}\right] \else [O\,{\sc ii}]\fi}
\newcommand{\OII}{\ifmmode \left[{\rm O}\,\textsc{ii}\right]\,\lambda3727 \else [O\,{\sc ii}]\,$\lambda3727$\fi}
\newcommand{\oiii}{\ifmmode \left[{\rm O}\,\textsc{iii}\right] \else [O\,{\sc iii}]\fi}
\newcommand{\OIII}{\ifmmode \left[{\rm O}\,\textsc{iii}\right]\,\lambda5007 \else [O\,{\sc iii}]\,$\lambda5007$\fi}

\newcommand{\NII}{\ifmmode \left[{\rm N}\,\textsc{ii}\right]\,\lambda6583 \else [N\,{\sc ii}]$\,\lambda6583$\fi}

\newcommand{\NeIII}{\ifmmode \left[{\rm Ne}\,\textsc{iii}\right]\,\lambda3968 \else [Ne\,{\sc iii}]$\,\lambda3968$\fi}

\newcommand{\NeV}{\ifmmode \left[{\rm Ne}\,\textsc{v}\right]\,\lambda3426 \else [Ne\,{\sc v}]$\,\lambda3426$\fi}

\newcommand{\HeII}{\ifmmode {\rm He}\,\textsc{ii}\,\lambda4686 \else He\,{\sc ii}$\,\lambda4686$\fi}

\newcommand{\sii}{\ifmmode \left[{\rm S}\,\textsc{ii}\right] \else [S\,{\sc ii}]\fi}

\newcommand{\SII}{\ifmmode \left[{\rm S}\,\textsc{ii}\right]\,\lambda6717,6731 \else [S\,{\sc ii}]$\,\lambda6717,6731$\fi}

\newcommand{  \OIIIuv   }{\ifmmode {\rm O}\,\textsc{iii}\,\lambda1663 \else O\,\textsc{iii}\,$\lambda1663$\fi}
\newcommand{  \oiv      }{\ifmmode {\rm O}\,\textsc{iv}  \else O\,\textsc{iv}\fi}
\newcommand{  \OIVuv    }{\ifmmode {\rm O}\,\textsc{iv}\,\lambda1402  \else O\,\textsc{iv}\,$\lambda1402$\fi}
\newcommand{  \OIVIR    }{\ifmmode {\rm O}\,\textsc{iv}\,25.9\,\mu {\rm m} \else O\,\textsc{iv}\,$25.9\,\mu$m\fi}
\newcommand{  \ovi      }{\ifmmode {\rm O}\,\textsc{vi}   \else O\,\textsc{vi}\fi}
\newcommand{  \Ovi      }{\ifmmode {\rm O}\,\textsc{vi}\,\lambda1035 \else O\,\textsc{vi}\,$\lambda1035$\fi}
\newcommand{  \nei      }{\ifmmode {\rm Ne}\,\textsc{i}   \else Ne\,\textsc{i}\fi}
\newcommand{  \neii     }{\ifmmode {\rm Ne}\,\textsc{ii}  \else Ne\,\textsc{ii}\fi}
\newcommand{  \NeiiIR   }{\ifmmode {\rm Ne}\,\textsc{ii}\,12.8\,\mu {\rm m} \else Ne\,\textsc{ii}\,$12.8\,\mu$m\fi}
\newcommand{  \neiii    }{\ifmmode {\rm Ne}\,\textsc{iii} \else Ne\,\textsc{iii}\fi}
\newcommand{  \neiv     }{\ifmmode {\rm Ne}\,\textsc{iv}  \else Ne\,\textsc{iv}\fi}
\newcommand{  \nev      }{\ifmmode {\rm Ne}\,\textsc{v}   \else Ne\,\textsc{v}\fi}
\newcommand{  \NevIR    }{\ifmmode {\rm Ne}\,\textsc{v}\,24.3\,\mu {\rm m} \else Ne\,\textsc{v}\,$24.3\,\mu$m\fi}
\newcommand{  \nevi     }{\ifmmode {\rm Ne}\,\textsc{vi}  \else Ne\,\textsc{vi}\fi}
\newcommand{  \mgi      }{\ifmmode {\rm Mg}\,\textsc{i}   \else Mg\,\textsc{i}\fi}
\newcommand{  \mgii     }{\ifmmode {\rm Mg}\,\textsc{ii}  \else Mg\,\textsc{ii}\fi}
\newcommand{  \MgII     }{\ifmmode {\rm Mg}\,\textsc{ii}\,\lambda2798 \else Mg\,\textsc{ii}\,$\lambda2798$\fi}
\newcommand{  \siii     }{\ifmmode {\rm S}\,\textsc{iii} \else S\,\textsc{iii}\fi}
\newcommand{  \siv      }{\ifmmode {\rm S}\,\textsc{iv}  \else S\,\textsc{iv}\fi}
\newcommand{  \sili     }{\ifmmode {\rm Si}\,\textsc{i}   \else Si\,\textsc{i}\fi}
\newcommand{  \silii    }{\ifmmode {\rm Si}\,\textsc{ii}  \else Si\,\textsc{ii}\fi}
\newcommand{  \Siliv    }{\ifmmode {\rm Si}\,\textsc{iv}  \else Si\,\textsc{iv}\fi}
\newcommand{  \SilIVuv  }{\ifmmode {\rm Si}\,\textsc{iv}\,\lambda1400  \else Si\,\textsc{iv}\,$\lambda1400$\fi}

\newcommand{  \caii     }{\ifmmode {\rm Ca}\,\textsc{ii}   \else Ca\,\textsc{ii}\fi}
\newcommand{  \feii     }{\ifmmode {\rm Fe}\,\textsc{ii}  \else Fe\,\textsc{ii}\fi}
\newcommand{  \feiii    }{\ifmmode {\rm Fe}\,\textsc{iii} \else Fe\,\textsc{iii}\fi}

\newcommand{ \Lhb   }{\ifmmode L\left(\hb\right) \else $L\left(\hb\right)$\fi}
\newcommand{ \fwhb  }{\ifmmode {\rm FWHM}\left(\hb\right) \else FWHM(\hb)\fi}
\newcommand{ \Lha   }{\ifmmode L\left(\ha\right) \else $L\left(\ha\right)$\fi}
\newcommand{ \fwha  }{\ifmmode {\rm FWHM}\left(\ha\right) \else FWHM(\ha)\fi}
\newcommand{ \Lmg   }{\ifmmode L\left(\mgii\right) \else $L\left(\mgii\right)$\fi}
\newcommand{ \fwmg  }{\ifmmode {\rm FWHM}\left(\mgii\right) \else FWHM(\mgii)\fi}
\newcommand{ \Lciv  }{\ifmmode L\left(\civ\right) \else $L\left(\civ\right)$\fi}
\newcommand{ \fwciv }{\ifmmode {\rm FWHM}\left(\civ\right) \else FWHM(\civ)\fi}
\newcommand{ \fwhm  }{\ifmmode {\rm FWHM} \else FWHM\fi} 
\newcommand{ \voff  }{\ifmmode v_{\rm off} \else $v_{\rm off}$\fi} 

\newcommand{ \mumg  }{\ifmmode \mu\left(\mgii\right) \else $\mu\left(\mgii\right)$\fi}
\newcommand{ \fmg   }{\ifmmode f\left(\mgii\right) \else $f\left(\mgii\right)$\fi}
\newcommand{ \muciv }{\ifmmode \mu\left(\civ\right) \else $\mu\left(\civ\right)$\fi}
\newcommand{ \fciv  }{\ifmmode f\left(\civ\right) \else $f\left(\civ\right)$\fi}
\newcommand{  \auvo     }{\ifmmode \alpha_{\nu,{\rm UVO}} \else $\alpha_{\nu,{\rm UVO}}$\fi}
\newcommand{  \Ledd     }{\ifmmode L_{\rm Edd} \else $L_{\rm Edd}$\fi}
\newcommand{  \lamLlam  }{\ifmmode \lambda L_{\lambda} \else $\lambda L_{\lambda}$\fi}
\newcommand{  \lLl      }{\ifmmode \lambda L_{\lambda} \else $\lambda L_{\lambda}$\fi}
\newcommand{  \nuLnu    }{\ifmmode \nu L_{\nu} \else $\nu L_{\nu}$\fi}
\newcommand{  \nLn      }{\ifmmode \nu L_{\nu} \else $\nu L_{\nu}$\fi}
\newcommand{  \Luv      }{\ifmmode L_{1450} \else $L_{1450}$\fi}
\newcommand{  \Lop      }{\ifmmode L_{5100} \else $L_{5100}$\fi}
\newcommand{  \lLop     }{\ifmmode \log\left(\Lop/\ergs\right) \else $\log\left(\Lop/\ergs\right)$\fi}
\newcommand{  \Lthree   }{\ifmmode L_{3000} \else $L_{3000}$\fi}
\newcommand{  \lLthree  }{\ifmmode \log\left(\Lthree/\ergs\right) \else $\log\left(\Lthree/\ergs\right)$\fi}

\newcommand{\Fthree}{\ifmmode F_{3000} \else $F_{3000}$\fi}
\newcommand{\fuv}{\ifmmode f_{\lambda}\left(1450{\rm \AA}\right) \else $f_{\lambda}\left(1450 {\rm \AA}\right)$\fi}
\newcommand{\fthree}{\ifmmode f_{\lambda}\left(3000{\rm \AA}\right) \else $f_{\lambda}\left(3000{\rm \AA}\right)$\fi}
\newcommand{\fH}{\ifmmode f_{\lambda}\left(1.65\micron\right) \else
$f_{\lambda}\left(1.65\micron\right)$\fi}

\newcommand{\fbol}{\ifmmode f_{\rm bol} \else $f_{\rm bol}$\fi}
\newcommand{\fbolwv}{\ifmmode f_{\rm bol}\left(\lambda\right) \else $f_{\rm bol}\left(\lambda\right)$\fi}
\newcommand{\fbolopt}{\ifmmode f_{\rm bol}\left(5100{\rm \AA}\right) \else $f_{\rm bol}\left(5100{\rm \AA}\right)$\fi}
\newcommand{\fbolthree}{\ifmmode f_{\rm bol}\left(3000{\rm \AA}\right) \else $f_{\rm bol}\left(3000{\rm \AA}\right)$\fi}
\newcommand{\fboluv}{\ifmmode f_{\rm bol}\left(1450{\rm \AA}\right) \else $f_{\rm bol}\left(1450{\rm \AA}\right)$\fi}

\newcommand{  \mbh      }{\ifmmode M_{\rm BH} \else $M_{\rm BH}$\fi}
\newcommand{  \lmbh     }{\ifmmode \log\left(\mbh/\Msun\right) \else $\log\left(\mbh/\Msun\right)$\fi} 
\newcommand{  \lledd    }{\ifmmode L/L_{\rm Edd} \else $L/L_{\rm Edd}$\fi}
\newcommand{  \Lbol     }{\ifmmode L_{\rm bol} \else $L_{\rm bol}$\fi}
\newcommand{  \lbol     }{\ifmmode L_{\rm bol} \else $L_{\rm bol}$\fi}
\newcommand{  \lLbol    }{\ifmmode \log\left(\Lbol/\ergs\right) \else $\log\left(\Lbol/\ergs\right)$\fi} 
\newcommand{  \Lagn     }{\ifmmode L_{\rm AGN} \else $L_{\rm AGN}$\fi}
\newcommand{  \lagn     }{\ifmmode L_{\rm AGN} \else $L_{\rm AGN}$\fi}

\newcommand{  \tgrow     }{\ifmmode t_{\rm growth} \else $t_{\rm growth}$\fi}
\newcommand{  \tUni      }{\ifmmode t_{\rm Universe} \else $t_{\rm Universe}$\fi}

\newcommand{  \Mindot	}{\ifmmode \dot{M}_{\rm infall} \else $\dot{M}_{\rm infall}$\fi}
\newcommand{  \Mbhdot	}{\ifmmode \dot{M}_{\rm BH} \else $\dot{M}_{\rm BH}$\fi}
\newcommand{  \Maddot	}{\ifmmode \dot{M}_{\rm AD} \else $\dot{M}_{\rm AD}$\fi}

\newcommand{  \as	}{\ifmmode a_{\rm *} 		\else $a_{\rm *}$\fi}
\newcommand{  \avec	}{\ifmmode \vec{a}_{\rm *} 	\else $\vec{a}_{\rm *}$\fi}
\newcommand{  \re	}{\ifmmode \eta      	\else $\eta$\fi}

\newcommand{  \mseed    }{\ifmmode M_{\rm seed} \else $M_{\rm seed}$\fi}
\newcommand{  \mbul     }{\ifmmode M_{\rm Bulge} \else $M_{\rm Bulge}$\fi} 
\newcommand{  \mstar    }{\ifmmode M_{*} \else $M_{*}$\fi} 
\newcommand{  \mgal     }{\ifmmode M_{*} \else $M_{*}$\fi} 
\newcommand{  \mhost    }{\ifmmode M_{\rm Host} \else $M_{\rm Host}$\fi}
\newcommand{  \mm       }{\ifmmode M_{*}/M_{\rm BH} \else $M_{*}/M_{\rm BH}$\fi}
\newcommand{  \mmsmall  }{\ifmmode M_{\rm BH}/M_{*} \else $M_{\rm BH}/M_{*}$\fi}
\newcommand{  \mmlarge  }{\ifmmode M_{*}/M_{\rm BH} \else $M_{*}/M_{\rm BH}$\fi}
\newcommand{  \mmwp     }{\ifmmode \left(M_{*}/M_{\rm BH}\right) \else $\left(M_{*}/M_{\rm BH}\right)$\fi}
\newcommand{  \ml       }{\ifmmode M_{*}/L_{*} \else $M_{*}/L_{*}$\fi}
\newcommand{  \mlwp     }{\ifmmode \left(M_{*}/L\right) \else $\left(M_{*}/L\right)$\fi}
\newcommand{  \mlk      }{\ifmmode \left(M_{*}/L_{K}\right) \else $\left(M_{*}/L_{K}\right)$\fi}
\newcommand{  \sigs     }{\ifmmode \sigma_{*} \else $\sigma_{*}$\fi}
\newcommand{  \Reff     }{\ifmmode R_{\rm e} \else $R_{\rm e}$\fi}

\newcommand  {\RBLR}        {\hbox{$ {R_{\rm BLR}} $}}

\def \nustar {{\em NuSTAR }}

\def \swiftxrt {{\em Swift}/XRT\ }
\def \swiftxrtsh {{\em Swift}/XRT}
\def \swiftbat {{\em Swift}/BAT\ }
\def \swiftbatsh {{\em Swift}/BAT}
\def \chandra {{\em Chandra\ }}
\def \chandrash {{\em Chandra}}

\def \xmmsh{{\em XMM-Newton}}

\def \suzakush{{\em Suzaku}}

\def\kmps{\hbox{$\km\s^{-1}\,$}}

\newcommand{\bj}{\ifmmode b_{\rm J} \else $b_{\rm J}$\fi}

\newcommand{\iab}{\ifmmode i_{\rm AB} \else $i_{\rm AB}$\fi}

\newcommand{\jab}{\ifmmode J_{\rm AB} \else $J_{\rm AB}$\fi}
\newcommand{\hab}{\ifmmode H_{\rm AB} \else $H_{\rm AB}$\fi}
\newcommand{\kab}{\ifmmode K_{\rm AB} \else $K_{\rm AB}$\fi}

\newcommand{\jveg}{\ifmmode J_{\rm Vega} \else $J_{\rm Vega}$\fi}
\newcommand{\hveg}{\ifmmode H_{\rm Vega} \else $H_{\rm Vega}$\fi}
\newcommand{\kveg}{\ifmmode K_{\rm Vega} \else $K_{\rm Vega}$\fi}

\def\arcmin{\hbox{$^\prime$}}
\def\arcsec{\hbox{$^{\prime\prime}$}}

\newcommand{  \Chisq    }{\ifmmode \chi^{2} \else $\chi^{2}$}
\newcommand{  \nelec    }{\ifmmode n_{e} \else $n_{e}$\fi}     % electron density
\newcommand{  \nh       }{\ifmmode n_{H} \else $n_{H}$\fi}     % hydrogen density
\newcommand{  \Ncol     }{\ifmmode N_{col} \else $N_{col}$\fi} % column density
\newcommand{  \NH       }{\ifmmode N_{H} \else $N_{\rm H}$\fi}     % column density

\def\sun{\hbox{$\odot$}}
\def\arcmin{\hbox{$^\prime$}}
\def\arcsec{\hbox{$^{\prime\prime}$}}

\def\ion#1#2{#1$\;${\small\rm\@Roman{#2}}\relax}

\newcommand{\OIIIa}{\ifmmode \left[{\rm O}\,\textsc{iii}\right]\,\lambda4959 \else [O\,{\sc iii}]\,$\lambda4959$\fi}
\newcommand{\NIIa}{\ifmmode \left[{\rm N}\,\textsc{ii}\right]\,\lambda6548 \else [N\,{\sc ii}]\,$\lambda6548$\fi}
\newcommand{\SIIa}{\ifmmode \left[{\rm S}\,\textsc{ii}\right]\,\lambda6716 \else [S\,{\sc ii}]\,$\lambda6716$\fi}
\newcommand{\SIIb}{\ifmmode \left[{\rm S}\,\textsc{ii}\right]\,\lambda6732 \else [S\,{\sc ii}]\,$\lambda6731$\fi}
\newcommand{\NeVa}{\ifmmode \left[{\rm Ne}\,\textsc{v}\right]\,\lambda3346 \else [Ne\,{\sc v}]\,$\lambda3346$\fi}
\newcommand{\NeVb}{\ifmmode \left[{\rm Ne}\,\textsc{v}\right]\,\lambda3426 \else [Ne\,{\sc v}]\,$\lambda3426$\fi}
\newcommand{\NeIIIa}{\ifmmode \left[{\rm Ne}\,\textsc{iii}\right]\,\lambda3869 \else [Ne\,{\sc iii}]\,$\lambda3869$\fi}
\newcommand{\NeIIIb}{\ifmmode \left[{\rm Ne}\,\textsc{iii}\right]\,\lambda3968 \else [Ne\,{\sc iii}]\,$\lambda3968$\fi}
\newcommand{\Mgb}{\ifmmode \left{\rm Mg}\,\textsc{i}\right\,\lambda5175 \else Mg\,{\sc i}\,$\lambda5175$\fi}

\newcommand{\mgb}{\ifmmode \left{\rm Mg}\,\textsc{i}\right \else Mg\,{\sc i}\fi}

\newcommand{\Cahk}{\ifmmode \left[{\rm Ca H+K}\,\textsc{ii}\right\,\lambda3935,3968 \else Ca H+K$\,\lambda3935,3968$\fi}

\def\kmpssh{\hbox{$\km\s^{-1}$}}

\def\arcmin{{\mbox{$^{\prime}$}}}

\def\arcsec{{\mbox{$^{\prime \prime}$}}}

\def\ga{{\rm\thinspace gauss}}

\def\km{{\rm\thinspace km}}

\def\Lsun{\hbox{$\rm\thinspace L_{\odot}$}}

\def\pc{{\rm\thinspace pc}}

\def\s{{\rm\thinspace s}}

\newcommand{  \Ntotwithdup }{972} %Apr 10 mk
\newcommand{  \Ntot    }{641} %Apr 15 Kyuseok
\newcommand{  \Nmbh    }{473}%Kyuseok (from 472) Apr 15
\newcommand{  \Nmbhper    }{74\%}%updated by Kyuseok (from 73) Apr 15
 
\newcommand{  \Nvbroad    }{225}%Kyuseok (from 226) indicates decent measurements of Mbh (flag 1 or 2) from broad Hb (TN12), i.e., Nvbroad = Nflagonebenny + Nflagtwobenny Apr 15

\newcommand{  \Nxray    }{836}%updated by Kyuseok based on our G-sheet (general) (from 836) Apr 15
\newcommand{  \Ncomp}{77\%, \Ntot/\Nxray} % optical spectra divided by Nxray (confirmed by Kyuseok: 0.77 = 643/836) Apr 15

\newcommand{  \Nflagone  }{128}	% confirmed (Kyuseok, Apr 15)
\newcommand{  \Nflagtwo  }{73} % confirmed (Kyuseok, Apr 15)
\newcommand{  \Nfitppxf }{201} %updated by Kyuseok (from 200) Apr 15 ppxf fit
\newcommand{  \Nfitppxfper }{31.3\%} %updated by mk Apr 15 ppxf fit 201/642
\newcommand{  \Nfitppxfbroad }{13} %updated mk Apr 15, probably wrong osterbrock wrong
\newcommand{  \Nrever  }{39}
\newcommand{  \Nmegamaser }{8}
\newcommand{  \Nbroad  }{207} %updated by Kyuseok (from 206) Apr 15
\newcommand{  \Nnoisy  }{166} %updated by Kyuseok (from 156) Apr 15
\newcommand{  \Nhiz   }{19} % high-z updated mk Apr 13

\newcommand{  \NSDSS   }{142} % all telescopes updated Isabella Apr 10
\newcommand{  \NAAO    }{112} % updated mk Apr 13, from 115 since 3 are in high z.
\newcommand{  \NPalermo   }{62} % updated by Kyuseok (67: from 633, 2: from high-z, 67+2 = 69), updated by Isabella (BATID = 17 wrong counterpart) Removed 6 from Landi for referee.
\newcommand{  \NKPNO   }{52}
\newcommand{  \NSAAO   }{46}
\newcommand{  \NMasetti   }{37}
\newcommand{  \NPalomar}{35}
\newcommand{  \NGemini }{29}
\newcommand{  \NMarziani}{18}

\newcommand{  \NTorrealba}{14} % Update to 14 MK, note by Kyuseok(2: from 633, 12: from high-z.  Apr 13
\newcommand{  \NFAST   }{10}
\newcommand{  \NAPO    }{4}
\newcommand{  \NUH     }{5}
\newcommand{  \NLick   }{4}

\newcommand{\NBPTniisey  }{338} %updated by Kyuseok (from 289) Apr 13
\newcommand{\NBPTniiseyper  }{53\%} %updated by Kyuseok (from 52)

\newcommand{\NBPTniiagn }{61} %updated by Kyuseok (from 84) Apr 13
\newcommand{\NBPTniiagnper }{10\%}%updated by Kyuseok (from 15) Apr 13

\newcommand{\NBPTniilinerper }{4\%}

\newcommand{\NBPTsiisey  }{317}%updated by Kyuseok (from 279) Apr 13
\newcommand{\NBPTsiiseyper  }{50\%}

\newcommand{\NBPTniiweakline  }{65}
\newcommand{\NBPTniiweaklineper  }{10\%}% for NII how many have weak lines

\newcommand{\NBPToisey  }{242}%updated by Kyuseok (from 234) Apr 13
\newcommand{\NBPToiseyper  }{38\%}%updated by Kyuseok (from 42) Apr 13

\newcommand{\NBPTallthree  }{225}% updated by Kyuseok: Number of Seyfert classified by all three BPT diagnostics diagrams "at the same time": Apr 14
\newcommand{\NBPTallthreeper  }{35\%} % updated by Kyuseok: Apr 14
\newcommand{\NBPTallthreeagree  }{182}% updated by Kyuseok: Apr14 This is really the exclusion which is a Seyfert in all three diagrams.  I need to look this up.
\newcommand{\NBPTallthreeagreeper  }{81\%} % 
\newcommand{  \NSyoneclassper}{7\%} % 43/642, May 4
\newcommand{  \NSyonenine}{116} %Sy 1.9, May 4
\newcommand{  \NSyonenineweak   }{31} %Sy 1.9 with EW Halpha< 50, May 4
\newcommand{  \NSyonenineweakper}{27\%} %Sy 1.9 weak per, divided by total May 4
\newcommand{  \NSyonelownH    }{128} % Sy1 to 1.8 broad lines sources with nH<10^20 
\newcommand{  \NSyoneninelownH   }{36} % Sy 1.9 sources with nH<10^20

\newcommand{  \NSyonenhlimit   }{57\%, 128/223} % Sey 1-1.8, NH limit at 20 May 4, mk
\newcommand{  \Nnakedseytwo   }{6\%, 14/221} % Sey 2 with low NH
\newcommand{  \Nnhdivider   }{21.9} % dividing line where column density changes

\newcommand{  \NsuperEddper  }{13}
\newcommand{  \NSytwolowEdd  }{15}
\newcommand{  \NSyonelowEdd  }{13}

\def \swiftbat {{\em Swift} BAT\ }
\def \swiftbatsh {{\em Swift} BAT}
\newcommand {\nhunit} {cm$^{-2}$}

\newcommand {\ppxf}{{\sc pPxf}}

\shorttitle{BASS - I}
\shortauthors{Koss et al.}

\begin{document}
% %%%%%%%%%%%%%%%%%%%%%%%%%%%%%%%%%%%%%%%%%%%%%%%%%%%%%%%%%%%%%%%%%%%%%%%%%%%%%%%%%%%%%%%%%%%%%%%

\title{BAT AGN Spectroscopic Survey I: Spectral Measurements, Derived Quantities, and AGN Demographics}
\

\author{
Michael Koss\altaffilmark{1,2,16},
Benny Trakhtenbrot\altaffilmark{2,17},
Claudio Ricci\altaffilmark{3,4},
Isabella Lamperti\altaffilmark{1},
Kyuseok Oh\altaffilmark{1},
Simon Berney\altaffilmark{1},
Kevin Schawinski\altaffilmark{1},
Mislav Balokovi\'{c}\altaffilmark{5},
Linda Baronchelli\altaffilmark{1},
D. Michael Crenshaw\altaffilmark{6},
Travis Fischer\altaffilmark{6},
Neil Gehrels\altaffilmark{7,18},
Fiona Harrison\altaffilmark{5},
Yasuhiro Hashimoto\altaffilmark{8}, 
Drew Hogg\altaffilmark{9},
Kohei Ichikawa\altaffilmark{10},
Nicola Masetti\altaffilmark{11,12},
Richard Mushotzky\altaffilmark{9},
Daniel Stern\altaffilmark{13},
Ezequiel Treister\altaffilmark{3},
Yoshihiro Ueda\altaffilmark{14},
Sylvain Veilleux\altaffilmark{9},
Lisa Winter\altaffilmark{15}
}

\altaffiltext{1} 
{Institute for Astronomy, Department of Physics, ETH Zurich, Wolfgang-Pauli-Strasse 27, CH-8093 Zurich, Switzerland}
\altaffiltext{2} 
{Eureka Scientific, 2452 Delmer Street Suite 100, Oakland, CA 94602-3017, USA}
\altaffiltext{3} 
{Instituto de Astrof\'{\i}sica, Facultad de F\'{\i}sica, Pontificia Universidad Cat\'olica de Chile, Casilla 306, Santiago 22, Chile}

\altaffiltext{4}{Kavli Institute for Astronomy and Astrophysics, Peking University, Beijing 100871, China}
\altaffiltext{5} {Cahill Center for Astronomy and Astrophysics, California Institute of Technology, Pasadena, CA 91125, USA} 
\altaffiltext{6} {Department of Physics and Astronomy, Georgia State University, Astronomy Offices, One Park Place South SE, Suite 700, Atlanta, GA 30303, USA} 
\altaffiltext{7} {NASA Goddard Space Flight Center, Greenbelt, MD 20771, USA} 
\altaffiltext{8} {Department of Earth Sciences, National Taiwan Normal University, No. 88, Sec. 4, Tingzhou Rd., Wenshan District, Taipei
11677, Taiwan R.O.C} 
\altaffiltext{9} {Department of Astronomy and Joint Space-Science Institute, University of Maryland, College Park, MD 20742, USA}
\altaffiltext{10} {National Astronomical Observatory of Japan, 2-21-1 Osawa, Mitaka, Tokyo 181-8588, Japan}
\altaffiltext{11} {INAF-Istituto di Astrofisica Spaziale e Fisica Cosmica di Bologna, via Gobetti 101, 40129 Bologna, Italy}
\altaffiltext{12} {Departamento de Ciencias F\'{\i}sicas, Universidad Andr\'es Bello, Fern\'andez Concha 700, Las Condes, Santiago, Chile}
\altaffiltext{13} {Jet Propulsion Laboratory, California Institute of Technology, 4800 Oak Grove Drive, MS 169-224, Pasadena, CA 91109, USA} 
\altaffiltext{14} {Department of Astronomy, Kyoto University, Kyoto 606-8502, Japan} 
\altaffiltext{15} {Atmospheric and Environmental Research, 131 Hartwell Ave No. 4, Lexington, MA 02421, USA} 
\altaffiltext{16} {Ambizione fellow} 
\altaffiltext{17} {Zwicky fellow}
\altaffiltext{18}{Deceased}

\email{Mike.koss@eurekasci.com}

\begin{abstract}

We present the first catalog and data release of the \emph{Swift-BAT AGN Spectroscopic Survey} (BASS).  We analyze optical spectra of the majority of AGN (\Ncomp) detected based on their 14--195\,keV emission in the 70-month  \swiftbat all-sky catalog.  This includes redshift determination, absorption and emission line measurements, and black hole mass and accretion rate estimates for the majority of obscured and un-obscured AGN (\Nmbhper, \Nmbh/\Ntot) with 340 measured for the first time. With $\sim90\%$ of sources at $z<0.2$, the survey represents a significant advance in the census of hard-X-ray selected AGN in the local universe.   In this first catalog paper, we describe the spectroscopic observations and datasets, and our initial spectral analysis.    The FWHM of the emission lines show broad agreement with the X-ray obscuration ($\sim$94$\%$), such that Sy 1-1.8 have \NH$< 10^{21.9}$\,\nhunit, and Seyfert 2, have \NH$> 10^{21.9}$\,\nhunit.  Seyfert 1.9, however, show a range of column densities.   Compared to narrow line AGN in the SDSS, the X-ray selected AGN have a larger fraction of dusty host galaxies ($\ha/\hb>5$) suggesting these types of AGN are missed in optical surveys.  Using the \OIII/\Hbeta\ and \NII/\Halpha\ emission line diagnostic, about half of the sources are classified as Seyferts, $\sim15\%$ reside in dusty galaxies that lack an \Hbeta\ detection, but for which the upper limits on line emission imply either a Seyfert or LINER,  $\sim15\%$ are in galaxies with weak or no emission lines despite high quality spectra, and a few percent each are LINERS, composite galaxies, HII regions, or in known beamed AGN.
\end{abstract}
\keywords{galaxies: active --- galaxies: nuclei --- quasars: general --- black hole physics}

\section{Introduction}
\label{sec:intro}

A significant population of obscured active galactic nuclei (AGN) are expected from models and observations of the cosmic X-ray background (CXB) spectrum \citep[e.g.,][]{Comastri:1995:1, Treister:2009:110,Draper:2010:L99,Ueda:2014:104}. The dusty and molecular torus is thought to be responsible for this obscuration and is considered to be a region of 1--100\,pc size around the central accreting supermassive black hole (SMBH) with high column densities of $>$10$^{23}$\,cm$^{-2}$ that absorbs much of the soft X-ray ($<$10\,keV) to optical radiation from the central engine and reemits it in the infrared \citep[e.g.,][]{Pier:1992:99}. When the obscuring torus is blocking the line of sight, emission from the broad line region (BLR) is blocked as well. Even if the line of sight is blocked by the obscuring torus, some direct hard X-ray emission ($>$10\,keV) may still be visible because of the high penetration ability ($\approx$90\%, \NH$>5\approx 10^{23}$\,\nhunit) provided the line of sight column density is not heavily Compton-thick (\NH$>10^{25}$\,\nhunit).\\
    
    %The AGN unified model  attempts to explain all types of observed properties of AGN such as obscuration, as only differences in viewing angle (Antonucci 1993).  Thus, type 1 and type 2 Seyferts are thought to be the same objects, but seen from different orientations (Osterbrock 1978).  Support for the unification model has come from the fact that narrow line AGN have been found to have a BLR in polarized light that is hidden by obscuration but visible in polarized light (Antonucci 1985).  In this model, the essential parts of an AGN are a supermassive black hole (SMBH) with an accretion disk, obscuring torus,  and broad and narrow-line region gas.\\ Radio selection of AGN is also largely obscuration independent, though only $\sim 10$\% of AGN are radio loud \citep[e.g.,][]{Miller:1990:207,Stern:2000:1526}.

Nebular emission lines observed in optical spectra probe the physical state of the ionized gas in galaxies and thus can be used to trace the nuclear activity of, for example, a central SMBH or the instantaneous rate of star formation \citep{Osterbrock:1985:166}.
Emission-line ratios have been turned into powerful diagnostic tools, not just for individual galaxies, but for massive spectroscopic surveys. \citet{Baldwin:1981:5} first proposed the use of line diagnostic diagrams, which have subsequently been developed and refined in numerous studies \citep[e.g.][]{Veilleux:1987:295, Kewley:2001:37, Shirazi:2012:1043}.\\

While the narrow line region (NLR) provides a way to detect obscured AGN in optical surveys, a significant problem is the possible presence of dust in this region. Dust is thought to be destroyed in the BLR, but to extend throughout the NLR \cite[e.g.,][]{Mor:2009:298}. This dust will scatter and absorb radiation and substantially change the level of ionization, complicating AGN identification. An additional complication is that in some AGN, bursts of star formation can overwhelm the AGN photoionization signature \citep[e.g.,][]{Moran:2002:L71,Trump:2015:26}. Thus, the largest optical surveys that select AGN are often incomplete for nearby galaxies because of obscuration or difficulty in detecting lower-luminosity AGN in galaxies with significant star formation.\\

An all-sky survey in the ultra-hard X-ray band (14--195\,keV) provides an important new way to address several fundamental questions regarding black hole growth and AGN physics, using a complete sample of AGN.  
The Burst Alert Telescope (BAT) instrument on board the {\it Swift} satellite has surveyed the sky to unprecedented depth, increasing the all-sky sensitivity by a factor of $\approx$20 compared to previous satellites, such as {\it HEAO 1} \citep{Levine:1984:581}. 
This has raised the number of known hard X-ray sources by more than a factor of 20 to \Nxray\ AGN \citep{Baumgartner:2013:19}. 
The majority of the BAT AGN are nearby, with a median redshift of $z\simeq0.05$ (among the sources with previously known redshifts). 
This sample is particularly powerful since emission in the 14--195 keV BAT band is relatively undiminished up to obscuring columns of $>10^{24}$\,cm$^{-2}$. 
BAT is therefore sensitive to heavily obscured objects where even hard X-ray surveys (2--10\,keV) are severely reduced in sensitivity. 
As the brightest AGN in the sky above 10\,keV, BAT-detected AGN provide an important low redshift template as they have similar luminosities to AGN detected in deep, small area X-ray surveys which focus on higher redshift AGN \citep[$z>1$, see e.g.,][and references therein]{Brandt:2015:1}.
Finally, BAT-detected AGN are also of renewed interest because of a large \nustar\ snapshot program, targeting $>$200 obscured AGN from \swiftbat \citep[e.g.,][]{Balokovic:2014:111, Brightman:2015:41,Koss:2015:149, Koss:2016:85,Ricci:2016:5}.\\

Bright AGN at lower redshifts ($z<0.2$) offer the best opportunity for high sensitivity studies of the SMBH accretion rate which requires accurate measurements of the SMBH mass and AGN bolometric luminosity.  Nearby optical spectroscopic studies can provide estimates of the black hole mass through measurements of the velocity dispersions of either the BLR gas and/or the hosts' stars for the majority of AGN.  The emission in the BAT band provides an obscuration-free estimate of the bolometric luminosity. Hence the combination of optical spectroscopy and hard X-ray emission from the BAT band provides the accretion rate (in terms of the Eddington rate, \lledd) for a large sample of AGN. Black hole mass measurements using reverberation mapping \citep[e.g.,][]{Bentz:2009:199}, or OH megamesars \citep[e.g.,][]{Staveley-Smith:1992:725} offer more precise measurements of black hole masses; however, these techniques can only be applied to a small ($N<50$) AGN sample that lacks the uniform selection criteria to understand the AGN population as a whole. 
Thus, a study of the BAT AGN sample provides an excellent opportunity to study black hole growth in a large sample of uniformly selected nearby AGN.\\

Despite the improvement in sensitivity above 10 keV, previous studies of BAT AGN have been limited to relatively small samples ($N\approx10-100$ AGN). 
The initial 9-month BAT survey, which was based on observations from 2005-2006, was studied in two different follow-up programs \citep[e.g.,][]{Winter:2010:503,Ueda:2015:1}. 
Other BAT optical spectroscopic studies focused on smaller samples ($N=11-75$) of newly identified BAT counterparts, such as from the Palermo BAT catalogs \citep{Parisi:2009:1345,Parisi:2012:A101,Parisi:2014:A67}. 
Yet other studies focused on small samples of AGN identified with \textit{INTEGRAL} above 20\,keV, with some overlap with the BAT AGN  \citep[e.g.,][]{Masetti:2013:A120}. 
While these AGN studies made significant advances in counterpart identification, most did not provide measurements of black hole masses and none measured stellar velocity dispersions in obscured sources.  \\

The goal of this project, the \emph{Swift-BAT AGN Spectroscopic Survey}, or BASS, is to complete the first large ($N\simeq$500) sample of ultra-hard X-ray selected AGN with optical spectroscopy and measured black holes masses using the deep catalogs that detect many more faint AGN. 
This enables new insights into the nature and geometry of the obscuring torus and NLR; a measurement of black hole growth in a relatively complete sample of AGN; and serves as a low-redshift benchmark for deep X-ray surveys of distant AGN.  \\

In this first paper of a series, we define the AGN sample and provide the results of a set of measurements focusing on emission line diagnostics, black hole masses, and accretion rates. 
One of the early science results derived from our analysis---a comparison of X-ray to optical line emission---was presented in \citet{Berney:2015:3622}. 
In Section 2, we discuss the parent sample of \swiftbat selected AGN and the different optical telescopes, instrumental setups, and basic reduction procedures used to collect the spectroscopic dataset we use. 
We describe the host galaxy template fitting, emission line fitting, measurements of black hole masses and bolometric luminosity in Section 3. 
Finally, the overall AGN spectroscopic properties and initial scientific results and follow-up projects are described in Section 4.
Throughout this work, we use a cosmological model with $\Omega_{\Lambda}=0.7$, $\Omega_{\rm M}=0.3$, and $H_{0}=70$\,\kms\,Mpc$^{-1}$ to determine cosmological distances.
However, for the most nearby sources in our sample ($z<0.01$), we use the mean of the redshift independent distance in Mpc from the NASA/IPAC Extragalactic Database (NED), whenever available\footnote{The NASA/IPAC Extragalactic Database (NED) is operated by the Jet Propulsion Laboratory, California Institute of Technology, under contract with the National Aeronautics and Space Administration}.

\section{Parent Sample and Data}
\label{sec:samples_data}

In this section we discuss the parent X-ray AGN sample as well as the different databases and dedicated observations used to construct our spectroscopic dataset. 

% After 70 months\footnote{http://heasarc.gsfc.nasa.gov/docs/swift/results/bs70mon/}, the BAT survey has a sensitivity of  $1.03\times10^{-11}$\,erg\,cm$^{-2}$\,s$^{-1}$ over 50\% of the sky or $1.34\times10^{-11}$\,erg\,cm$^{-2}$\,s$^{-1}$ over 90\% of the sky. The BAT flux measurements are time averaged over the entire 70 months (spanning 2005-2012) of observations to enable detection of the faintest sources. 

\subsection{The 70-Month \swiftbat Catalog and X-ray Data}
\label{subsec:crossmatch}

The BAT survey is an all-sky survey in the ultra-hard X-ray range ($>10$\,keV) that, as of the first 70 months\footnote{http://heasarc.gsfc.nasa.gov/docs/swift/results/bs70mon/} of operation has identified 1210 objects \citep{Baumgartner:2013:19}. Because of the large positional uncertainty of BAT ($\approx2\arcmin$) higher angular resolution X-ray data for every source from {\it Swift}-XRT or archival data have been obtained, providing associations for 97\% of BAT sources. From this catalog, we exclude BAT sources lacking a soft X-ray counterpart, and those associated with galaxy clusters and Galactic sources. In addition, we exclude M82, which is a very nearby star forming galaxy detected by \swiftbat without the presence of an AGN.  This leaves \Nxray\ BAT-detected AGN.\\

Almost all of the host galaxy AGN counterparts used in this study (99\%, 632/641) are based on the published counterparts in \citet{Baumgartner:2013:19} paper which was based on \swiftxrtsh,  \xmmsh, \chandrash, \suzakush, and $ASCA$ followup.  In \citet{Baumgartner:2013:19} the host galaxy counterpart of the BAT AGN was determined using the brightest counterpart above 3 keV typically from \swiftxrtsh observations, within the BAT error radius ($\approx$3$\arcmin$.  In only nine cases we use updated galaxy counterparts based on subsequent published studies.  These differences are associated with the updated counterparts for SWIFTJ1448.7-4009 and SWIFTJ1747.7-2253 provided in \citet{Masetti:2013:A120}; SWIFTJ0634.7-7445, SWIFTJ0654.6+0700, and SWIFTJ2157.4--0615 found in \citet{Parisi:2014:A67}; SWIFTJ0632.8+6343, SWIFTJ0632.8+6343 and SWIFT J1238.6+0928 provided in the updated INTEGRAL catalog of \citet{Malizia:2016:19}, and finally SWIFTJ0350.1-5019, we use the counterpart from Ricci et al., submitted (ESO201-4).  Another issue is that some BAT AGN counterparts host dual AGN \citep{Koss:2016:L4}, however, the median value of the($L_{\rm \ 2-10 \ keV}^{\rm int}$) ratio between the dual AGNs is 11 \citep{Koss:2012:L22}, so the majority of the emission in the BAT detection is typically coming from a single AGN.  A full review of all of the AGN galaxy counterpart identifications and dual AGN is provided in Ricci et al., submitted.\\

We then cross-match this sample to the Roma Blazar Catalog (BZCAT) catalog \citep{Massaro:2009:691} and identify 11\% (96/\Nxray) of AGN as possibly beamed sources, such as blazars or flat spectrum radio quasars where Doppler boosting may amplify the non-thermal emission including the hard X-rays. As the Roma BZCAT authors note, classifications of beamed AGN in their catalog have significant uncertainty because only the very brightest AGN have the necessary polarimetric observations or detections of compact cores and superluminal motions using high resolution radio imaging combined with variability studies.  \\

All of the AGN in the BAT survey have been analyzed using X-ray observations spanning 0.3--195\,keV. This includes homogeneous model fitting using some of the best available soft X-ray data in the 0.3$-$10\,keV band from \xmmsh, \chandrash, \suzakush, or \swiftxrt and the 14--195\,keV band from \swiftbatsh. This analysis provides measurements of the obscuring column (\NH) and intrinsic X-ray emission ($L_{\rm \ 2-10 \ keV}^{\rm int}$).   Full details of the X-ray analysis and fitting measurements are provided in separate publications \citep[][Ricci et al., submitted.]{Ricci:2015:L13}.

\subsection{Optical Spectroscopic Data}

The goal of the BASS sample is to use the largest available optical spectroscopic sample of \swiftbat sources using dedicated observations and public archival data. In many cases, there were many duplicate observations of the same source.  In such cases, we select the single best spectrum for line measurements for each AGN in our catalog from a single telescope. 

\begin{figure}
\centering
\includegraphics[width=8.5cm]{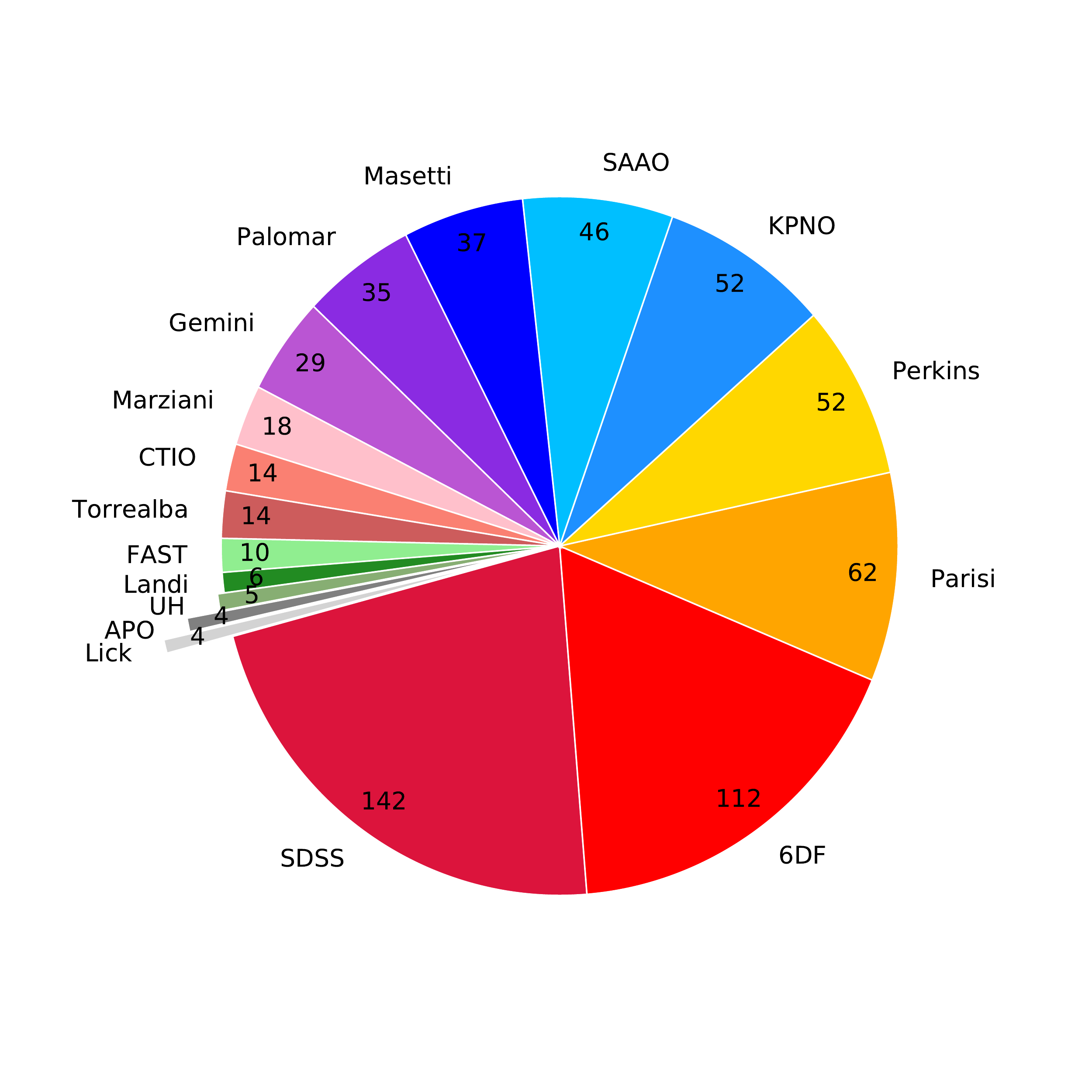}
\caption{Sources of \Ntot\ unique AGN spectra used for the BASS catalog taken from public surveys and targeted campaigns.}
\label{fig:pie_diff_tele}
\end{figure}

For line fitting, our first criterion was full coverage of the  spectral region covering all the lines from \Hbeta\ to \SII\ (i.e., 4800--7000\,\AA). We then selected based on signal-to-noise in the continuum for fitting of absorption lines and stellar population templates.  Each spectrum was visually inspected to avoid any cases that may be problematic for fitting (e.g., bad sky subtraction or noise spikes).   This process reduced the number of spectra from \Ntotwithdup\ spectra including duplicates to \Ntot\ unique spectra.  Of these \Ntot\ unique spectra, 33\% (209/641) were from targeted observations and 67\% (433/641) were part of archives or previously published papers.   A summary of all the observational setups is shown in Table~\ref{tab:observationssum} and Fig.~\ref{fig:pie_diff_tele}.  The main data products are available online \footnote{http://www.bass-survey.com}.\\

%For measurements of velocity dispersions where spectral resolution rather than wavelength coverage was critical a separate spectrum was sometimes used.

\begin{table*}
\begin{minipage}{\textwidth}
%\tabletypesize{\scriptsize}
\begin{center}
\caption{Summary of Instrumental Setups} 
\begin{tabular}{llllllll}
\tableline
\tableline
\multirow{2}{*}{Telescope}
& \multirow{2}{*}{Inst.}
& \multirow{2}{*}{Total}
& \multirow{2}{*}{Grating}
& Slit 
& Resolution
& Total range
& Velocity dispersion fit range \\
&
&
&
& Width (\arcsec)
& FWHM $[\AA$]
& $[\AA]$
& $[\AA]$

\\
\tableline
SDSS & SDSS & 142 & & 2,3 & 2.76 & 3900--8000 & 3900--7000 \\ 
UK Schmidt & 6dF & 112 & 600V,316R & 6.7& 5.75 & 3900--7000& 3900--7000 \\ 
Perkins 1.8m &Deveny& 52 &300&2& 5.43 & 3900--7500 & 3900--5500 \\ 
KPNO 2.1m & Goldcam & 36 &32&2& 7.4 & 4200--8800& 4400--5500  \\
& & 16 &26,35&2&3.3& 3900--8500& 3900--7000  \\  
SAAO 1.9m&300& 46 &7&2& 5.00 & 4000--7500 & 4600--7000  \\ 
Hale 200--inch & DBSP & 20 &600 &1.5& 4.4,5.8 & 3900--7000& 3900--5500, 8450--8700  \\
& & 15 &600 & 2 & 4.8,6.8 & 3900--7000& 3900--5500, 8450--8700  \\  
Gemini 8.1m &GMOS& 21 &B600& 1 & 4.84 & 4000--7000 &4000--7000 \\
&& 5 && 0.75 & 3.75 & 4300--7000 & 4300--7000  \\ 
&& 3 &&0.75& 3.72 & 4000--7000 &4000--7000 \\ 
%&& 1 && 1 & 5.00 & 3900--6500 & 3900--6500  \\ \tableline
CTIO 1.5m &R--C& 10 &26,35& 2 & 4.30 & 3900--7500 &3900--7000  \\ 
&& 2 &47& 2 & 3.10 & 5200--7500 & 5200--7000  \\ 
&& 2 &36& 2 & 2.20 & 4500--5500 &4500--5500  \\ 
Tillinghast	1.5m&FAST&	\NFAST & 300 & 3 & 5.5 & 3700--7500 & \\ 
UH 2.2m &SNIFS& \NUH&300&2.4&5.80 & 3200--7000 &3900--5500  \\ 
APO 3.5m & DIS & \NAPO & B400,R300 & 1.5 & 7.00 & 3900--7000 &3900--5600  \\ 
Shane 3m &Kast& \NLick&600,830&2& 4.0,3.2 & 3900--7000 & 3900--7000  \\ 

\tableline
\end{tabular}
\label{tab:observationssum}
\end{center}
\end{minipage}
\end{table*}

\subsubsection{Archival Public Data}
\label{subsubsec:sdss}

A large fraction of the data we use are drawn from several large public catalogs of optical spectra. 
Here we report on the number of best spectra that were used for each AGN in the catalog. The largest was from the Sloan Digital Sky Survey \cite[][]{York:2000:1579}, with \NSDSS\ sources from data release 12 \citep[DR12,][]{Alam:2015:12}. 
For \NAAO\ additional sources, the spectral measurements are based on archival optical spectra obtained as part of the final data release for the 6dF Galaxy Survey \citep[6dFGS,][]{Jones:2009:683}. The main characteristics of the 6dFGS survey are reported in \citet[][]{Jones:2004:747}. We note that, unlike the SDSS spectra, the 6dF spectra are not flux calibrated on a nightly basis and therefore we have not used them for broad line black hole mass measurements.  
\\

We also used publicly available spectra from smaller compilations of AGN, as long as they were flux and wavelength calibrated and the spectral resolution  was well determined. 
The spectral resolutions were measured based on the FWHM of sky lines in the spectra when possible. For \NSAAO\ sources, the spectral measurements are based on optical spectra obtained using the SAAO telescope, in an effort to study all of the earlier 9 month survey of BAT AGN \citep{Ueda:2015:1}. 
We used \NMasetti\ spectra from several optical spectroscopic studies of newly identified AGN from \textit{INTEGRAL} which overlap with the BAT sample  in this study \citep{Masetti:2004:L41,Masetti:2006:11,Masetti:2006:21, Masetti:2006:547,Masetti:2006:1139,Masetti:2008:113, Masetti:2008:121,Masetti:2010:A96, Masetti:2012:A123,Masetti:2013:A120}. 
For \NMarziani\ spectra, we used the low redshift AGN atlas of \citet{Marziani:2003:199}, which was obtained using several 2m-class telescopes. 
Optical spectra of \NTorrealba\ high redshift AGN in the BAT sample were from obtained through the Monitoring of Jets in Active Galactic Nuclei with VLBA Experiments (MOJAVE) project, targeting AGN selected at 2cm \citep{Torrealba:2012:9}. \NFAST\ sources are based on flux-calibrated optical spectra of broad line AGN observed with FAST \citep{Landt:2007:282}.  We also used spectra obtained to followup the Palermo BAT catalog, which has produced its own BAT AGN catalog, which has significant overlap with this sample \citep[$N=\NPalermo$;][]{Parisi:2009:1345,Parisi:2012:A101, Parisi:2014:A67}.  Finally, we use 6 sources from an early study by \citet{Landi:2007:109} from the first 3-month BAT catalog \citep{Markwardt:2005:L77} which are also detected in the 70-month BAT sample. \\

\subsubsection{Targeted Spectroscopic Observations}
\label{subsec:new_data}

The other source of spectra for our analysis are dedicated spectroscopic observational campaigns of BAT AGN, taken over the last several years, using a variety of telescopes and instruments. 
In terms of the data reduction and analysis, however, we have maintained a uniform approach. 
All the spectra were processed using the standard tasks in \texttt{IRAF} for cosmic ray removal, 1-d spectral extraction, wavelength, and flux calibrations. The spectra were all taken as longslit observations, except for observations taken with an IFU on the University of Hawaii (UH) 2.2m telescope. In all cases, the spectral resolutions listed were measured based on the FWHM of sky lines in the spectra or arc lines taken for wavelength calibration. The spectra were flux calibrated using standard stars, which were typically observed once per night.   
Basic information about these observations are given in Table~\ref{tab:observationssum}, and the following provides a more detailed account of these dedicated campaigns.
\\

We had several large, multi-year programs on small 1-2m telescopes. 
For \NKPNO\ sources, the spectral measurements were taken with the KPNO 2.1m telescope and the GoldCam spectrograph, through a 2\arcsec\ slit. We used two different setups for observing. The first set of observations was from \citet{Winter:2010:503} and used the grating 26new, which covers 3660--6140\,\AA\ on the blue side, and grating 35, which covers 4760--7240\,\AA\ on the red side. Both these setups had a spectral resolution of 3.3\,\AA. Additionally, a separate Goldcam program (PI M. Koss) used a single lower dispersion grating, grating 32, which covered a larger wavelength range than the higher dispersion gratings (4280--9220\,\AA), at a spectral resolution of 6.7\,\AA. There were also programs using the Perkins 1.8m telescope and DeVeny Spectrograph at the Lowell Observatory and CTIO 1.5m RC spectrograph (PI M. Crenshaw).\\
   
We also had optical spectroscopic programs using larger telescopes. 
For \NGemini\ sources, the spectral measurements were obtained with the Gemini North and South telescopes, using the twin Gemini Multi-Object Spectrograph (GMOS) instruments. The GMOS observations took place between 2009 and 2012, as part of nine different observing programs (P.I. M.\ Koss, E.\ Treister, and K.\ Schawinski). In this study we use data from Gemini programs GN-2009B-Q-114, GN-2010A-Q-35, GN-2011A-Q-81, GN-2011B-Q-96, GN-2012A-Q-28, GN-2012B-Q-25, GS-2010A-Q-54, and GS-2011B-Q80. 
Most of these programs focused on dual AGN, but covered the brighter BAT sources with the slit aligned with the secondary galaxy nucleus. We used two spectral setups for observations. The majority of targets were observed using the B600-G5307 grating with a 1$\arcsec$ slit in the 4300--7300\,\AA\ wavelength range, providing a spectral resolution of 4.8\,\AA. The Gemini/GMOS \texttt{IRAF} pipeline was used for wavelength calibration, spectro-spatial flat-fielding, cosmic ray removal, and flux calibration.  \\

We used the Palomar Double Spectrograph (DBSP) on the Hale 200-inch Hale telescope for \NPalomar\ targets. These AGN were observed as part of the \nustar BAT snapshot program, focusing mostly on Seyfert 2 AGN (P.I. F.\ Harrison and D.\ Stern). The observations were performed between 2012 October and 2015 February. The majority of observations were taken with the D55 dichroic and the 600/4000 and 316/7500 gratings using a 1.5\arcsec\ slit, providing resolutions of 4.4\,\AA\ and 5.8\,\AA, respectively. 
    
    %The typical seeing was 1\arcsec and signal-to-noise for these \NGemini\ spectra was $S/N\sim70$.  

Finally, we had smaller programs using the 3.5m Apache Point telescope (APO, \NAPO\ sources), the 3m Shane telescope at the Lick observatory (\NLick\ sources), and the UH 2.2m telescope (\NUH\ sources). 
The APO observations used the B400 and R300 gratings with a 1.5\arcsec\ slit, providing wavelength coverage from 3570--6230\,\AA\ in the blue and 5190--9810\,\AA\ in the red and resolution of 7\,\AA. The Lick observatory Kast spectrograph was used with blue and red coverage between 3900--7000\,\AA\ and spectral resolution of 4\,\AA.  On the UH telescope, we used the SuperNova Integral Field Spectrograph (SNIFS). SNIFS has a blue (3000--5200\,\AA) and red (5200--9500\,\AA) channel, with a resolution of 5.8\,\AA\ in the blue and 8.0\,\AA\ in the red. The SNIFS reduction pipeline, {\tt SNURP}, was used for wavelength calibration, spectro-spatial flat-fielding, cosmic ray removal, and flux calibration \citep{Bacon:2001:23,Aldering:2006:510}. A sky image was taken after each source image and subtracted from each Integral Field Unit (IFU) observation. The extraction aperture was 2.4\arcsec\ in diameter.
%The typical seeing was 1\arcsec and signal-to-noise for these \NUH\ spectra was $S/N\sim30$.

\section{Spectroscopic Measurements}
\label{sec:spec_analysis}

We performed three separate sets of spectral measurements with the \Ntot\ BASS spectra. The general properties of each AGN and the details of the optical spectra are presented in Table~\ref{tab:general_info}.
In the first step, each AGN host galaxy was fit using galaxy stellar templates (Section \ref{subsec:ppxf}) and the velocity dispersion was measured when possible (Section \ref{subsec:ppxf_veldisp}). 
The emission lines were then fit (Section \ref{subsec:narrow_lines}) using narrow components and broad components when needed.  
Black hole masses in AGN with broad emission lines were measured by a more detailed fit to the spectral regions that include the broad \Hbeta\ and/or broad \Halpha\ lines, or the \MgII\ and \CIV\ lines in high redshift sources (Section \ref{subsec:bhmass_hbeta}). 
Finally, we estimated the bolometric luminosity (\Lbol) from the X-ray luminosity to estimate the accretion rates (Section \ref{subsubsec:fbol}).

\subsection{Galaxy Template Fitting}
\label{subsec:ppxf}

We use the penalized PiXel Fitting software \citep[{\tt pPXF};][]{Cappellari:2004:138} to measure stellar kinematics and the central stellar velocity dispersion (\sigs). This method operates in pixel space and uses a maximum penalized likelihood approach for deriving the line-of-sight velocity distribution (LOSVD) from kinematic data \citep[][]{Merritt:1997:228}. As a first step, the \ppxf\ code (version 5.1.9) creates a model galaxy spectrum $G_{\rm mod}\left(x\right)$ by convolving empirical stellar population models by a parametrized LOSVD. Then it determines the best-fitting parameters of the LOSVD by minimizing the value of $\chi ^2$, which measures the agreement between the model and the observed galaxy spectrum over the set of reliable data pixels used in the fitting process.  Finally, \ppxf\ uses the ``best fit spectra'' to calculate the velocity dispersion and associated uncertainty from the absorption lines.\\

The \ppxf\ code uses a large set of single stellar populations to fit each galaxy spectrum.  We used the templates from the Miles Indo-U.S. Catalog (MIUSCAT) library of stellar spectra \citep{Vazdekis:2012:157}. The MIUSCAT library of stellar spectra contains $\approx$1200 well-calibrated stars covering the spectral region of 3525--9469\,\AA\ at a spectral resolution of 2.51\,\AA\ (FWHM).  These spectra are then computed into stellar libraries with an IMF slope of 1.3, and the full range of metallicities ($M/H=-2.27$ to $+0.40$) and ages (0.03--14 Gyr).  These templates have been observed at higher spectral resolution (FWHM=2.51\,\AA) than the AGN observations and are convolved in \ppxf\ to the spectral resolution of each observation before fitting.  \\  
% The MIUSCAT stellar populations models are based on the Indo-U.S., CaT, and MILES empirical stellar libraries. 
%In Table~\ref{tab:observationssum}, we list the spectral resolution values and wavelength ranges used for fitting the stellar component.

  We fitted the spectra in the wavelength region 3900--7000\,\AA, if this entire range was covered by the given spectrum. This range covers the \Cahk\ and \Mgb\ absorption features. For some spectra the velocity dispersion was estimated only based on the \mgb\ absorption line, because of a lack of blue wavelength coverage (e.g., spectra taken with KPNO and grating 32, Perkins, and some Gemini gratings). To avoid complications with discontinuities and dispersion changes in spectra which have both blue and red setups, only the blue channel spectra (3800--5500\,\AA) were used to estimate velocity dispersion, focusing on the Ca H+K and \mgb\ absorption lines. The only exception is the dual-channel SDSS data, where the full range was fit. Whenever available, we also fitted the \caii\ triplet spectral region (8450--8700\,\AA).\\

 %At 6863.00/(1+z)\,\AA\ there is a strong sky absorption line that can lead pPXF to overestimate the value of \sigs . Therefore for these spectra we decided to fit only until 5500\,\AA.\\

%We also de-redshifted the spectra to the rest-frame wavelength before calling pPXF. The reason is that, if the galaxy has a significant redshift value ($z >\ $0.03), pPXF will apply a large velocity shift to match the template to the galaxy spectrum. Consequently the initial value for the velocity would be large, and this can cause pPXF to stop. If the galaxy spectrum is brought to the rest-frame, this problem is solved. The maximum redshift value for the galaxies in our sample was 0.597, so we needed to de-redshift the spectra. To de-redshifted the spectra, we divided the wavelength $\lambda$ by (1+z) and we multiplied also the flux density by (1+z). In this way, the flux is conserved.

%To use the pPXF method we need to fix some parameters: the wavelength region of the emission lines masks and their width, the [OIII] velocity offset and the wavelength range to fit.\\   We follow the emission line masking procedure used in \citet{Oh:2011:13}.

We modified the {\tt ppxf-kinematics-example-sdss} code to measure stellar kinematics in our sample. Table \ref{tab:emlinesmask} shows the emission and bright night sky lines that were masked. The code automatically applies a mask for several bright emission lines: H$\delta$, H$\gamma$, H$\beta$, \OIII, \OI, \NII, H$\alpha$, and \SII.  We masked sky lines at $\lambda$5577, $\lambda$6300, $\lambda$6363, and $\lambda$6863. We also mask the region around the Ca H $\lambda$3968 line, because of overlap with the H$\epsilon$ $\lambda$3970 and the \NeIIIb\ emission lines \cite[see, e.g.,][]{Greene:2005:721}. 
We have also masked the region surrounding the Na\,{\sc I} line, since it may be affected by interstellar absorption. Finally, we masked regions affected by sky emission lines.
The width of these emission lines masks was set to 2400 \kms. For the broad-line AGN, we use a wider mask (3200 \kms) for the Balmer lines (H$\alpha$, H$\beta$, H$\gamma$, H$\delta$), in order to mask the broad emission components. This allowed us to measure the velocity dispersion for some AGN with broad emission in \Halpha, but narrow \Hbeta. \\

%The wavelength ranges affected by the sky emissions are given in the rest-frame, so when we de-redshifted the spectrum, we also divided the wavelength position of the sky emission lines by (1+z).

%Flag 1
% Number of AGN per flag

\subsubsection{Velocity Dispersion Measurements}
\label{subsec:ppxf_veldisp}

\begin{figure}
\centering
\includegraphics[width=8.5cm]{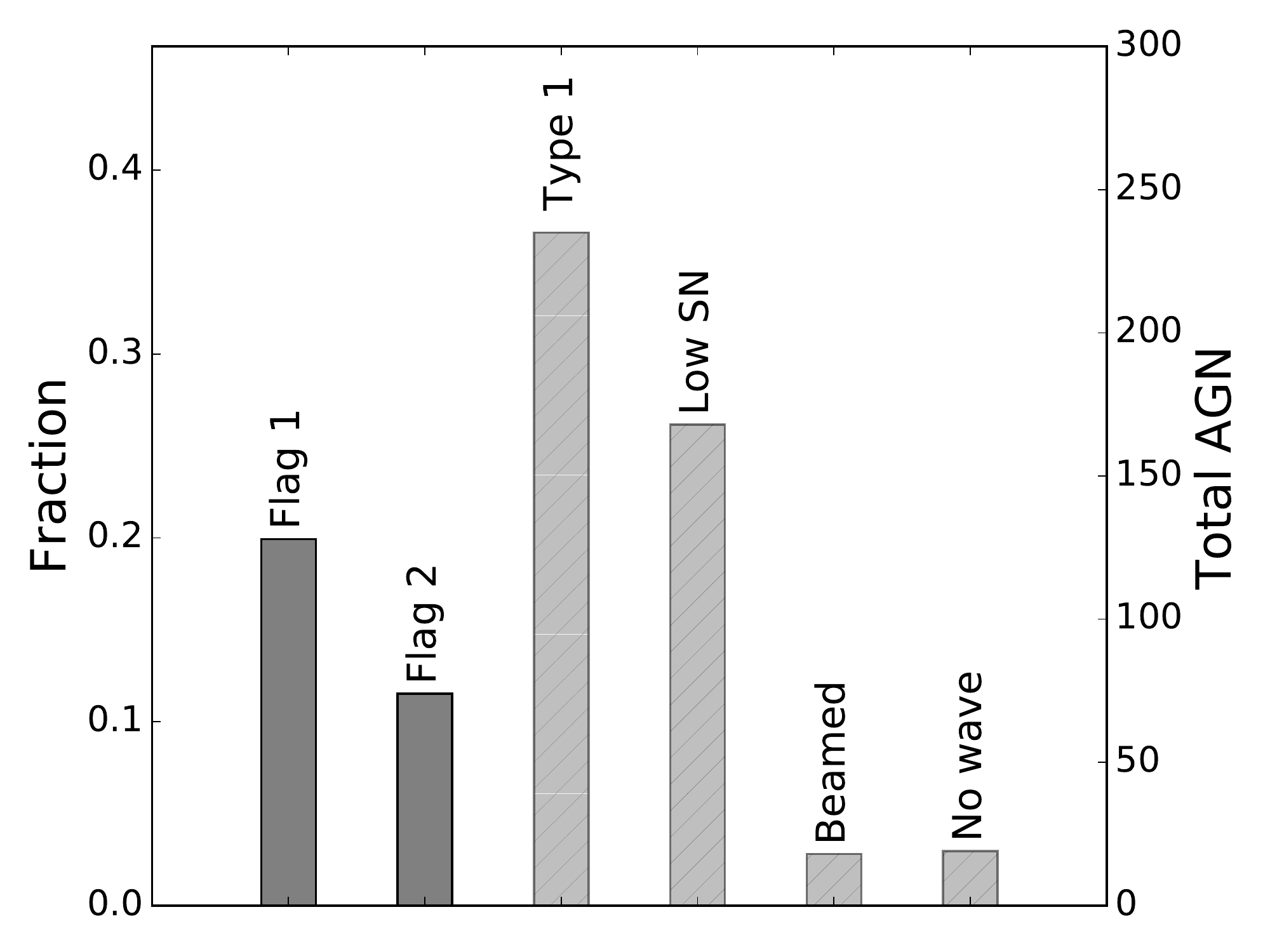}
\caption{Measurements of velocity dispersion from our sample using \ppxf.  For 31\% of spectra ($\Nfitppxf$/641), we have a robust measurement of \sigs. $\Nflagone$ spectra have a good fit and small error in the value of \sigs\ (flag 1). $\Nflagtwo$ spectra have worse fits, but the value of \sigs\  is sufficiently good based on visual inspection (flag 2).  The remaining categories (dashed histograms) have features that prevented measurements.  For $\Nbroad$ spectra we could not measure $\sigs$ because of AGN contamination to the stellar continuum. For $\Nnoisy$ spectra, the signal-to-noise level of the continuum or the grating resolution were not sufficient to measure the absorption lines. A small number of of sources were blazars with featureless continuum.  Finally, we also had a small number of sources with no coverage of key absorption line features such as Ca H+K, \mgb, or the \caii\ triplet regions.}
\label{fig:pieppxf}
\end{figure}

Fig.~\ref{fig:pieppxf} summarizes the results of the stellar velocity dispersion measurements and individual measurements are found in Table~\ref{tab:ppxf}.  $\Delta\sigma_*$ denotes the error in \sigs.  
We were able to achieve a reliable velocity dispersion measurement, with $\Delta\sigs <60\,\kms$, in $\Nfitppxf$ of the $\Ntot$ (\Nfitppxfper) galaxies in our sample. 
For these AGN, the stellar continuum was subtracted prior to emission line fitting (discussed below).
For most AGN with broad \Hbeta, the AGN continuum contaminated the host galaxy and stellar absorption lines, limiting reliable \sigs\ measurements to only \Nfitppxfbroad\ type 1.2--1.8 AGN.  Additionally, in $\Nnoisy$ sources the signal-to-noise ratio and spectral resolution were too low to robustly identify the absorption lines. \\

Four authors (MK, BT, SB, and IL) visually inspected the fit of the stellar continuum and absorption lines, and assigned a quality flag to each spectrum. 
In our sample we have $\Nflagone$ spectra with flag 1 and $\Nflagtwo$ spectra with flag 2, which both designate reliable measurements. More details and examples of these quality flags are given in the Appendix (Section \ref{sec:template}). 
Generally, the \sigs\ uncertainty measured by \ppxf\ for flag 1 fits is typically small ($\langle \Delta \sigs \rangle\simeq12\,\kmpssh$). 
Flag 2 fits have somewhat worse quality fits, judged from our visual inspection, consistent with \ppxf\ measurements ($\langle \Delta\sigs \rangle\simeq27\,\kmpssh$), but the Ca H+K and \mgb\ absorption lines are still well fit. \\

For AGN with reliable measurements of \sigs\ we calculated the black hole mass, \mbh, using the $\mbh-\sigs$ relation. 
We use the relation from \citet{Kormendy:2013:511}:
\begin{equation}
\log\left(\frac{\mbh}{\Msun}\right) = 4.38\times \log\left(\frac{\sigs}{200\,\kms}\right) + 8.49 \,\, .
\end{equation}
The slope of this relation is shallower than the slope of the relation from \citet{McConnell:2013:184}, who report a value of 5.64, and is consistent with the slope of the relation from \citet{Gultekin:2009:198}. 
A small number of sources have direct measurements of black hole masses, either from reverberation mapping (\Nrever) or OH megamasers (\Nmegamaser), which we have adopted and tabulated whenever available.

%\begin{equation}
%\log{\left(\frac{M}{M_\odot}\right)}=4.24\cdot \log{\left(\frac{\sigs}{200 \text{ km/s}}\right)} + 8.12
%\end{equation} 
%The differences depend on the properties of the black holes in the sample (redshift range, mass range, galaxies properties), on the size of the sample and on the method used to to measure M$_{BH}$ (\toref\ Rerrarese and Merrit (2000), G\"ltekin (2009), McConnell and Ma (2013), Kormendy and Ho (2013)).  

% Since this is an empirical relation, different studies have derived slightly different version of this relation.First we used the relation from McConnell and Ma (2013):
%\begin{equation}
%\log{\left(\frac{M}{M_\odot}\right)}=5.64\cdot \log{\left(\frac{\sigs}{200 \text{ km/s}}\right)} + 8.32
%\end{equation} 
%This relation is one of the most recent but the value of the slope (5.64) is quite high and therefore can lead to very high M$_{BH}$ for high values of \sigs. 

%pie-chart

\subsection{Emission Line Measurements}
\label{subsec:narrow_lines}

We fit emission lines in our sample of optical spectra using an extensive spectroscopic analysis toolkit for astronomy, {\tt PySpecKit}, which uses a Levenberg-Marquardt algorithm for spectral fitting \citep{Ginsburg:2011:1109.001}. All emission line fits were visually examined by five authors (MK, BT, SB, KS, and IL) to verify proper fitting, and to adjust some subtle parameters.  We implement separate methods for fitting sources with only narrow lines and for sources with broad lines. 
For narrow-line sources, we first fit and subtract a host stellar component, to remove the galaxy continuum and stellar absorption features, as described in Section \ref{subsec:ppxf}. 
We then separately fit three spectral regions, focusing on the \oii\ (3300--4000\,\AA), \Hbeta\ (4650--5050\,\AA), and \Halpha\ (6250--6770\,\AA) emission lines. 
All measurements for narrow line sources are listed in Table~\ref{tab:OII_complex}, Table~\ref{tab:Hb_complex}, and  Table~\ref{tab:Ha_complex} for the \OII, \Hbeta, and \Halpha\ regions measurements, respectively. The emission line classifications are provided in Table~\ref{tab:BPT}.For broad line sources, the properties are listed in Table~\ref{tab:broad} for both \Hbeta\ and \Halpha.\\

For 6DF spectra, the survey applied a single calibration to all spectra to convert measured counts to the correct spectral shape, but a nightly flux calibration was not applied.  For 6DF emission line measurements, we have calculated a rough flux calibration factor for 6DF spectra based on 12 overlapping spectra in distant AGN ($z>0.05$) with the SDSS.  This factor assumes 1 ct is equal to $7.69\times10^{-17} erg s^{-1} cm^{-2} AA^{-1}$.      \\

In each of the three spectral fitting regions, we adopt a power-law fit (1st order) to account for the (local) AGN continuum and a series of Gaussian components to model the emission lines. We define an emission line detection when we reach a $S/N>3$ over the line with respect to the noise of the adjacent continuum, and otherwise list upper detection limits.  To estimate errors in line fluxes and widths (in terms of FWHM), we use a simple re-sampling procedure that adds noise based on the error spectrum and reruns the fitting procedure 10 times. The (fractional) flux uncertainty for the \oiii\ emission line is typically less than 1\%.  We use the narrow Balmer line ratio (\ha/\hb) to correct for dust extinction, assuming an intrinsic ratio of $R_{\rm V}=3.1$ and the \citet{Cardelli:1989:245} reddening curve.   In the case of a \hb\ non-detection, we assume the 3$\sigma$ upper limits for the extinction correction.  When neither \hb\ or \ha\ is detected we present the fluxes as measured.  \\

For the \Hbeta\ spectral region, we fit the \HeIIop, \Hbeta, \OIIIa, and \OIII\ lines. The widths of the narrow lines are tied with an allowed variation of $\pm$500\,\kms. The central wavelength of the narrow line region is defined by a joint fit of all the narrow lines where the wavelength separation of all lines is tied.  The intensity of \OIIIa\ relative to \OIII\ is fixed at the theoretical value of 2.98 \citep{Storey:2000:813} and the intensity of \NIIa\ relative to \NII\ is set to the theoretical value of 2.96 \citep{Acker:1989:44}. 
For the \Hbeta\ complex, we use the 4660--4750\,\AA\ (except around \heii) and 5040--5200\,\AA\ regions for continuum determination.   
Within the \Halpha\ spectral region, we fit the \OI, \NIIa, \Halpha, \NII, and \SIIa, and \SIIb\ lines. Here too, the widths of the narrow lines are tied, with an allowed variation of $\pm$ 500\,\kms and the systemic redshift is determined from all narrow lines.  In the case of a non-detection of the narrow \Halpha\ or \SII\ line (i.e., due to a very strong broad \Halpha\ component or weak narrow emission lines), we use the FWHM of \OIII\ to constrain the widths of the narrow lines in the \Halpha\ region.  The relative strengths of \NIIa\ and \NII\ lines are fixed at 1:2.94. 
To estimate the continuum for the \Halpha\ complex, we use the wavelength regions 5800--6250\,\AA\ and 6750--7000\,\AA.  
Finally, within the \oii\ spectral region, we fit the \NeVa, \NeVb, \OII, \NeIIIa, and \NeIIIb\ lines. The continuum around the \oii\ spectral region is usually more complicated to fit due to a non-linear shape or because it lies near the blue edge of the wavelength coverage. To fit this blue continuum, we use the region between 3300\,\AA\ and 4000\,\AA\ except for small regions surrounding the emission lines themselves ($\pm$ 1000\,\kms).\\

For sources with broad \hb, we use the fitting procedure described in detail in \citet[see Appendix C1 therein]{Trakhtenbrot:2012:3081}, and here we provide only a brief description of the key spectral components. 
The AGN spectrum is first fitted with a linear (pseudo-) continuum, based on two narrow ($\pm$10\,\AA) continuum bands, typically around 4440\,\AA\ (or 4720\,\AA) and 5110\,\AA.  
Next, a broadened and shifted iron emission template \cite[][]{Boroson:1992:109} is fitted to, and subtracted from, the continuum-free spectrum. 
Then, we fit the remaining emission lines with a set of Gaussian profiles. In particular, the narrow components of \hb, [O\,{\sc iii}]\,$\lambda4959$ and [O\,{\sc iii}]\,$\lambda5007$ are fitted with a single Gaussian profile, while the broad components of \heii\ and \hb\ are described by two Gaussian components (each). As described in \cite{Trakhtenbrot:2012:3081}, the widths of the narrow components are tied among different emission lines, primarily to allow a robust decomposition of the narrow and broad components that make up the \hb\ emission line profile. The FWHM of the broad \hb\ line is measured from the (reconstructed) best-fit model of the broad component.\\

For sources with broad \ha, we use a fitting procedure that involves several progressively complicated steps, depending on the complexity of the emission lines. 
We start with allowing two Gaussian components for the \Halpha\ emission lines: a narrow component (with $\fwhm < 1000\,\kms$) and a broad component ($\fwhm > 1000\,\kms$), both with the wavelength centered at the \ha\ line. 
A visual inspection is made (MK, BT, KO, and IL) to determine whether a more complex fit is required, this time using multiple Gaussians that are also allowed to be shifted. If the fit quality is still poor, we use the width of the \oiii\ line (from the \hb\ complex fitting procedure) as an additional constraint on the narrow components in the \Halpha\ spectral region.  Examples of emission line fits are given in the Appendix.

\subsection{Broad Line Black Hole Mass Measurements}
\label{subsec:bhmass_hbeta}

%%%%%%%%%%%%%%%%%%%%%%%%%%%%%%%%%%%%%%%%%%%%%
%First, we subtracted from each spectrum the best-fit pPXF model, to remove the host contamination.}

\begin{figure}
\centering
\includegraphics[width=8.5cm]{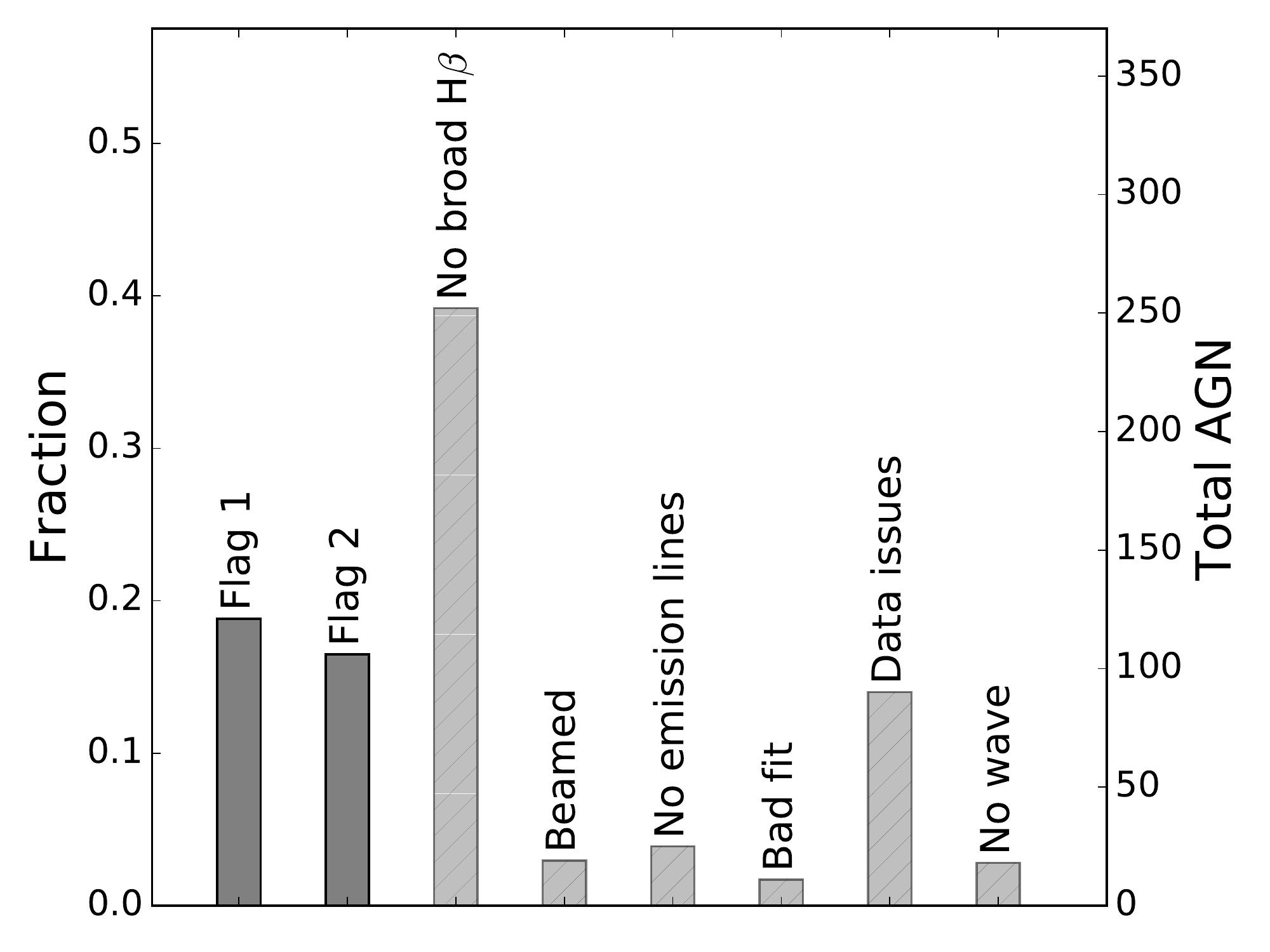}
\caption{Summary of broad line H$\beta$ black hole mass measurements using fitting procedure described by \citet{Trakhtenbrot:2012:3081}  which includes fitting Fe~II lines.   For 35\% ($\Nvbroad$/641), we have a measurement of black hole mass using broad \Hbeta\ lines.} ``Flag 1" represents an excellent fit based on visual inspection. ``Flag 2" spectra have a good fit based on visual inspection.  The remaining categories (dashed histograms) have features that prevented measurements.  ``Beamed AGN" are sources which are known to be beamed and lack emission lines.  ``No emission lines" indicates the galaxy has a high quality spectra but no Balmer lines are detected. ``No wave" indicates sources typically high redshift which lack coverage of \Hbeta.  Finally, in a small fraction the data quality was low or a very poor fit was obtained.    
\label{fig:pie_Hbeta}
\end{figure}

For sources with broad Balmer lines, we estimated the black hole masses (\mbh) through virial, ``single-epoch'' prescriptions, which are in turn based on the $\RBLR-L$ relation obtained through reverberation mapping of low-redshift AGN \citep[e.g.,][]{Kaspi:2000:631,Bentz:2006:133}.  The virial mass estimators we used for broad \Hbeta\ are known to suffer from systematic uncertainties of about 0.3 dex \cite[see, e.g.,][and references therein]{Shen:2013:61,Peterson:2014:253}.\\
%\citep[e.g.,]{
For sources with broad \hb\ we used the same prescription used in \cite{Trakhtenbrot:2012:3081}, which uses the continuum and line emission parameters for virial estimates of \mbh:
\begin{equation}
  \mbh\left(\hb\right)=1.05\times 10^8  
			\left(\frac{\Lop}{10^{46}\,\ergs}\right)^{0.65} 
			\left[\frac{\fwhb}{10^3\,\kms}\right]^2 \,\, \Msun \,\, ,
\label{eq:M_Hb}
\end{equation}
where \Lop\ is the the monochromatic luminosity at rest-frame 5100\,\AA, $\lambda L_{\lambda}$(5100\AA), measured from the best-fit model of the \hb\ region. As mentioned above, \fwhb\ is measured from the entire (best-fit) broad profile. Although the fitting procedure is executed automatically for the large data-set studied here, we note that we visually inspected all the \hb\ fitting results (including more than one spectrum per source), and applied minor manual adjustments to provide satisfactory fits to the data.  We also stress that we did not apply the \hb\ fitting code to spectra with poor absolute flux calibration (i.e., those from the 6DF survey).  A summary of these fits is found in Fig.~\ref{fig:pie_Hbeta}.\\

\begin{figure}
\centering
\includegraphics[width=8.5cm]{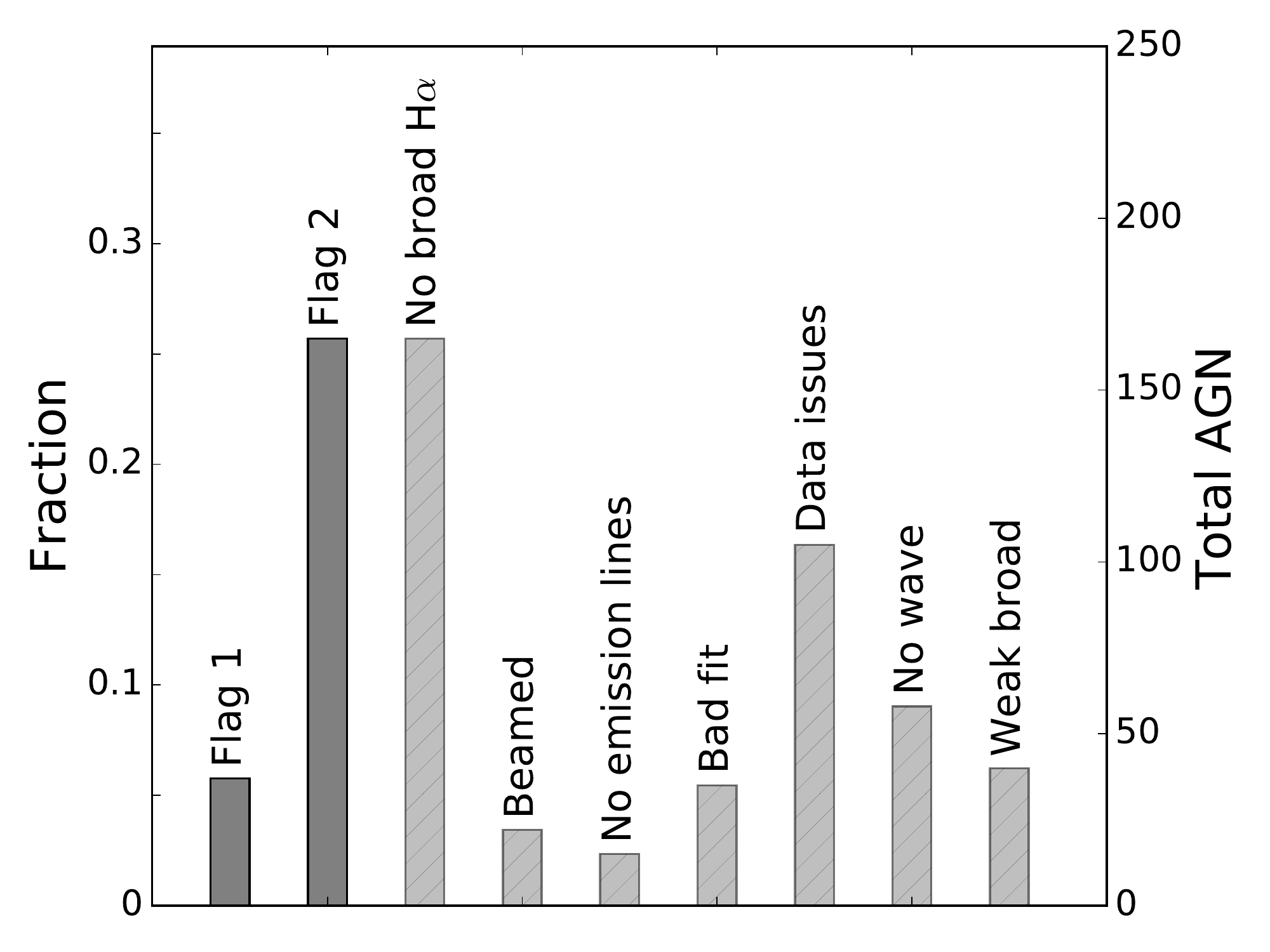}
\caption{Summary of broad line H$\alpha$ black hole mass measurements.  For 38\% (242/641), we have a measurement of black hole mass using broad \Halpha\ lines.  Categories are as in Fig. \ref{fig:pie_Hbeta}.  }
\label{fig:pie_Halpha}
\end{figure}

For sources with broad \Halpha\ lines we used the prescription of \cite{Greene:2005:122}:
\begin{equation}
  \mbh\left(\ha\right)=1.3\times 10^6  
			\left(\frac{L_{\ha}}{10^{42}\,\ergs}\right)^{0.57} 
			\left[\frac{\fwha}{10^3\,\kms}\right]^{2.06} \,\, \Msun \,\, ,
\label{eq:M_Ha}
\end{equation}
where $L_{\ha}$ is the integrated luminosity of the broad component of the \ha\ line, determined from the best-fitting model.
This prescription is therefore mostly unaffected by host light. 
Since the \Halpha-related prescription (Eq.~\ref{eq:M_Ha}) is based on a \emph{secondary} calibration of a $\RBLR-\Lha$ relation, it carries somewhat larger systematic uncertainties (compared with the \hb-based one). 
However, it can be applied to Seyfert 1.9 AGN without broad \Hbeta\ lines and it may perform better for sources that have high levels of stellar contamination and/or extinction.  A summary of the results of the fitting is provided in Fig.~\ref{fig:pie_Halpha}.  
We found about a quarter of Seyfert 1.9 (\NSyonenineweakper, \NSyonenineweak/\NSyonenine) have weak broad \Halpha\ lines (EW$<$50\,\AA; see general discussion of Seyfert sub-classes in Section~\ref{subsec:agn_types}). 
More details on these objects are given in the Appendix.

For \Nhiz\ high-redshift sources ($z\sim1-3.3$), the available optical spectra include either the \MgII\ or \CIV\ broad emission lines.
The emission complexes around these two lines were fitted using dedicated procedures, described in detail in \citet[see Appendices C2 and C3 therein]{Trakhtenbrot:2012:3081}. 
These take into account the (blended) emission features from iron and \HeIIuv\ transitions (in the case of \mgii\ and \civ, respectively).  For both broad lines, each of the doublet features is modeled with two broad Gaussians. We assume no narrow-line contribution to these transitions \cite[see][and references therein]{Trakhtenbrot:2012:3081}.

We used the best-fit models of the broad emission lines, together with the adjacent continuum luminosities, to estimate \mbh\ in these \Nhiz\ high-redshift sources.
For the \mgii\ line, we used the prescription presented by \cite{Trakhtenbrot:2012:3081}, which is calibrated against \hb-based mass estimates using a larger sample of SDSS quasars for which both lines are available. 
An identical prescription was also independently derived by \cite{Shen:2011:45}.  
For the \civ\ line, we used the prescription presented by \cite{Vestergaard:2006:689}. We note that \civ-based estimates of \mbh\ are known to be considerably less reliable \citep[e.g.,][]{Denney:2012:44} than those based on lower-ionization transitions, perhaps due to significant contribution from non-virialized BLR gas motion to the emission line profile \cite[see detailed discussion in][and references therein]{Trakhtenbrot:2012:3081}. 
We therefore advise that the \civ-based determinations of \mbh\ provided here may carry large uncertainties and, possibly, systematic biases. 
At such high redshifts, however, they provide the only estimate of \mbh, in lieu of NIR spectroscopy of the other emission lines.

The spectral and derived parameters of the \Nhiz\ high-redshift sources are listed in Table~\ref{tab:high_z}.  We finally note that the \swiftbat sample of AGN probably includes several other high-redshift (and therefore high-luminosity) sources, for which an optical spectrum is not available within the large optical surveys we use, and/or the effects of beaming are not yet well understood.  
We plan to address this population in a separate publication.
% end hi-z colored section 

%%%%%%%%%%%%%%%%%%%%%%%%%%%%%%%%%%%%%%%%%%%%%

\subsection{Bolometric Luminosity}
\label{subsubsec:fbol}

We estimated the bolometric luminosity of the AGN in our sample (\Lbol) from the observed X-ray luminosity measured by the \swiftbat\ survey in the energy range 14--195\,keV. First, we divide the 14--195\,keV luminosity by 2.67 to convert to the intrinsic 2--10\,keV luminosity, following \citet{Rigby:2009:1878} which is based on scaling the \citet{Marconi:2004:169} templates to higher X-ray energies.  We then used the median bolometric correction from \citet{Vasudevan:2009:1553} of the BAT sample which resulted in a factor of 8 between \Lbol\ and $L_{14-195\ \text{keV}}$.  More advanced luminosity-dependent bolometric corrections will be examined in future studies.

%We consider the relation between $L_{bol}$ and $L_{2-10\text{keV}}$ :
%\begin{equation}
%\log (L_{bol})=\log (k\cdot L_{2-10\text{keV}}) 
%\end{equation} 
%The value of $k$ depends on the Eddington ratio $\lambda = L_{bol}/L_{Edd}$ and can be iteratively calculated in the following way:

% \begin{table}[htbp]
%\centering
%\begin{tabular}{c|l}

%Eddington ratio & Bolometric correction k\\ \hline  
%$\lambda<0.02$ & $k = 15$ \\ 
%$0.02<\lambda<0.2$& $\log{\left(k\right)}= \log{\left(4\right)} \log{\left(\lambda\right)} + 2.1989$\\ 
%$\lambda>0.2$  & $k = 60$ \\ 
%\end{tabular}
%\label{}
%\end{table}

%%%%%%%%%%%%%%%%%%%%%%%%%%%%%%%%%%%%%%%%%%%%%%%%%%%%%%%%%%%%%%%%%%%%%%%%%%%%%%%%%%%%%%%%%%%%%%%%%%%%%%%%%%%%%%%%%%%%%%%%%%%%%%%%%%%%%%%

%%%%%%%%%%%%%%%%%%%%%%%%%%%%%%%%%%%%%%%%%%%%%%%%%%%%%%%%%%%%%%%%%%%%%%%%%%%%%%%%%%%%%%%%%%%%%%%%%%%%%%%%%%%%%%%%%%%%%%%%%%%%%%%%%%%%%%%

\section{Results}
\label{sec:results}

In this Section we present the X-ray luminosity using our new measurements of redshifts (Section~\ref{subsec:x-rayred}). We proceed with AGN classification based on the narrow and broad lines, and then compare it with the classification of un-obscured sources based on X-ray data (i.e., using the column density \NH; Section \ref{subsec:agn_types}). 
Next, we provide optical emission line classifications for all the AGN in the survey (Section \ref{subsec:bpt_results}).  We then compare our demographics of BASS X-ray selected AGN to SDSS-selected AGN (Section \ref{subsec:sdsscompare_results}).  We also present the distributions of black hole masses, bolometric luminosities, and accretion rates (Section \ref{subsec:bhmass_results}).  Finally, we discuss the variety of unusual AGN we have identified in this large survey (Section \ref{subsec:unusual_agn}). 

\subsection{Redshift Distribution}
\label{subsec:x-rayred}
\begin{figure*}[hbtp]
\centering
\includegraphics[width=\textwidth]{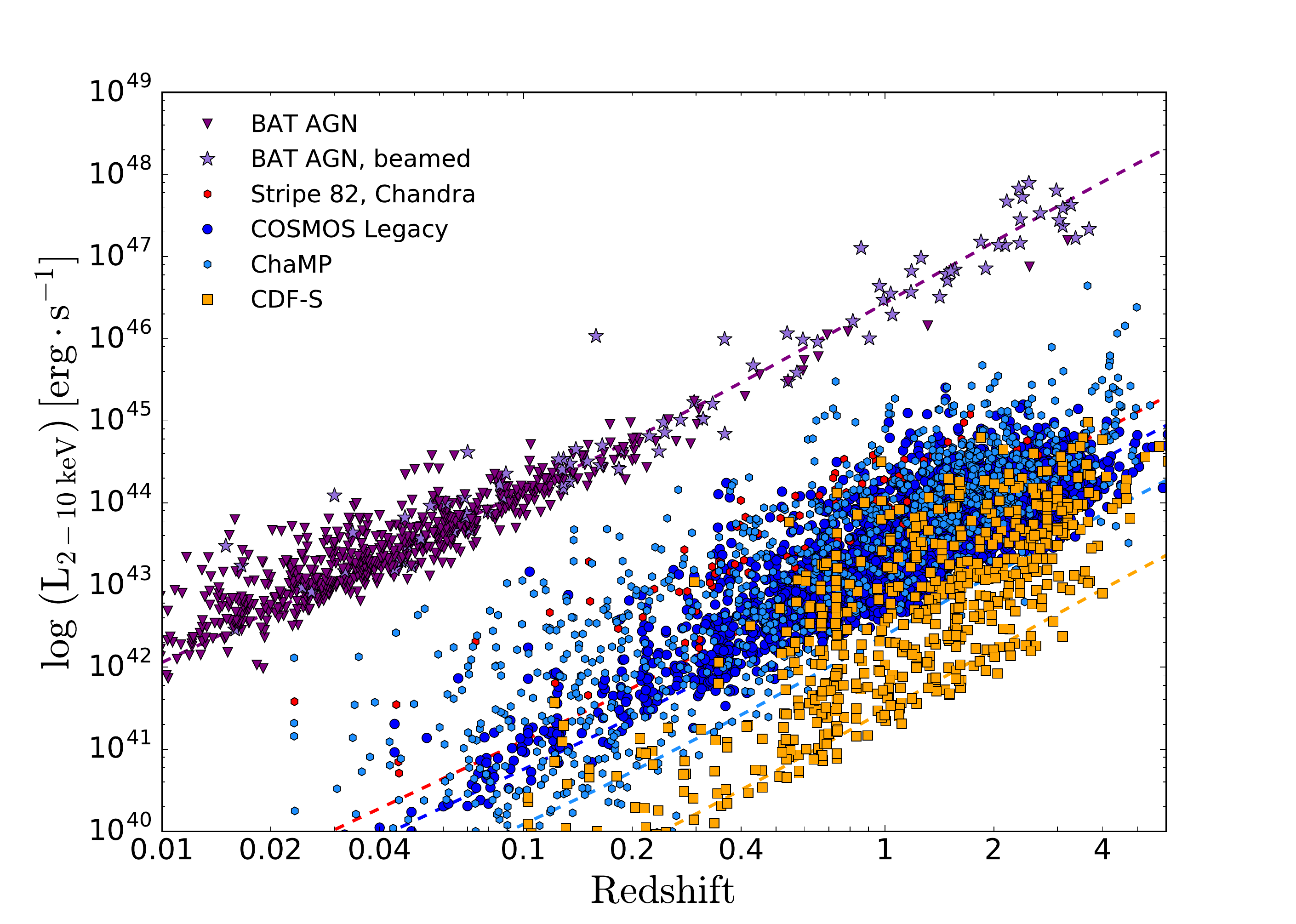}
\caption{Distribution of the restframe hard X-ray luminosity based on the 14-195 keV emission of the BAT AGN.  We used a ratio of 2.67 \citep{Rigby:2009:1878} to estimate the 2--10\,keV luminosity from the 14--195\,keV luminosity.  We have included other deeper X-ray surveys for comparison.  For other X-ray surveys we have assumed $\Gamma$=1.5 to estimate the intrinsic luminosity for obscured and unobscured X-ray sources. The BAT AGN redshift are measured from BASS using narrow emission lines, or NED when no spectra are available. The red dashed line shows the flux limit of BAT over 90$\%$ of the sky ($1.34\times 10^{-11}$\,\ergcms).  The unbeamed AGN in the BASS sample tend to span the moderate to high luminosity end of the X-ray luminosity function (XLF) at all redshifts.  Samples from deeper published surveys, such as the \chandra Deep Field South (CDF-S), tend to sample a lower luminosity range of the X-ray luminosity function.}
\label{Lum_vs_redshift}
\end{figure*}

\begin{figure*}[hbtp]
\centering
\subfigure{\includegraphics[width=8.2cm]{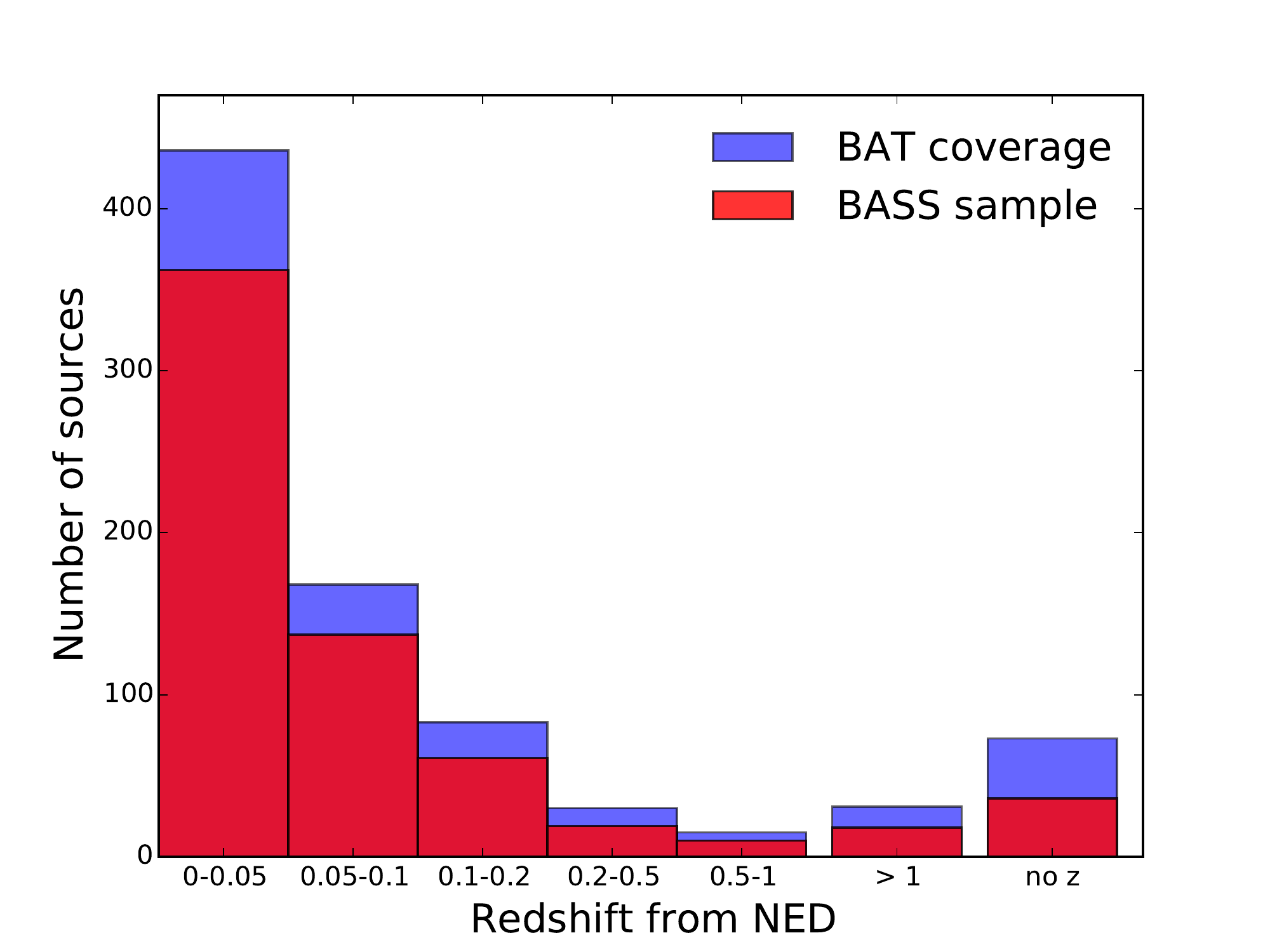}}
\subfigure{\includegraphics[width=8.2cm]{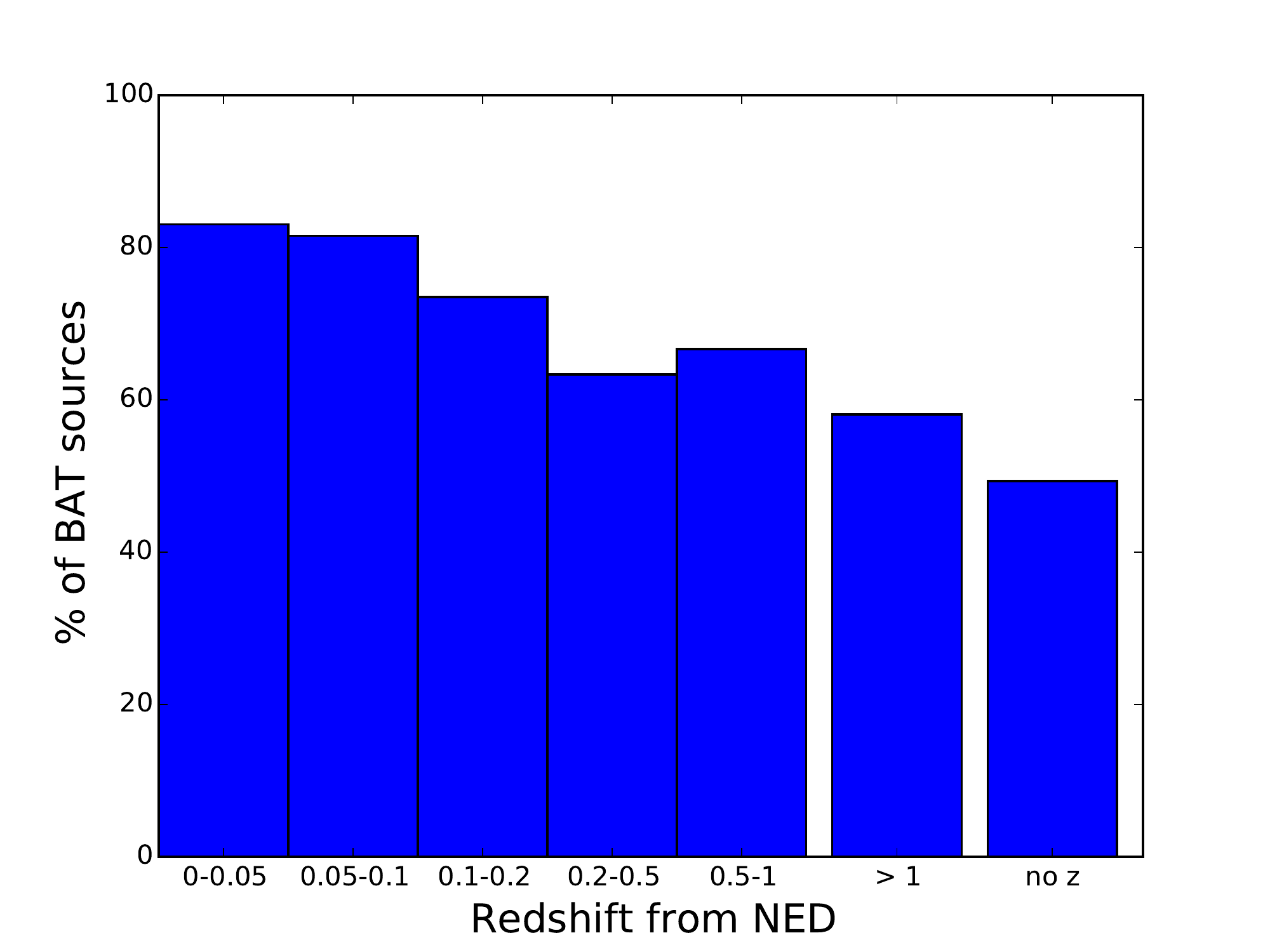}}
\caption{Summary of spectroscopic coverage with redshift for the BASS sample.  {\em Left:} number of sources with spectra (red) compared to the number of AGN in the BAT catalog (blue) at different redshift intervals based on NED.     The right panel shows the percentage of BAT AGN covered by the BASS sample in redshift intervals.  The majority of the spectra are low redshift ($z<$0.05), consistent with the BAT sample.  The spectroscopic completeness is 75\% but falls with increasing redshift because of the increasing fraction of faint sources and beamed sources. }
\label{BASS_coverage}
\end{figure*}

We begin by showing a plot of the X-ray luminosities in the entire \swiftbat AGN sample, using our new redshift measurements or those from NED when the spectra are not available (Fig. \ref{Lum_vs_redshift}). We also show several other, deep X-ray surveys, for comparison. 
The majority ($\approx90\%$) of BAT-detected AGN are nearby ($z<0.2$). Their X-ray luminosities are similar to AGN found in deeper surveys because of the larger survey area.   Our survey finds 46 new redshifts for sources without measurement in NED.  This leads to a redshift completeness of 96\% (803/836) for the 70-month BAT AGN catalog.  The newly measured sources are at similar redshifts (median $z=0.049$) as the sources in the rest of the sample (median $z=0.038$).  A summary of the spectroscopic coverage is in Fig. \ref{BASS_coverage}.

Only three sources with redshifts measured in BASS show significant differences with those in NED. 
QSO B0347-121 is listed at $z=0.18$ in NED; however, our measurements of the 6DF spectrum clearly show emission lines at $z=0.032$, in agreement with the redshift tabulated in the 6DF catalog. 
ESO 509-IG 066 NED02 is listed in NED as $z=0.0446$; however, our measurements of the 6DF spectrum clearly show emission lines at $z=0.034$, again in agreement with the 6DF catalog.  1RXS J090915.6+035453 is listed as $z=3.20$ in NED, but the lines are clearly redshifted  to $z=3.28$ based on our measurements, and in agreement with the SDSS catalog measurement.

\clearpage

\subsection{Narrow and Broad Line Classification}
\label{subsec:agn_types}

\begin{figure}
\centering
\includegraphics[width=8.5cm]{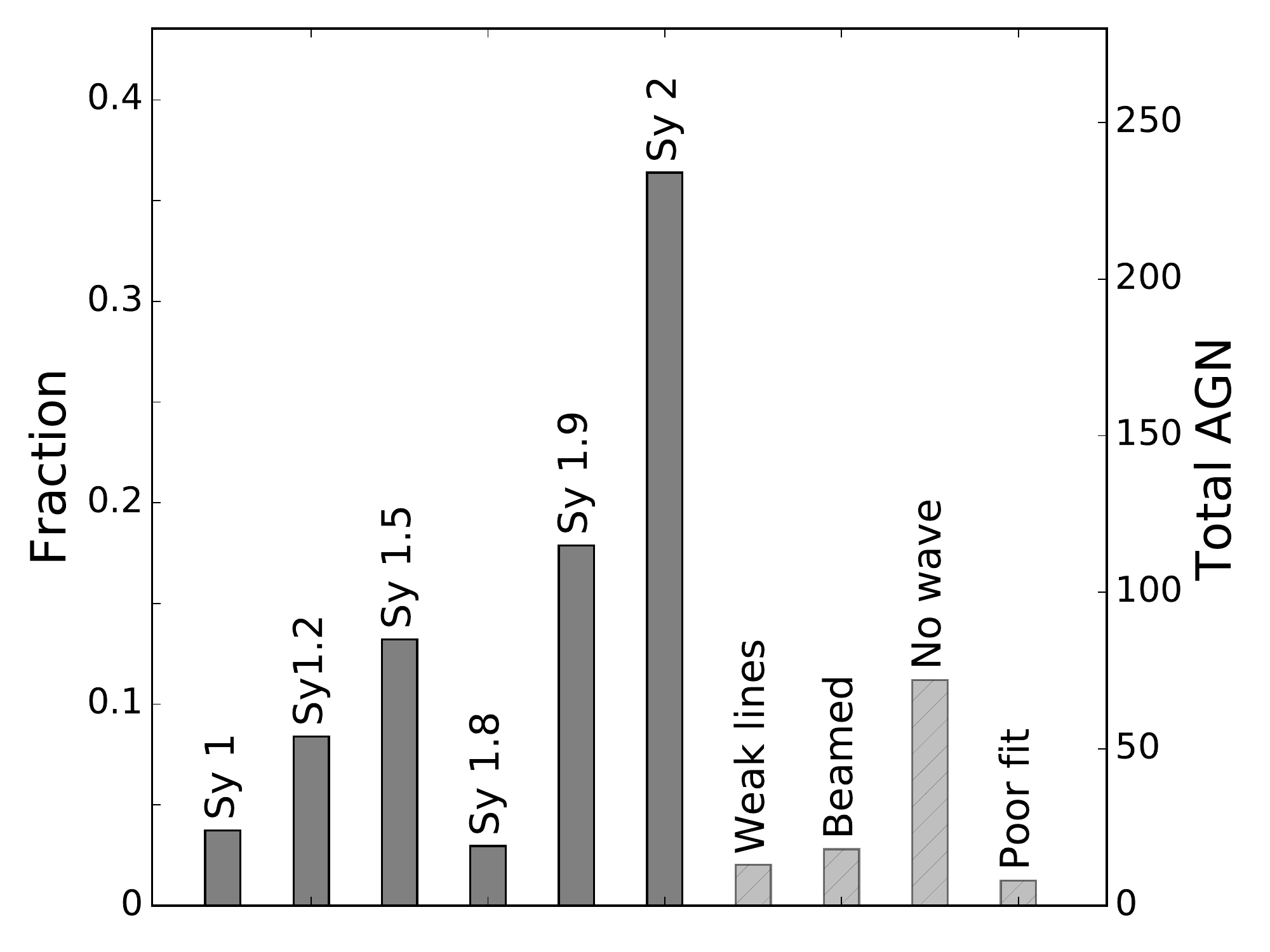}
\caption{Dichotomy of BAT AGN depending on the presence and strength of broad emission lines. For 84\% of spectra (539/641), we have measurements of Seyfert types.   A Seyfert 1.9 classification is a source with a narrow \Hbeta\ line and broad \Halpha\ line. The remaining categories (dashed histograms) have features that prevented measurements.  ``Weak lines" refers to objects without the presence of any Balmer emission lines in the spectra despite high quality spectra. We use the quantitative classifications for Seyfert types (1, 1.2, 1.5, 1.8) based on \citet[][]{Winkler:1992:677} using the total flux of \oiii\ and \Hbeta. }
\label{fig:pie_Seyfert}
\end{figure}

We first classify the BAT AGN depending on the presence and strength of broad emission lines \citep[e.g.,][]{Osterbrock:1981:462}.  A Seyfert 1.9 classification is a source with a narrow \Hbeta\ line and broad \Halpha\ line.  We use the quantitative classifications for Seyfert sub-classes (1, 1.2, 1.5, and 1.8) based on \citet[][]{Winkler:1992:677} using the total flux of \oiii\ and \Hbeta. 
A summary of the results of this classification can be found in Fig.~\ref{fig:pie_Seyfert}.  About half of the sources are Seyfert 2 or Seyfert 1.9, and about a quarter Seyfert 1.2 and 1.5.  A small fraction are true Seyfert 1 (\NSyoneclassper) and only two are Seyfert 1.8 sources.\\

We compare our Seyfert types (Sy 1, Sy 1.2, Sy 1.5, Sy 1.8, Sy 1.9, and Sy 2) to the most recent 13th edition Veron-Cetty catalog of AGN \citep{Veron-Cetty:2010:A10}.  Only a minority (36\%, 230/641) of the BASS sample is classified in this catalog by Seyfert type.  Using this subsample of 230 we find that the majority, 89\% (206/230), agree with the Seyfert type classification with the Veron-Cetty catalog. These include 77\% (177/230) which show exact type agreement or are listed as a unspecified Sy 1  in the Veron Cetty catalog and found to be Sy 1, Sy 1.2, Sy 1.5, or Sy 1.9 in BASS.  Another 13\% (29/230) are Sy 1, but are listed as Sy 1.2 or Sy 1.5 and vice versa in BASS.

The remaining 10\% (24/230) show disagreement among Seyfert type between BASS and Veron-Cetty.  The majority of these are listed as Sy 1.9 in our sample but are found to be Sy 2 in Veron-Cetty 67\% (16/24) most likely because our of higher quality spectra.  There are no examples of Sy 1.9 in the Veron-Cetty catalog which are found to be Sy 2 in our catalog.\\

     Another six sources are Sy 1 unspecified in Veron-Cetty, but Sy 2 in our catalog. For PKS 0326-288, the Veron-Cetty reference \citep{Mahony:2011:2651}  lists the source as a narrow line source in agreement with our classification. MCG +02-21-013 has no references in the Veron-Cetty catalog for the Seyfert type, but NED lists both a Sy 1 and Sy 2 in agreement with our classification.  NGC 4992 has no references in the Veron-Cetty catalog for the Seyfert type, but is classified as a Sy 2 in a recent paper by \citet{Smith:2014:112} in agreement with our classification.  A recent paper on MCG+04-48-002 \citep{Koss:2016:L4} lists it as a Sy 2 in agreement with our classification while it has no references in the Veron-Cetty catalog for the Seyfert type.  NGC 5231 is listed as a Sy 2 to match our classification in recent work \citep{Parisi:2012:A101}.  Finally, 2MASX J10084862-0954510 is shown to be a Sy 1 in \citet{Bauer:2000:547} with broad Balmer lines which is very different from our spectra and may be a case of variability.\\

Finally, two sources are listed as Sy 1.0 in Veron-Cetty, but are listed as Sy 1.9 in the BASS catalog.  The reference for ESO 198-024 as a Sy 1.0 in Veron-Cetty is based on imaging variability \citep{Winkler:1992:659} in the optical rather than spectroscopy and could be consistent with our Sy 1.9 classification.  Finally,  the reference in Veron-Cetty for a Sy 1.0 classification \citep{Guainazzi:2000:113} lists the source as a Sy 1.9 which is the same as our classification though the authors mention the source may have changed from a narrow line AGN to a Sy 1.9.\\

In summary, we find broad agreement (89\%) with past studies of Seyfert type classification from the Veron-Cetty catalog. The major difference is that 16 sources in the BASS catalog are now classified as Sy 1.9 because of broad lines detected in \Halpha\ for the first time.  Finally, 8 sources show a different Seyfert type possibly due to variability which will be further studied in future publications.  \\

% Pie chart for the Osterbrock classification of the sources

\begin{figure*}
\centering
\includegraphics[height=8cm]{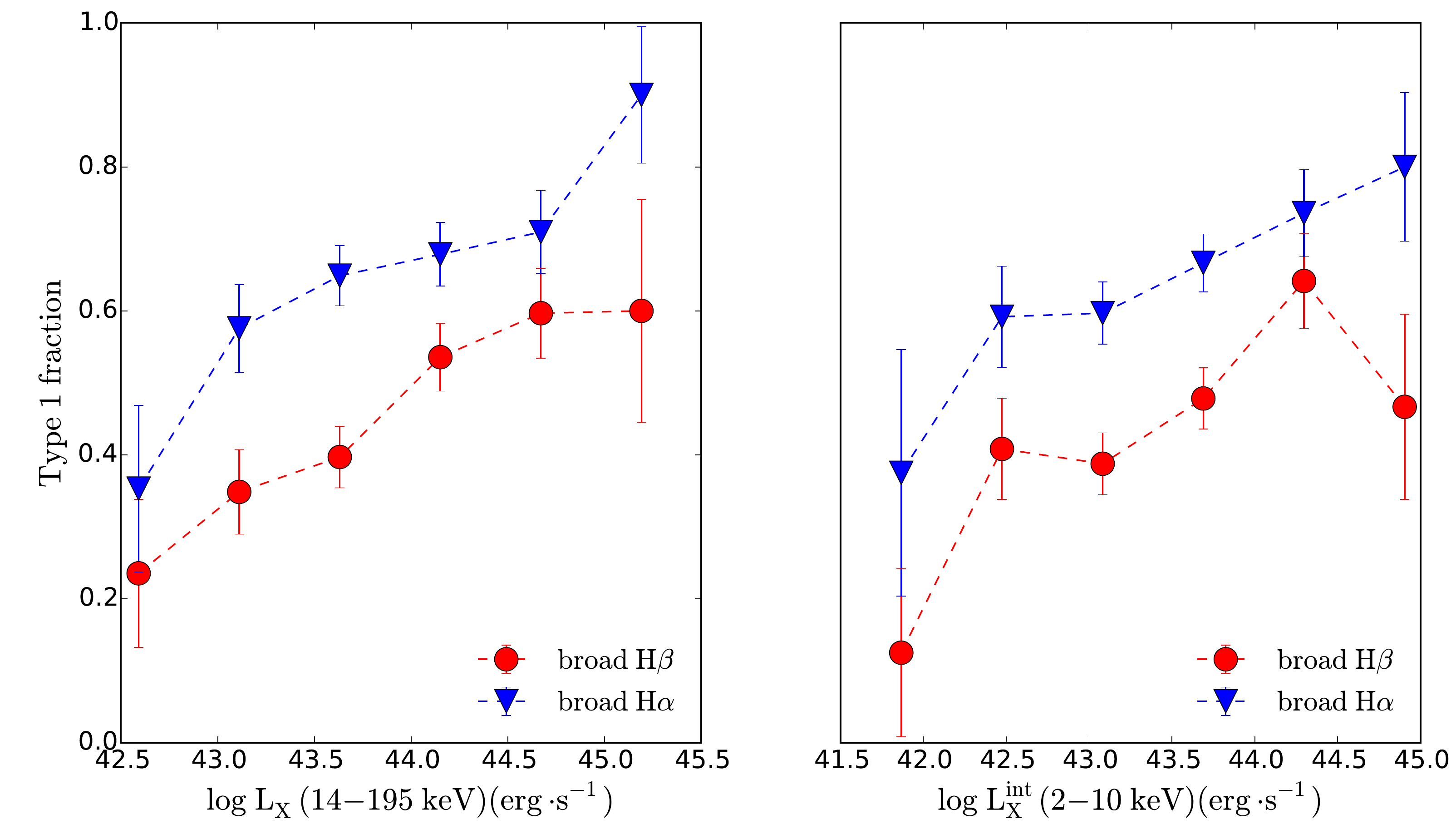}
\caption{The fraction of AGN classified as Type 1, based on the presence of broad Balmer lines, vs.\ X-ray luminosity (14--195\,keV in the left panel, and 2--10\,keV in the right one). 
Error bars are calculated using the binomial distribution.  There is a general increase in the fraction of broad line AGN with X-ray luminosity, using both \Halpha\ and \Hbeta.}
\label{fig:type1_frac}
\end{figure*}

In Figure~\ref{fig:type1_frac} we present the fraction of sources with broad \Halpha\ and/or \Hbeta\ lines, plotted against X-ray luminosity. For both the 14--195\,keV and 2--10\,keV luminosities we find a general increase in type 1 fraction with increasing luminosity, as has been found in past studies using broad \hb\ \citep[see e.g.,][and references therein]{Merloni:2014:3550}. We find that AGN with broad \Halpha\ lines are consistently more common, by about 10--20\%, than AGN with \Hbeta\ lines, across a wide range of X-ray luminosity. \\ 
%There is a consistent increase across a range of X-ray luminosity ($\approx10^{42}-10^{45}$\ergpssh).\\

\begin{figure*}
\centering
\includegraphics[height=10cm]{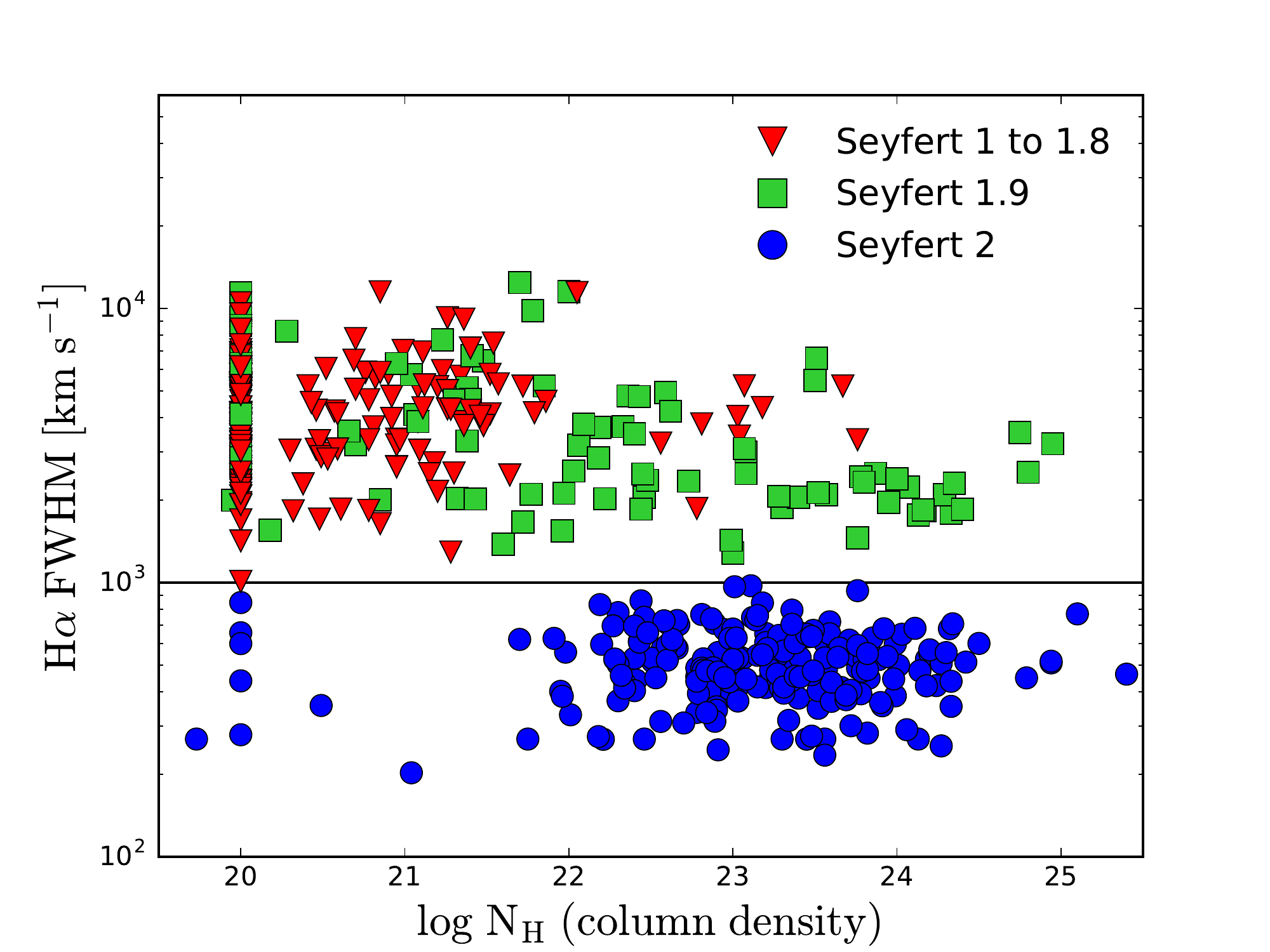}
\caption{FWHM of \Halpha\ as a function of column density. Colors represent the Osterbrock classification. In the sample, \NSyonelownH\ Seyfert 1 to 1.8 and \NSyoneninelownH\ Seyfert 1.9s have a column density consistent with a lower limit (\NH$=10^{20}$ \nhunit).  We note that there are only two Seyfert 1.8 in the sample, so the broad line sample is dominated by Seyfert 1-1.5 types.  We find a general agreement between the X-ray column density and the presence of broad lines with \NH$<10^{22}$ \nhunit\ as a dividing line between most type 1 and type 2 sources.  However, Seyfert 1.9 show the full range of column densities.}
\label{fig:nH}
\end{figure*}

Finally, in Fig.~\ref{fig:nH} we plot the FWHM of \Halpha\ as a function of the column density derived from the X-rays (\NH).  
We note that roughly half (\NSyonenhlimit) of Seyfert 1-1.8 broad-line AGN with column density measurements have only upper limits on \NH, at $10^{20}$ \nhunit\ corresponding to being unobscured. \\
 
The FWHM of the emission lines show broad agreement with the X-ray obscuration ($\sim$94$\%$), such that Seyfert sub-types 1, 1.2, 1.5, and 1.8 have \NH$< 10^{21.9}$\,\nhunit, and Seyfert 2, have \NH$> 10^{21.9}$\,\nhunit.  Seyfert 1.9, however, show a range of column densities.\\

Additionally, a small fraction of Seyfert 2 sources (\Nnakedseytwo), have X-ray obscuration below \NH$=10^{\Nnhdivider}$ \nhunit.  We note however that Seyfert 1.9 sources, which have evidence of a broad line in \Halpha, but not \Hbeta, span the full range of column densities from unobscured to Compton-thick (i.e., $\NH<10^{24}$~\nhunit).

% Pie chart for Hbeta broad line measurements

\clearpage

\subsection{Emission Line Classification}
\label{subsec:bpt_results}

\begin{figure*}
\centering
\includegraphics[height=18cm]{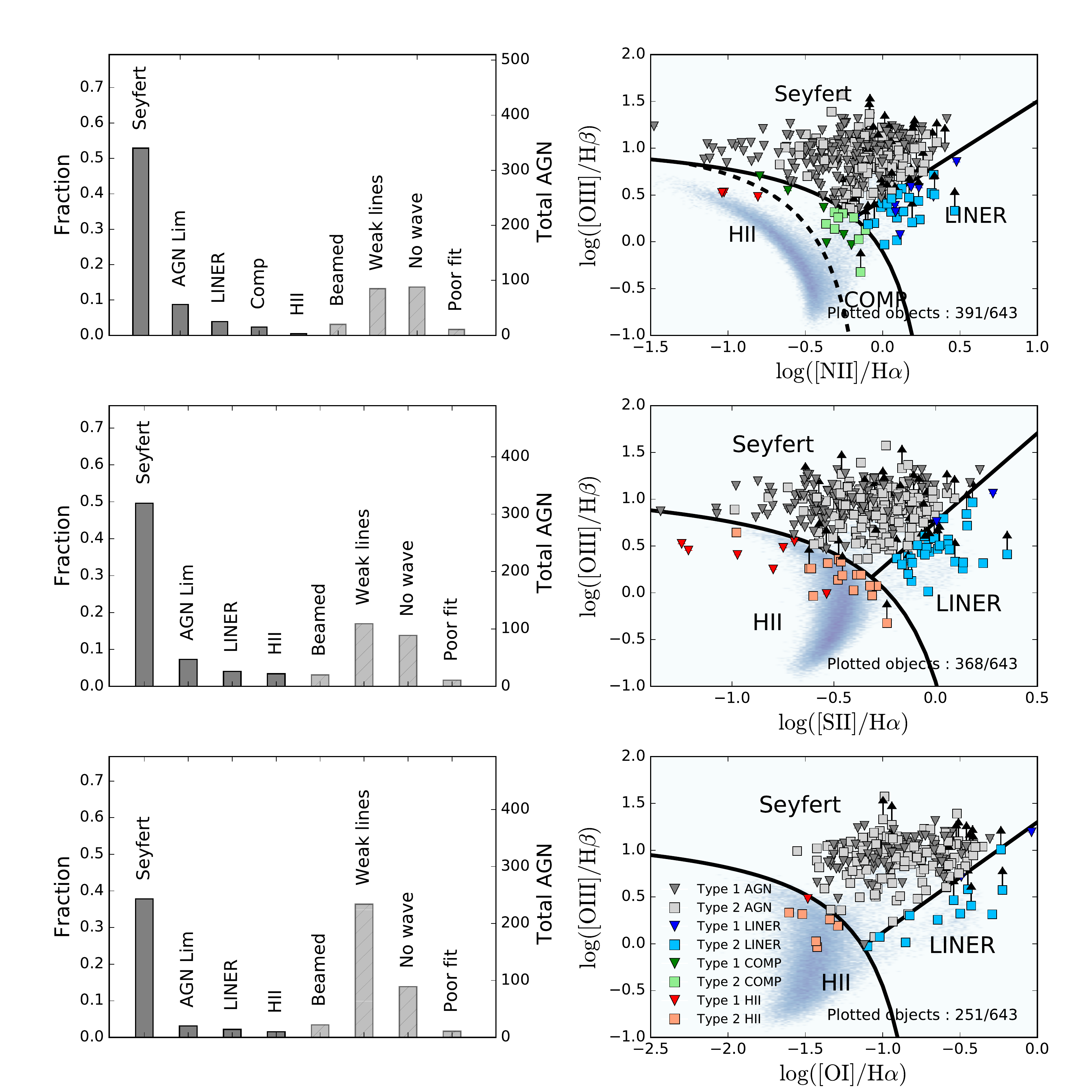}
\caption{Classification of the sample using line diagnostics diagrams \citep{Kewley:2006:961}.  {\it Left Column}: histograms for the entire sample. ``AGN lim" refers to objects which have an \Hbeta\ upper limit either in the Seyfert or in the LINER region.  The remaining categories (dashed histograms) have features that prevented measurements.  ``Weak lines" refers to objects with high SNR spectra which lack sufficiently strong emission line measurements to be placed on the diagram.  ``Beamed AGN" refers to any sources crossmatched to the blazar and beamed AGN catalogs.   We note that most beamed AGN in our sample are not blazars and have emission lines and are classified using emission line diagrams.  ``Poor fit" refers to AGN with complex emission line profiles that were poorly fit using out automated routine.  Finally, ``no wave" refers to objects lacking sufficient wavelength coverage to measure the needed emission lines.   {\it Right Column}:  Line diagnostic diagrams for sources with sufficient measurable emission lines to be classified using line diagnostic diagrams. The grey area represents the distribution of the SDSS sample. Narrow line objects are shown with squared and broad line objects (broad \Hbeta\ detection) with triangles. The arrows represent upper limits and lower limits.  We find good agreement between all three classical line diagnostic diagnostics in the fraction of Seyferts, LINERS, composite galaxies, and HII regions with \NII\ showing the largest number of line measurements and \OI\ the fewest.  }
\label{BPTnormal}
\end{figure*}

We use the emission line diagnostics of \citet{Veilleux:1987:295}, revised by \citet{Kewley:2006:961}. We classify each AGN using the \OIII/\Hbeta\ vs.~\NII/\Halpha, \SII/\Halpha, and \OI/\Halpha\ diagnostics (Fig.~\ref{BPTnormal}).  For the \NII\ diagnostic, we further separate the star forming (HII) galaxies and composite galaxies, and separate AGN into LINERs and Seyferts \cite[following][]{Schawinski:2007:512}.  Finally, we also apply the \OIII/\OII\ and \HeII/\Hbeta\ diagnostics, defined by \cite{Shirazi:2012:1043}.\\

We find that roughly half of BAT AGN are found in the Seyfert region of the \nii\ diagram (\NBPTniiseyper, \NBPTniisey/\Ntot).  The next largest sub-group are sources without an \Hbeta\ detection (\NBPTniilinerper, \NBPTniisey/\Ntot), though the detection limits imply either a Seyfert or LINER AGN classification. About 15\% of sources have weak lines and, despite high signal to noise optical spectra, lack enough emission line measurements for line diagnostic diagrams. The remaining categories of LINERs, composites galaxies, and HII classifications are rare, with only a few percent of BAT AGN found in each.  A few percent of sources also have complex emission line profiles where a good fit to the emission lines was not obtained. Finally, about 10\% of sources lack sufficient wavelength coverage, because of the instrumental setup or their high redshift.  \\

The \SII\ diagnostic shows a very similar distribution, though a few percent lower fraction of Seyferts (\NBPTsiiseyper, \NBPTsiisey/\Ntot), due to the weaker \sii\ line (and limited $S/N$ of the data), and a larger fraction of HII regions.  For the \oi\ diagnostic, the line is markedly weaker than the \nii\ and \sii\ line, and therefore identifies somewhat fewer Seyferts (\NBPToiseyper, \NBPToisey/\Ntot), and has about double the number of sources that lack emission line detection.  For sources with line detections in all three diagnostics (\NBPTallthreeper, \NBPTallthree/\Ntot), we find good agreement in the AGN classification across the diagnostics (\NBPTallthreeagreeper, \NBPTallthreeagree/\NBPTallthree).   \\

\begin{figure*}
\centering
\includegraphics[height=14cm]{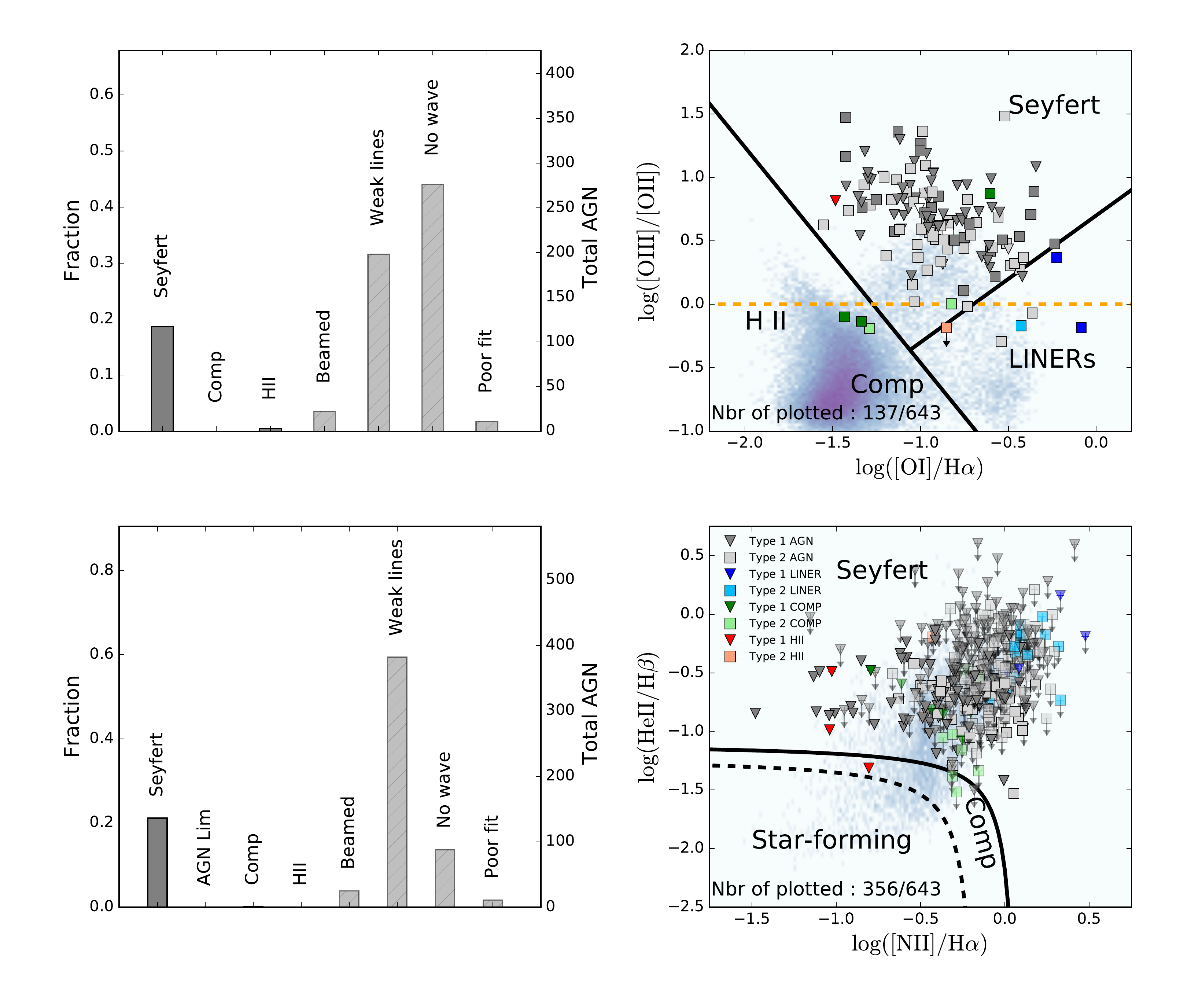}
\caption{Classification of the sample using the \OIII/\OII~ vs.~\OI/\Halpha  ~and \HeII/\Hbeta~ vs.~\NII/\Halpha ~diagnostic diagrams \citep{Shirazi:2012:1043} using the same scheme as the previous line diagnostic figure.  Compared to the traditional diagnostics (\NII, \SII, and \OI), these two diagnostics fail to classify the majority objects because of the difficulty detecting the \HeII\ line and the lack of blue coverage in most spectra for the \OII\ line.  }
\label{BPTblue}
\end{figure*}

We also classify the sample using the \OIII/\OII~ vs.~\OI/\Halpha  ~and \HeII/\Hbeta~ vs.~\NII/\Halpha ~diagnostic diagrams (Fig.~\ref{BPTblue}).  Compared to the more commonly used diagnostics (i.e., \nii, \sii, and \oi), these two diagnostics are not efficient in classifying the majority of AGN in our sample, because of the difficulty detecting the \heii\ line and the lack of blue coverage in most spectra for the \oii\ line.

%\begin{figure*}
%\centering
%\includegraphics[height=16cm]{pie.pdf}
%\caption{Classification of the sample using the BPT diagrams (resp. \NII, \SII, and \OI\ ) showed on the right. 'AGN lim' means the objects have an $H_{\beta}$ upper limit either in the Seyfert or in the LINER region. In the BPT diagrams we removed the objects classified as blazars and the ones with $z>0.4$. The grey area represents the SDSS sample. Narrow line objects are shown with squared and broad line objects (broad $H_{\beta}$ detection) with triangles. The arrows represent upper limits and lower limits.}
%\label{BPT}
%\end{figure*}

%he \NII diagram shows 361/527 objects, the \SII diagram 349/527 objects and the \OI diagram 244/527 objects.

\subsection{Comparison to Optical Emission Line Selected AGN from the SDSS}
\label{subsec:sdsscompare_results}

\begin{figure*}
\centering
\includegraphics[width=8.5cm]{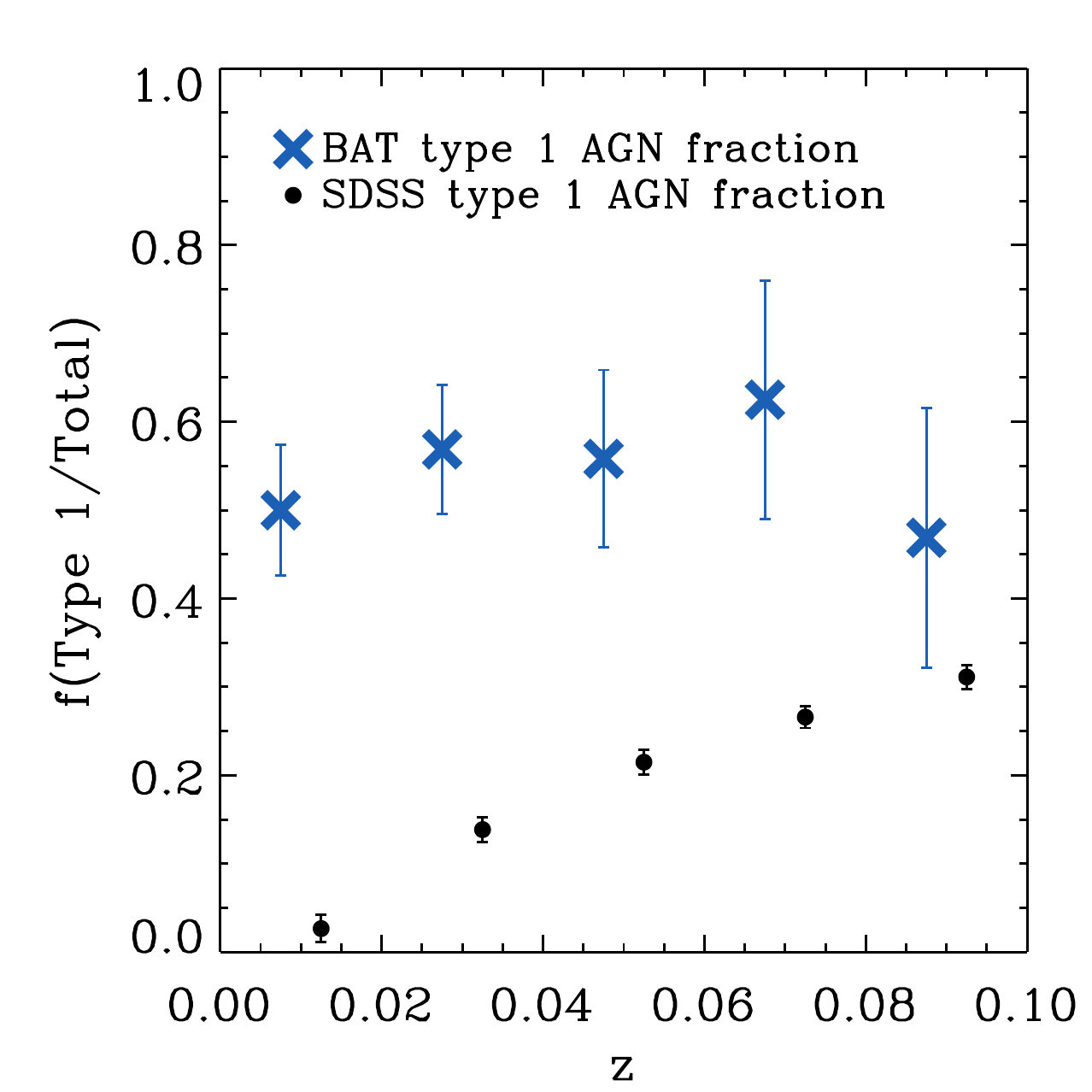}
\includegraphics[width=8.5cm]{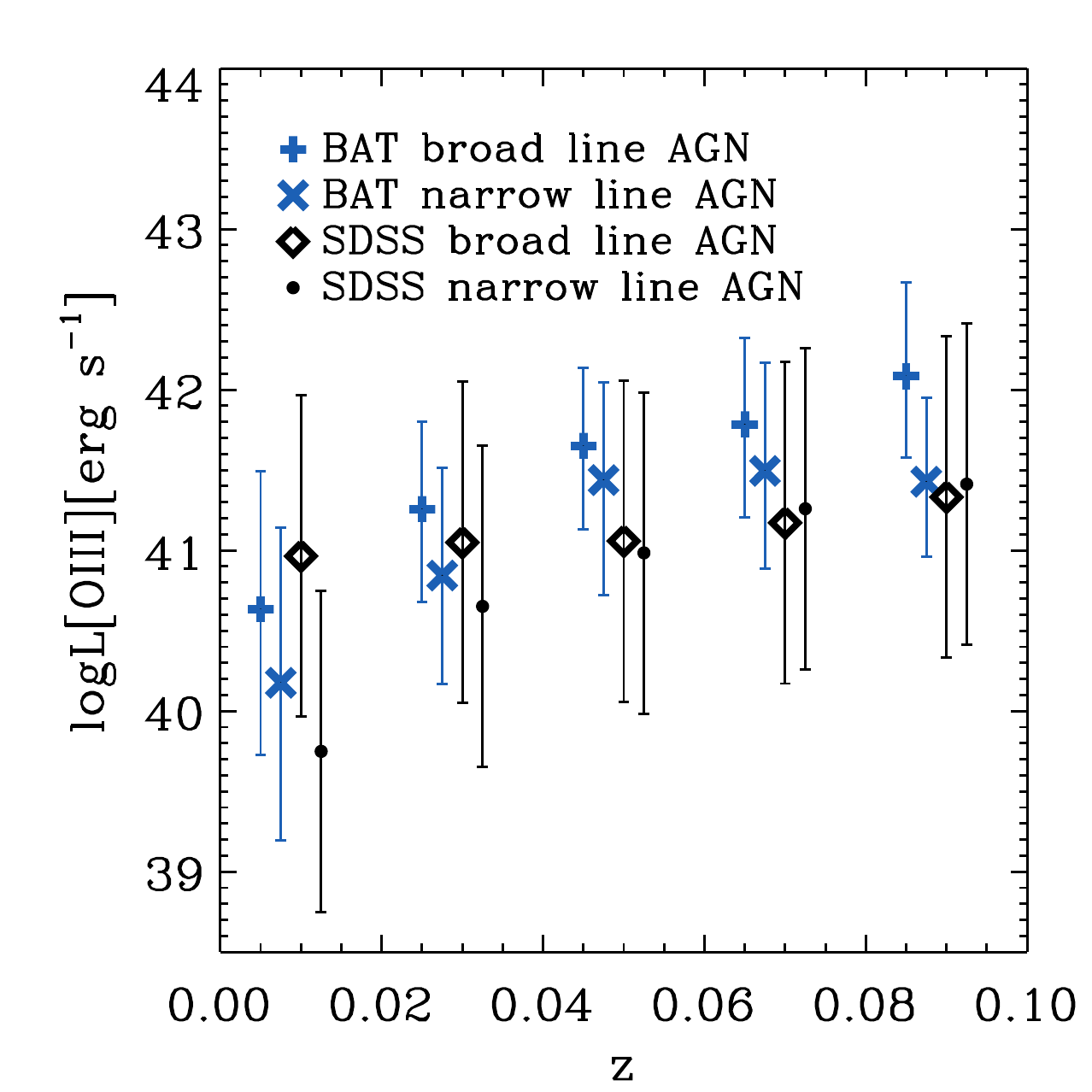}
\includegraphics[width=8.5cm]{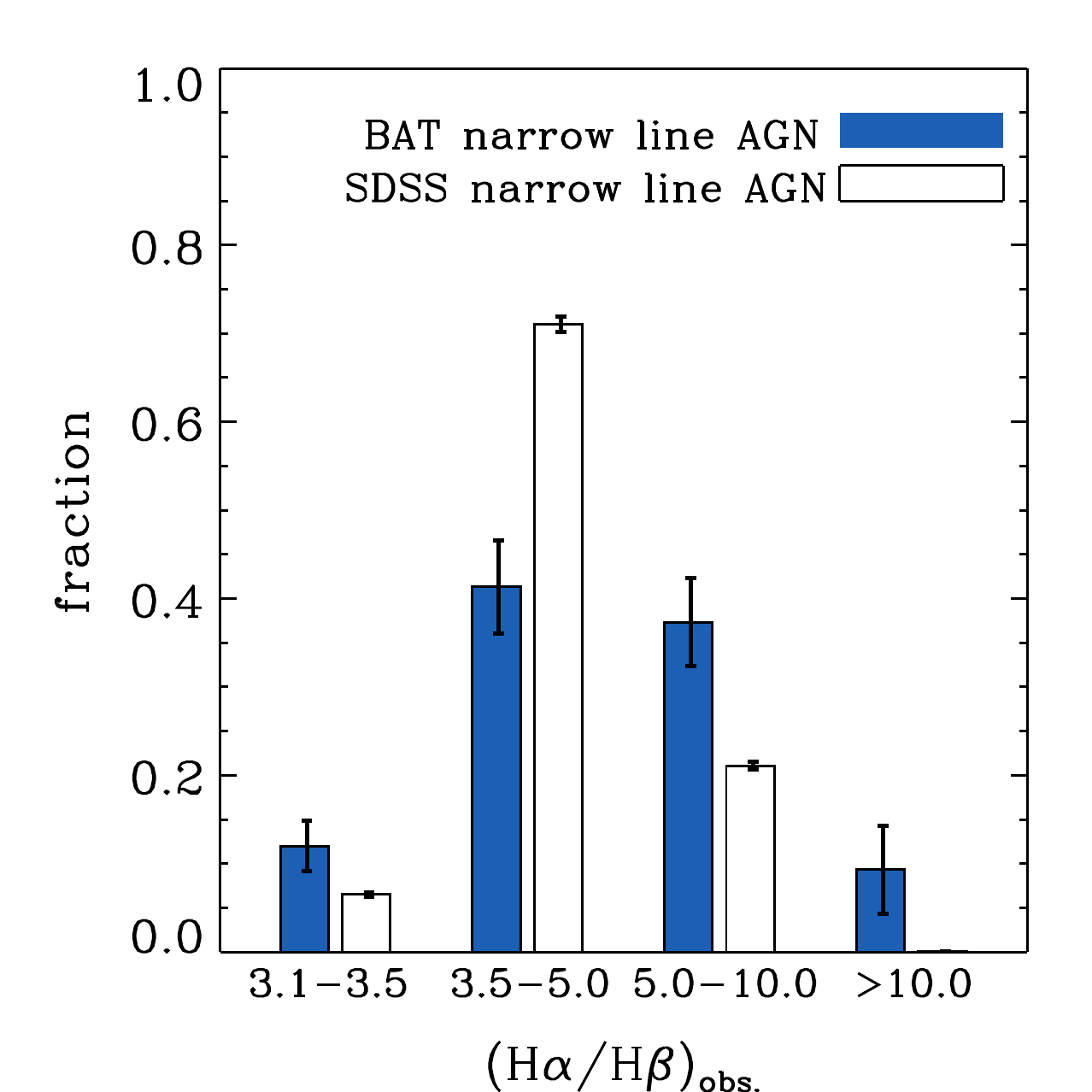}
\includegraphics[width=8.5cm]{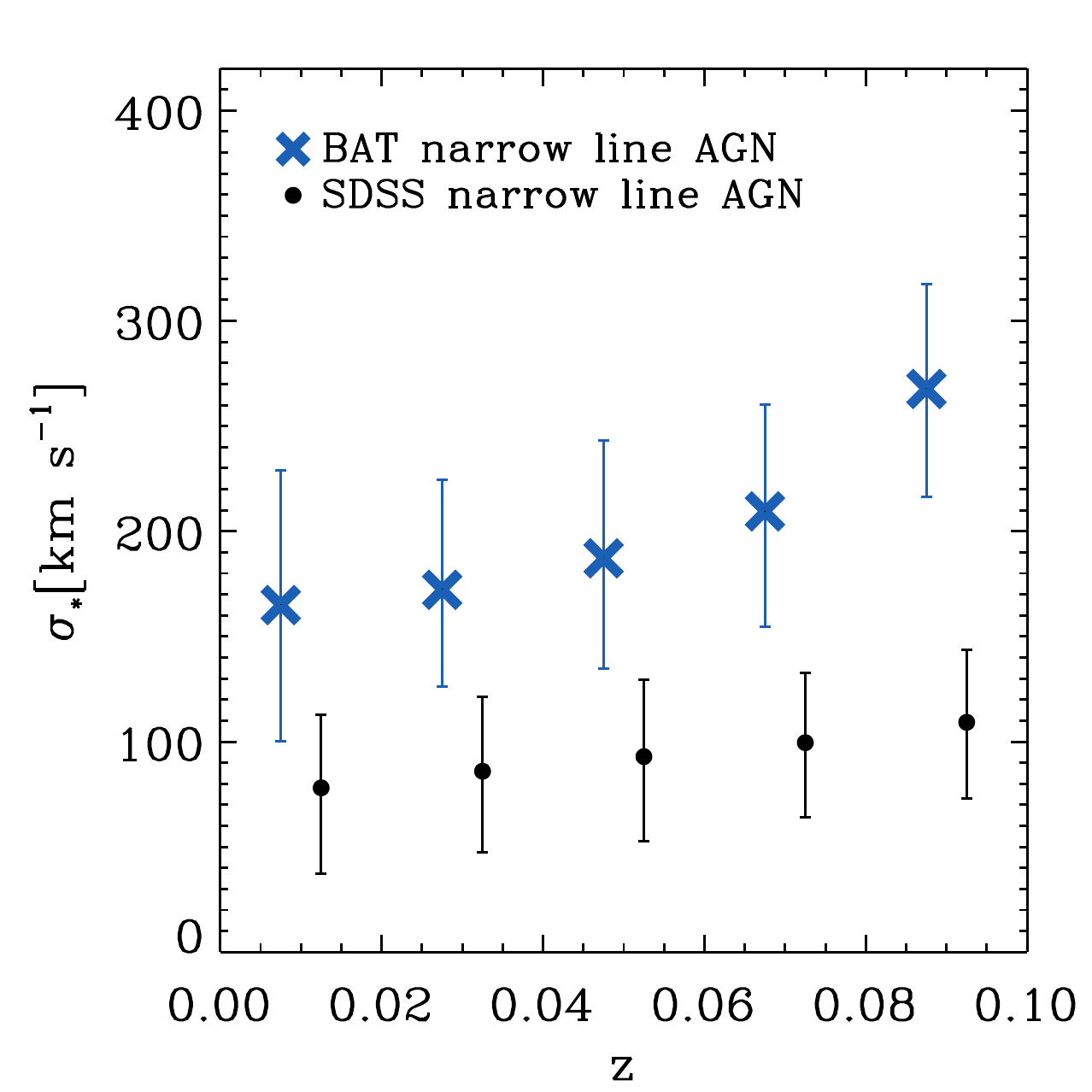}
\caption{Comparison of BASS X-ray selected AGN to optically selected Seyferts in the SDSS from the OSSY catalog \citep{Oh:2011:13}.  Median and 1$\sigma$ distribution are shown with 0.02 redshift bins.  \emph{Upper left}: Type 1 AGN fraction vs. redshift.  The BASS sample is fairly constant with redshift whereas the SDSS evolves strongly. \emph{Upper right}: \oiii\ luminosity vs. redshift.  The type 1 AGN and type 2 AGN in BASS have higher \oiii\ luminosity than SDSS AGN.   \emph{Bottom left}: Balmer decrement as compared to the SDSS narrow line AGN sample.  Narrow line BAT-detected AGN have a larger fraction of AGN in dustier galaxies (\ha/\hb$>$5). \emph{Bottom right}: Velocity dispersion vs. redshift.  BAT AGN tend to have larger velocity dispersion than SDSS-selected AGN, consistent with the fact that the SDSS Seyfert 2 AGN sample is 30 times larger than the BAT AGN.   }
\label{f1_z}
\end{figure*}

We perform a comparison of the demographics of the BASS X-ray selected AGN to optically selected Seyferts in the SDSS, based on the OSSY catalog \citep{Oh:2011:13, Oh:2015:1}.  The results are shown in Fig.~\ref{f1_z}. We find that the BASS X-ray selected AGN show a relatively constant type 1 to type 2 fraction of $\sim$55\% over the redshift range of $z<0.1$, while the fraction in SDSS AGN is much lower and furthermore shows a strong dependence on redshift (2\%--30\%).\footnote{The OSSY catalog classifies AGN as type 1 or type 2 solely based on the presence of a broad \Halpha\ emission line \cite[see details in][]{Oh:2015:1}.} 
The \oiii\ luminosities of BAT AGN are higher, on average, than those of SDSS AGN, for both Seyfert 1 and Seyfert 2 AGN. 
The BASS narrow line AGN show a larger number of sources with high Balmer decrements (\Halpha/\Hbeta$>5$), compared to SDSS AGN. This can be clearly understood by the requirement to have robust detections of all relevant emission lines for SDSS AGN to be classified as such, which the hard X-ray selection of BASS AGN overcomes.
Finally, the average stellar velocity dispersions of the BAT narrow line AGN ($191\pm17\,\kms$) are significantly higher  than those of narrow line SDSS AGN ($101\pm10\,\kms$), and show a stronger redshift dependence.

% note by Kyuseok
% Few numbers are changed: 
% 1) "We find that the X-ray selected AGN show a relatively constant type 1 to type 2 fraction of 55\% at a range of redshifts...." (previously it was 60\%)
% 2) "Finally, the velocity dispersions of the BAT narrow line AGN are much larger ($\approx$200$\,\kmpssh$)..." (previously it was ($\approx$100$\,\kmpssh$))
% Question: how much larger number of AGN in the SDSS explains the trend of higher velocity dispersion in the BASS samples? 

\subsection{Black Hole Mass and Accretion Rate Distribution}
\label{subsec:bhmass_results}

\begin{figure*}
\centering
\subfigure{\includegraphics[width=0.485\textwidth]{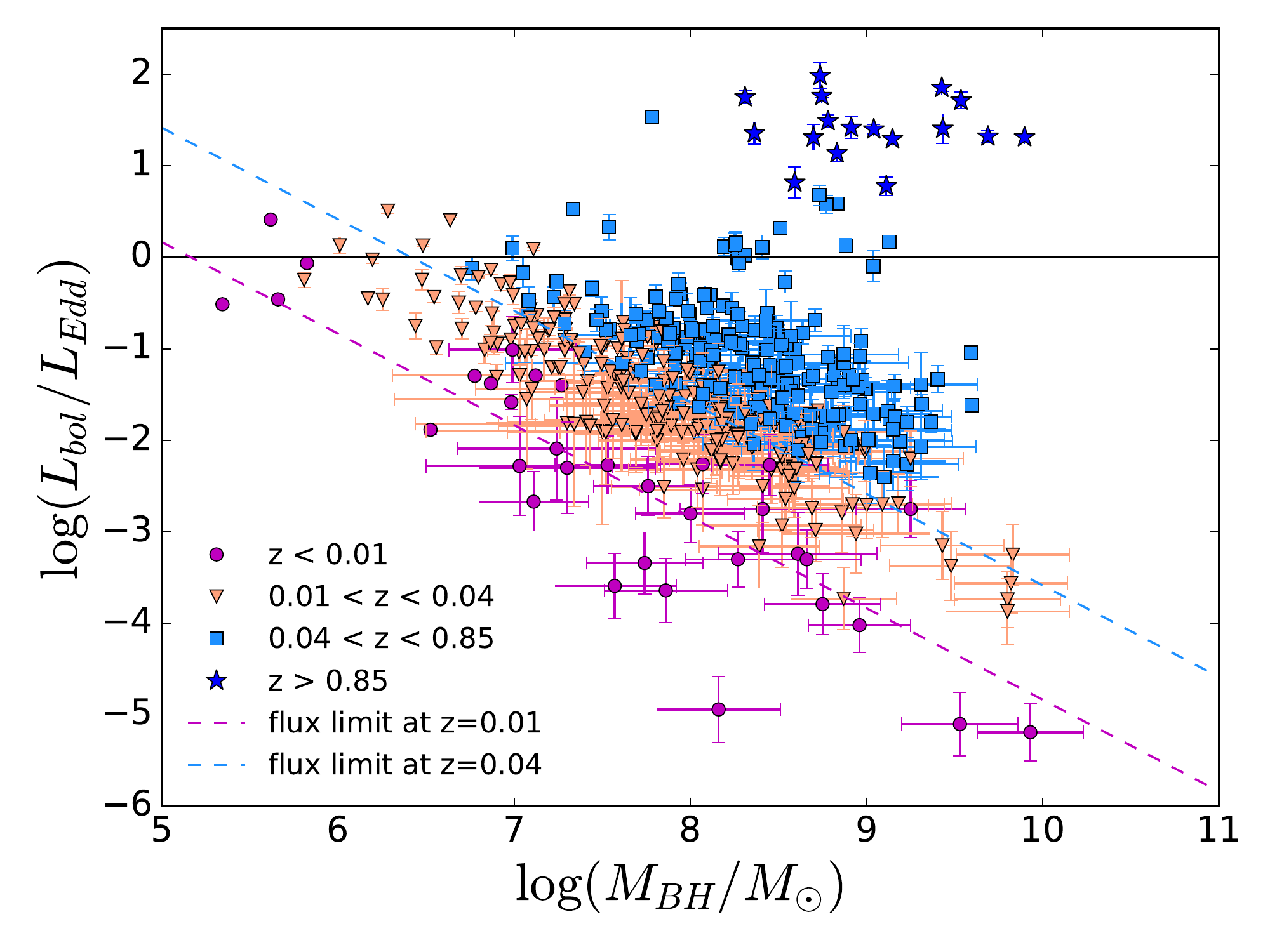}}
\hspace{0.1cm}
\subfigure{\includegraphics[width=0.485\textwidth]{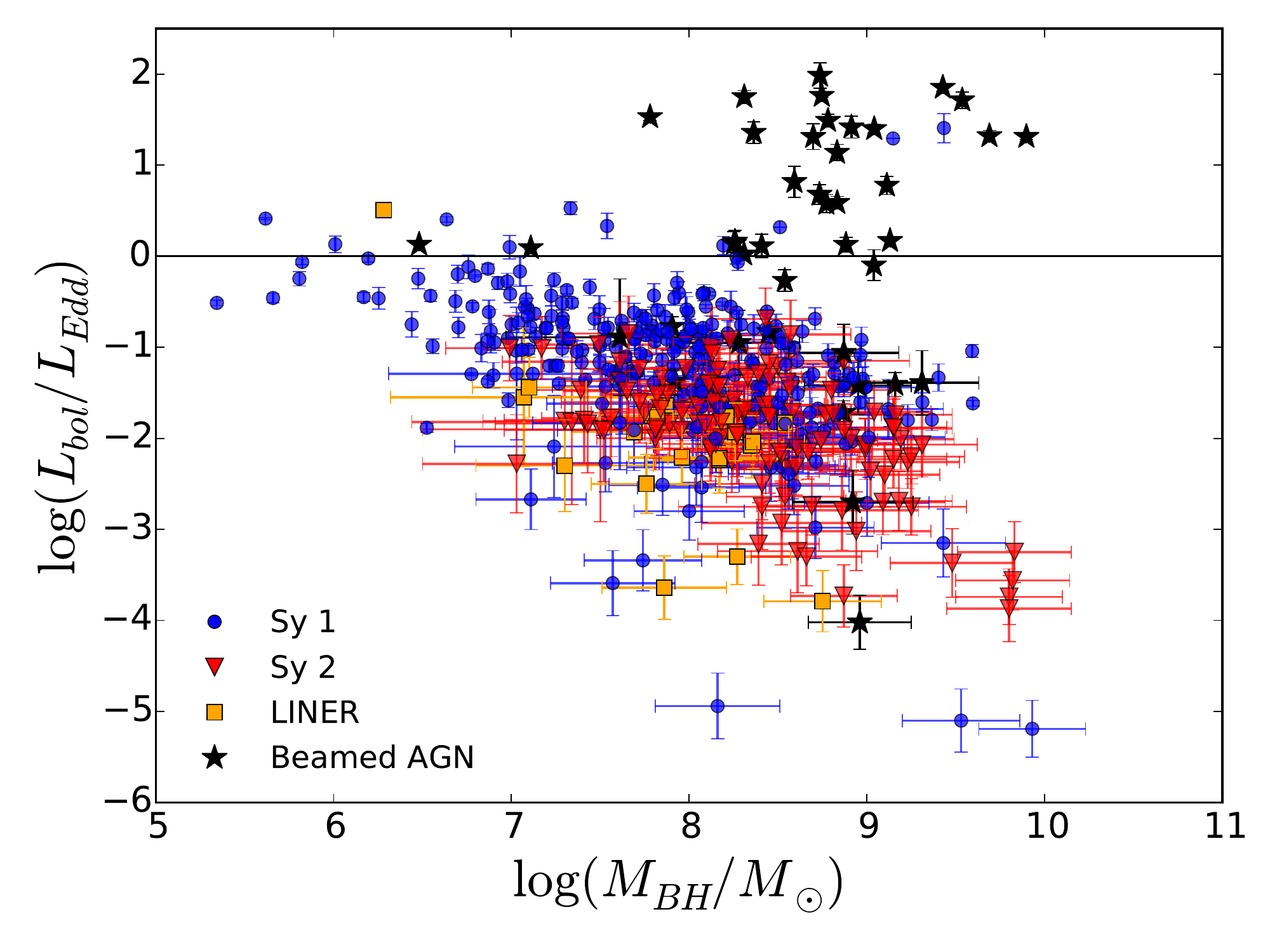} }
% Histograms for Sy 1, Sy2, LINER, beamed 
%\vfill
\vspace*{0.1cm}
\subfigure{\includegraphics[width=0.32\textwidth]{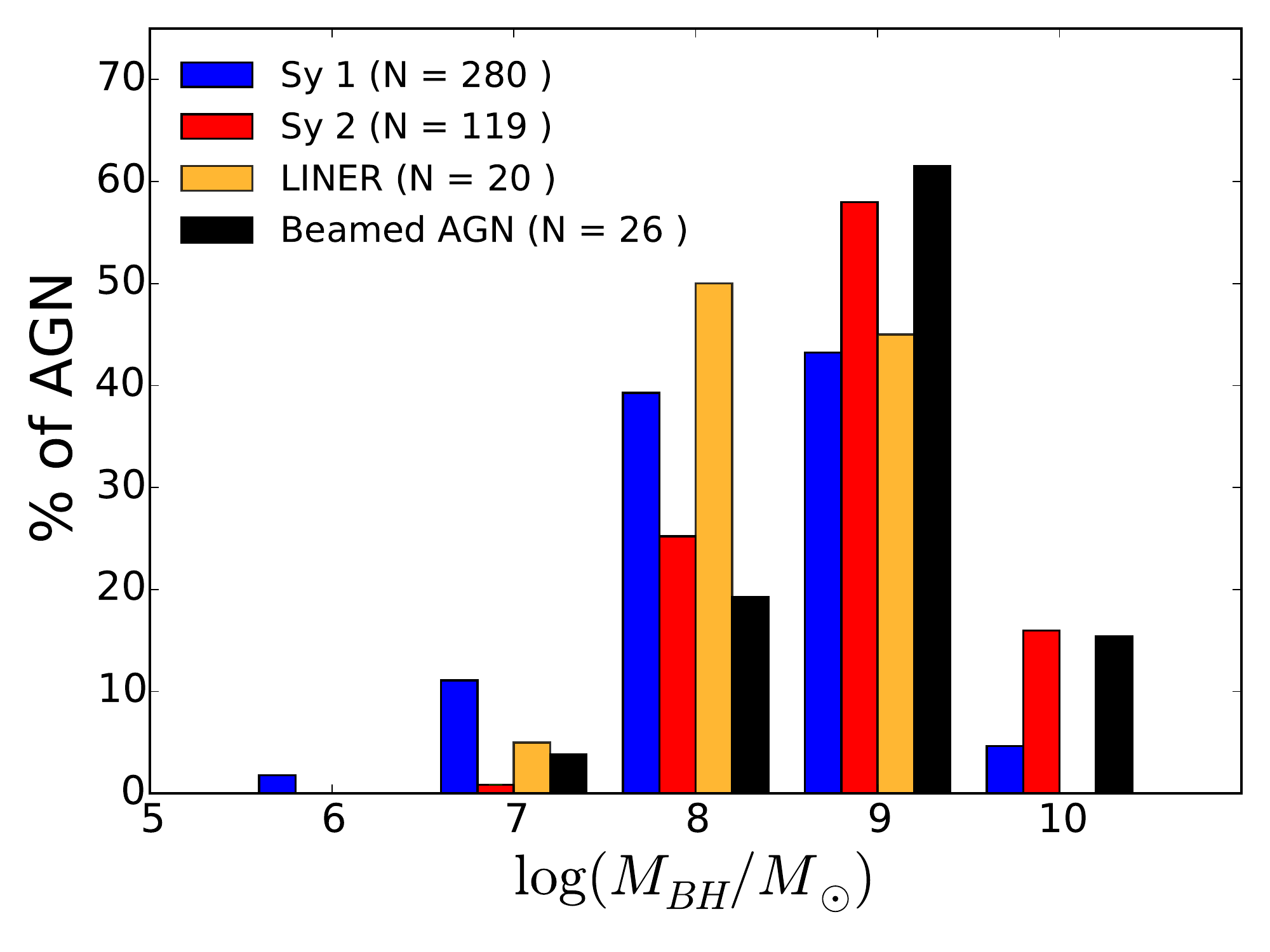}}
%\subfigure{\includegraphics[width=5cm]{histogram_Edd_ratio_class_all_Sy12.pdf}}
\subfigure{\includegraphics[width=0.32\textwidth]{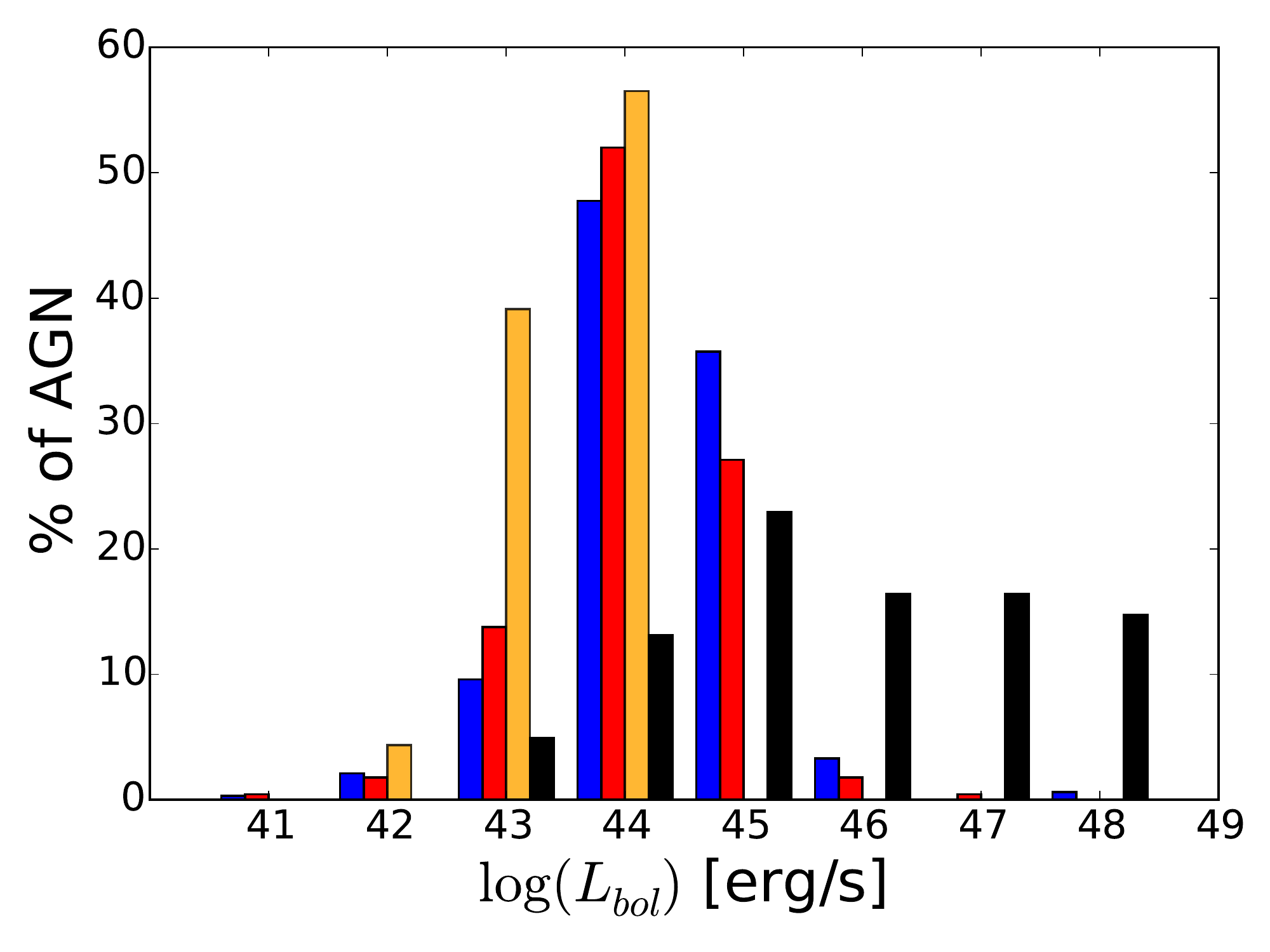}}
\subfigure{\includegraphics[width=0.32\textwidth]{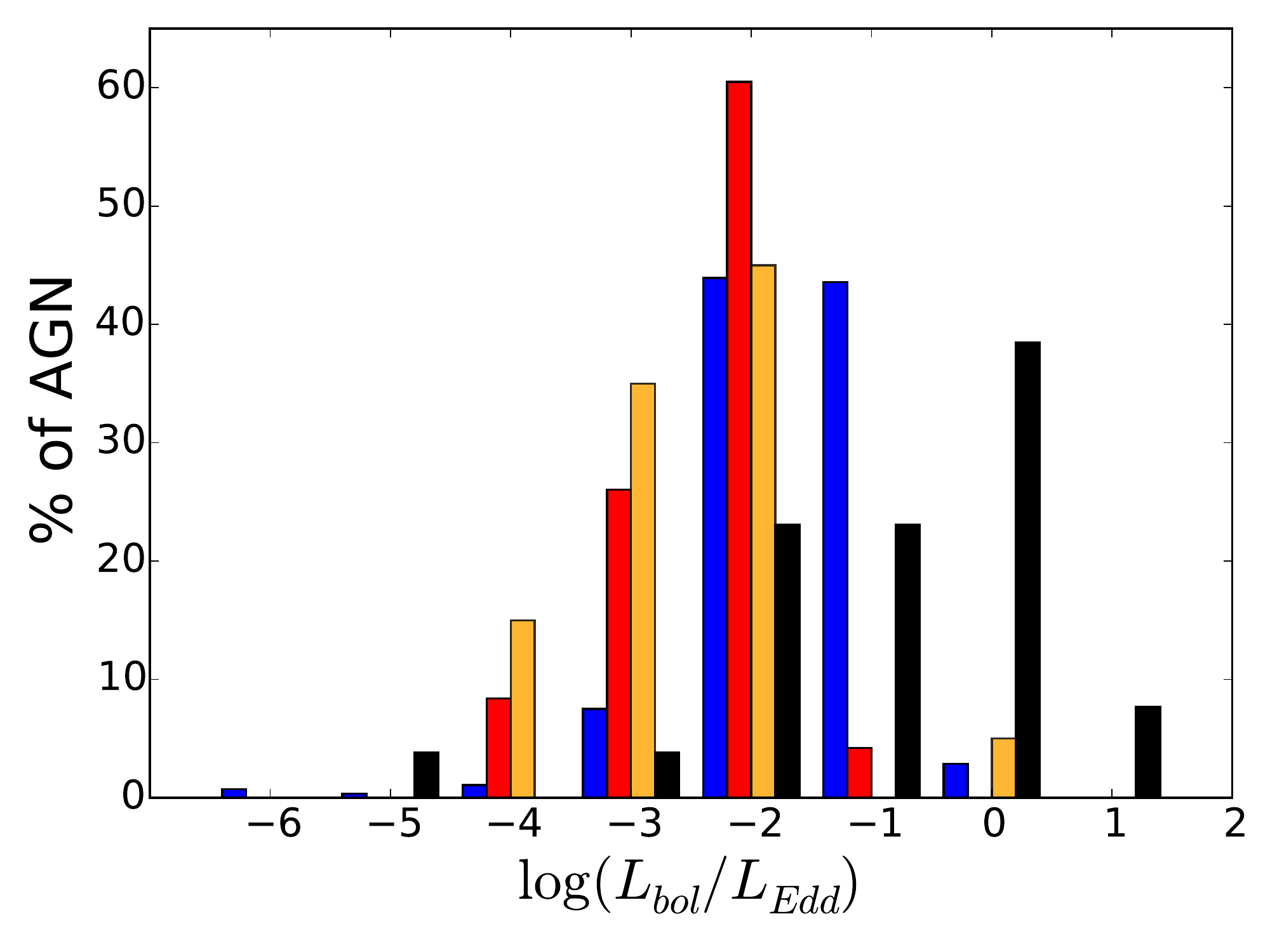} }
\caption{The top left panel shows the distribution of AGN Eddington ratio (\lledd) and black hole mass (\mbh) with redshift. The top right panel shows the distribution of the Eddington ratio and black hole mass with AGN type.  The bottom panels show histograms of black hole mass (left), bolometric luminosity (center), and Eddington ratio (right) with AGN type.}
\label{Histo_type12}
\end{figure*}

Using the black hole mass estimated from velocity dispersion and broad Balmer lines, we find the the black hole masses of our BASS AGN range between $10^{5.4}-10^{10}\, \Msun$. Figure~\ref{Histo_type12} shows the distribution of \mbh\ in different redshift ranges and for Seyfert 1-1.9, Seyfert 2, LINERs, and beamed AGN. 

We use the median and median absolute deviation (MAD) to compare the populations because of the spread over several orders of magnitude.  The median and MAD are $M_{\mathrm{BH}}=(5.8\pm8.5)\times10^7$\Msun\ for $z<0.01$, $M_{\mathrm{BH}}=(6.4\pm8.1)\times10^7$\Msun\ for $0.01<z<0.04$, and $M_{\mathrm{BH}}=(1.2\pm1.5)\times10^8$\Msun\ for $0.04<z<0.85$, and $M_{\mathrm{BH}}=(7.4\pm6.4)\times10^8$\Msun\ for $z>0.85$.  An Anderson-Darling test indicates the distributions of black hole masses at $z<0.01$ and $0.01<z<0.04$ consistent with being drawn from the same population, but those at $0.04<z<0.85$ and $z>0.85$ are drawn from the same population at less than the 1\% level, consistent with their higher median black hole masses. 

The higher median black hole masses found for high-redshift AGN is likely a selection effect driven by the fixed survey flux limit.
  
The median and MAD values are 
$\mbh=(1.9\pm2.6)\times 10^8\,\Msun$ for Seyfert 1-1.9; 
$\mbh=(1.9\pm2.1)\times 10^8\,\Msun$ for Seyfert 2; 
$\mbh=(1.7\pm2.1)\times 10^8\,\Msun$ for LINERs; and 
$\mbh=(6.0\pm6.0)\times 10^8\,\Msun$ for beamed AGN.  
An Anderson-Darling test indicates the distributions of black hole masses of Seyfert 1-1.9, Seyfert 2, and LINERS are consistent with being drawn from the same population, but the likelihood that beamed AGN are drawn from the same population is less than the 1\% level.  \\

 While we do not find any significant difference between the black hole mass distributions of Seyfert 1-1.9 and Seyfert 2, we note that our survey has systematic biases against the smaller black holes ($\mbh<10^7\,\Msun$) in Seyfert 2 AGN. 
Specifically, the velocity dispersion measurements for Seyfert 2 are limited by the instrumental broadening in lower spectral resolution setups. Typical instrumental resolutions are between $2-7$ \AA FWHM, (corresponding to limiting black hole masses of $\mbh=10^6-10^8\,\Msun$). 
We have recently been granted two filler programs with  VLT/XSHOOTER (Oh et al., prep) that will further address this issue, as the spectral resolution would be sufficient to measure limiting black hole masses of $\mbh=10^5\,\Msun$. 
A final issue is that for galaxies with a significant rotation component, \sigs\ measured from a single aperture spectrum can vary by up to $\sim$20\%, depending on the size of the adopted extraction aperture \citep{Kang:2013:26}. \\

%{\color{red}\todo\ in principle the factor should be 1.5 for solar metalicity...\todo}

%We use a median and median absolute deviation (MAD) to compare the populations because of the spread over several orders of magnitude.

%We find that the typical black-hole mass of our sample is a factor of four smaller than typical BAT-detected AGN ($M_{\mathrm{BH}}=(1.3\pm0.4)\times10^7$\Msun\ vs.  $M_{\mathrm{BH}}=(5.1\pm0.4)\times10^7$\Msun) where the error refers to the MAD 1$\sigma$ error. 

We combined the \mbh\ estimates with the estimates of \Lbol, derived from the BAT X-ray luminosity, to calculate the Eddington ratios of the BASS AGN, \lledd\,\ergs (where $L_{\rm Edd}\equiv1.3\times10^{38}\,\left(\mbh/\Msol\right)$).  The maximum value of the bolometric luminosity of our sample is $\Lbol = 10^{48.5}\,\ergs$. The AGN with higher \Lbol\ have, in general, higher \lledd, but there are also some AGN with relatively high bolometric luminosity ($ \Lbol > 10^{45}\,\ergs$) and low Eddington ratio ($\lledd < 0.01$). The sources with the highest bolometric luminosity ($\Lbol > 10^{47}\,\ergs$), however, do have high accretion rates  ($\lledd > 0.1$).
Conversely, several of the most massive BHs in unbeamed AGN in our sample ($\mbh >10^{9.7}\,\Msun$) also have the lowest accretion rates.  Regarding the redshift distributions, the median and MAD Eddington ratios are $\lledd=0.005\pm0.008$ for $z<0.01$, $\lledd=0.026\pm0.036$ for $0.01<z<0.04$, $\lledd=0.047\pm0.062$ for $0.04<z<0.85$, and $\lledd=1.46\pm2.16$ for $z>0.85$.  An Anderson-Darling test indicates each of the redshift distributions of Eddington ratios are each drawn from the same population at the less than 1\% level, consistent with their steadily increasing medians. These properties of our sample are not surprising, given the flux limited (and low redshift) nature of our sample. 
\\

Regarding the Eddington ratios among different Seyfert types, we do find that
Seyfert 2 AGN have, in general, lower Eddington ratios because  Seyfert 1 AGN have higher bolometric luminosities. The peak of the distribution for Seyfert 2 is at $\lledd\simeq 0.01$.  For the Seyfert 1 AGN, the peak of the distribution is between $\lledd= 0.01$ and $0.1$, with a small number (\NsuperEddper) of unbeamed sources above the Eddington limit,  $\lledd \gtrsim 1$.  The median and MAD are $\lledd=0.10\pm0.09$ for Seyfert 1-1.9, $\lledd=0.014\pm0.014$ for Seyfert 2, and $\lledd=0.009\pm0.012$ for LINERs, and $\lledd=1.46\pm2.16$ for beamed AGN.  

We note the Eddington ratios of beamed AGN are derived from the observed luminosities (and masses), and have not been corrected for beaming angle.  
An Anderson-Darling test indicates the distributions of Eddington ratios for Seyfert 2 AGN and LINERs are consistent with being drawn from the same population. However, the likelihood that the Eddington ratios of type 1 and beamed AGN are drawn from the same population is less than 1\%, consistent with their much larger medians.  It is interesting to note that using the observed Eddington ratios is highly efficient at separating beamed AGN from unbeamed sources because typical beamed AGN are above the Eddington limit.  \\

We also found sources with extremely low accretion rates ($\lledd <\ 0.001$). There are \NSytwolowEdd\ type 2 AGN and \NSyonelowEdd\ type 1 AGN with Eddington ratios $\lledd <\ 0.001$. In general, type 1 AGN are found to have high accretion rates, $\lledd>0.01$ \cite[e.g.,][]{Nicastro:2000:L65, Yuan:2004:724,Trump:2011:60,Elitzur:2014:3340}, and therefore it is quite surprising to find type 1 sources with such a low value of the Eddington ratio. \\

%We investigate also the distribution of \lledd\ among the \Nliner\  LINERs in our sample. LINERs generally have $\lambda < 0.01$ and the distribution peaks at $\lledd = 0.001$. This is in part due to the fact that LINERs have lower bolometric luminosities, generally below about $10^{45}\,\ergs$.\\

%  Eddington ratio vs. MBH

\subsection{Unusual AGN}
\label{subsec:unusual_agn}

\begin{figure*}[h!]

\includegraphics[width=0.485\textwidth]{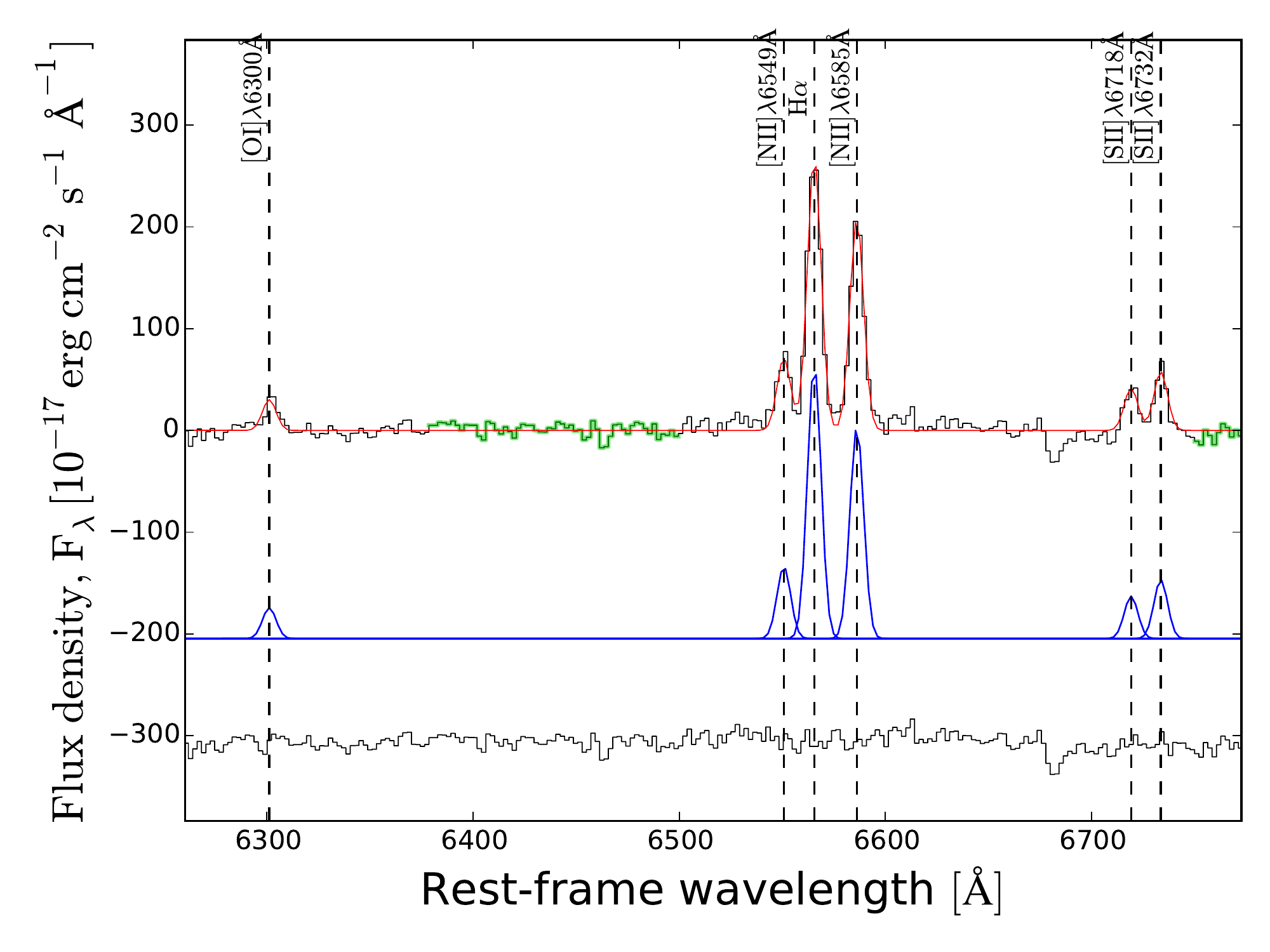}
%\hspace{0.1cm}
\includegraphics[width=0.485\textwidth]{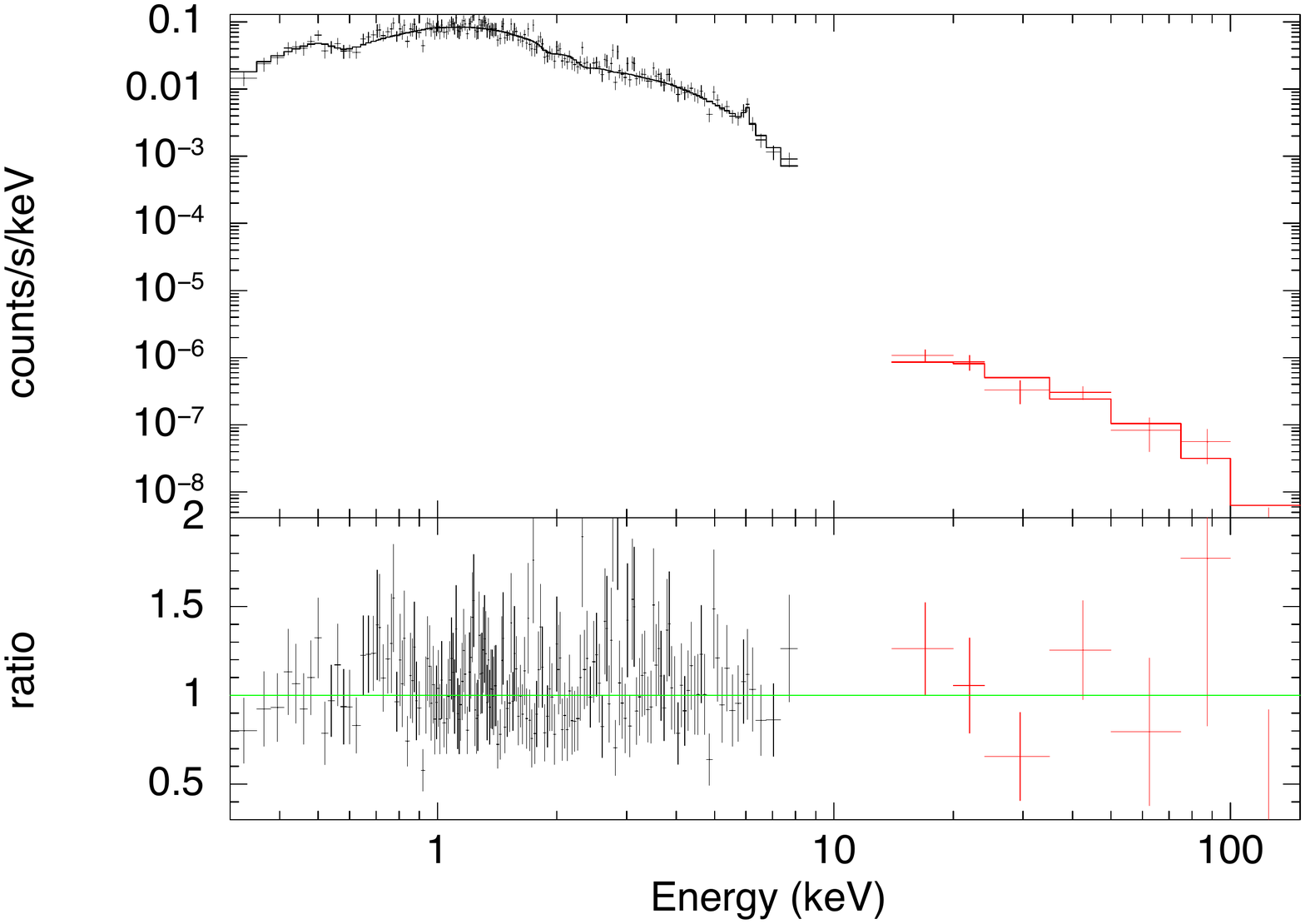}\\
\vfill
\includegraphics[width=0.49\textwidth]{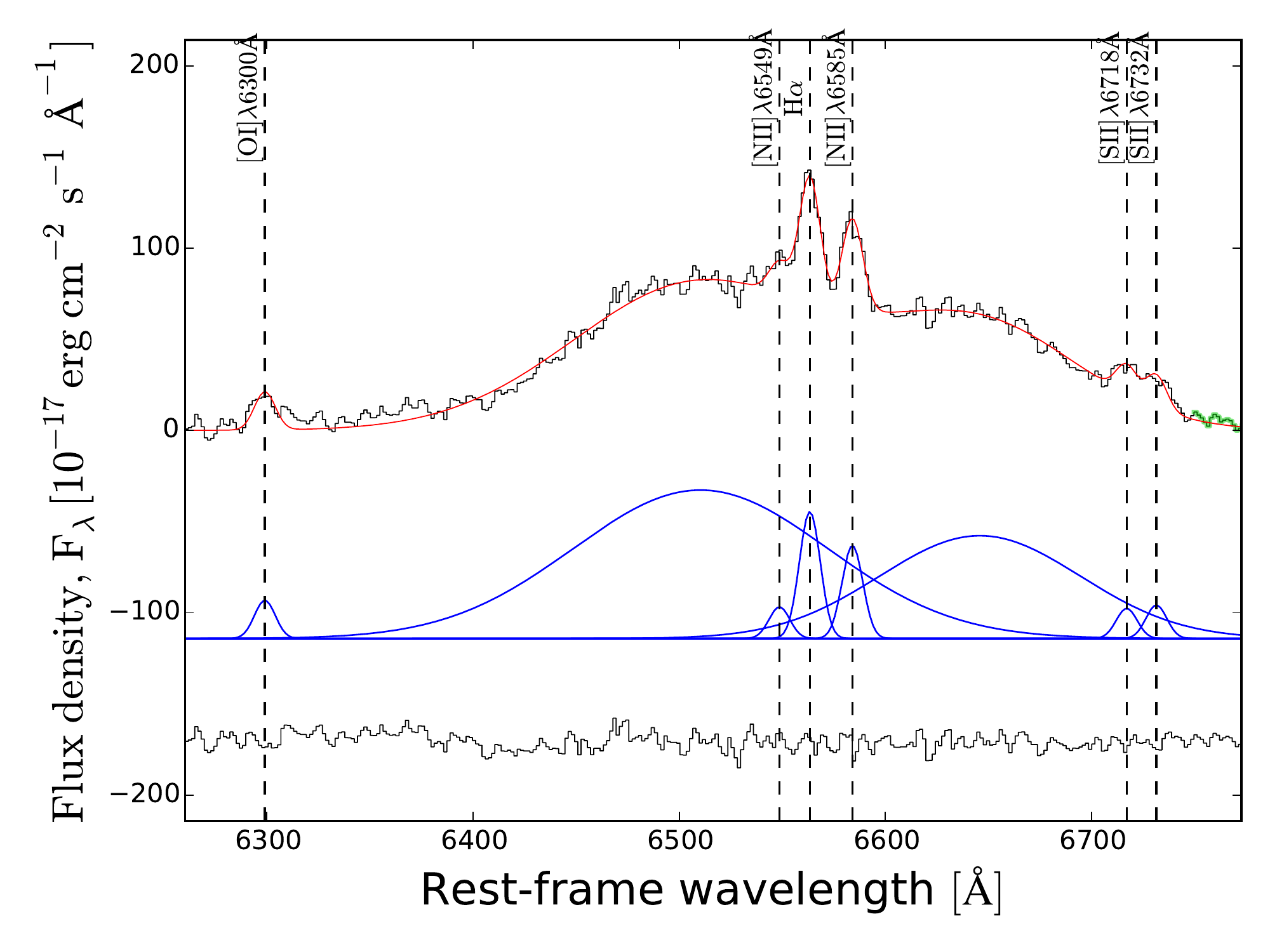}
\includegraphics[width=0.5\textwidth]{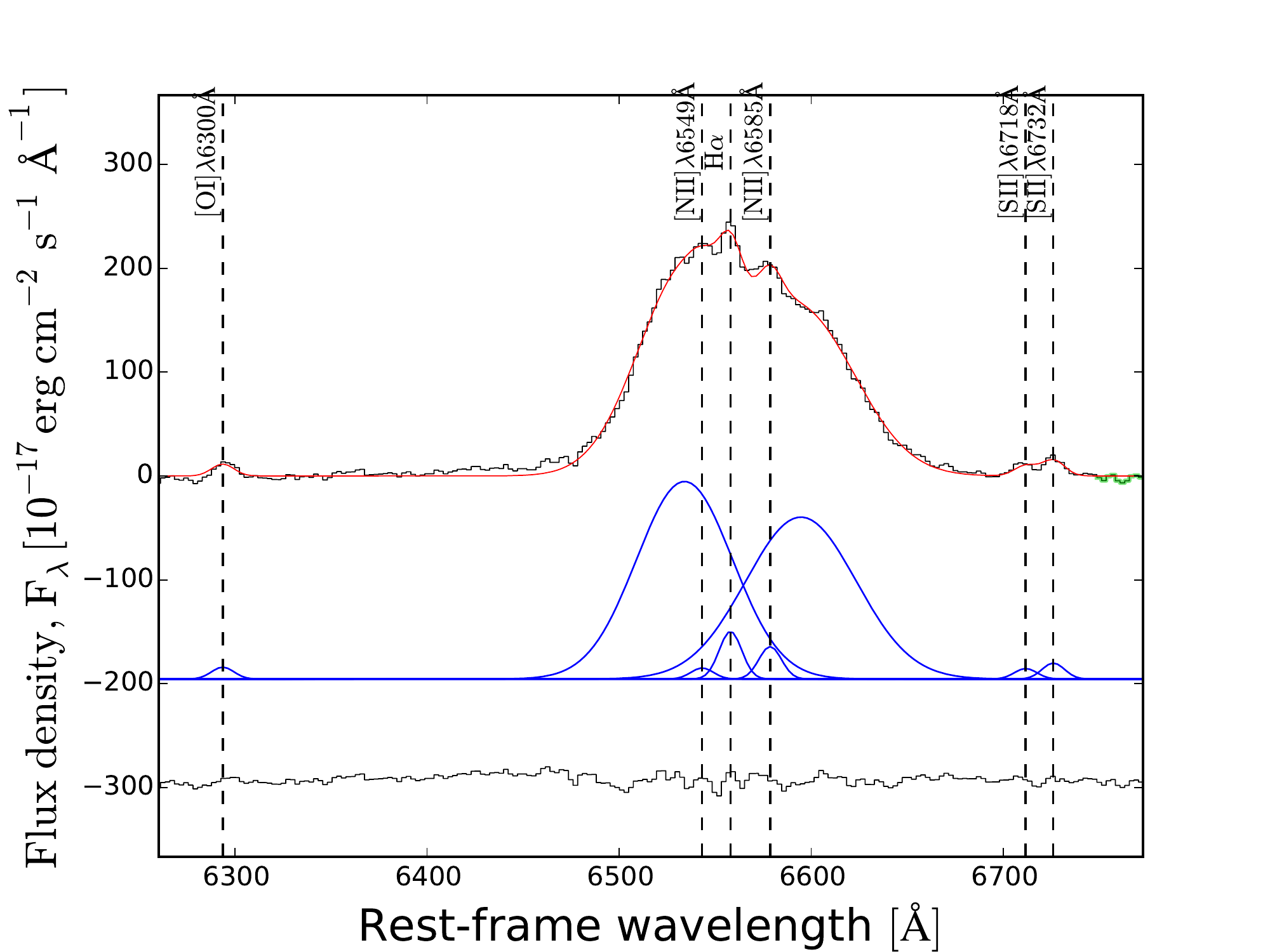}\\
\vfill
\includegraphics[width=0.49\textwidth]{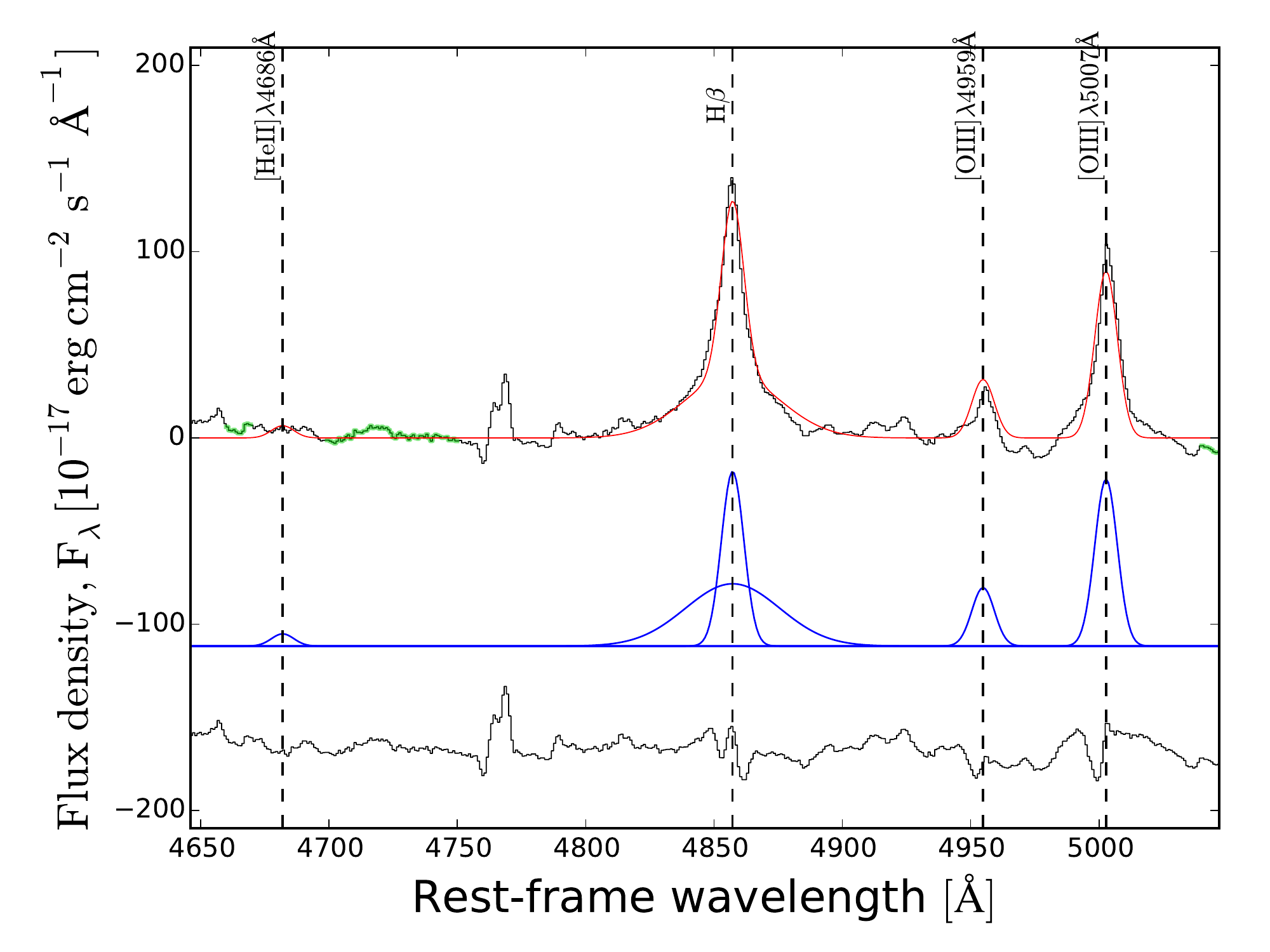}
\includegraphics[width=0.49\textwidth]{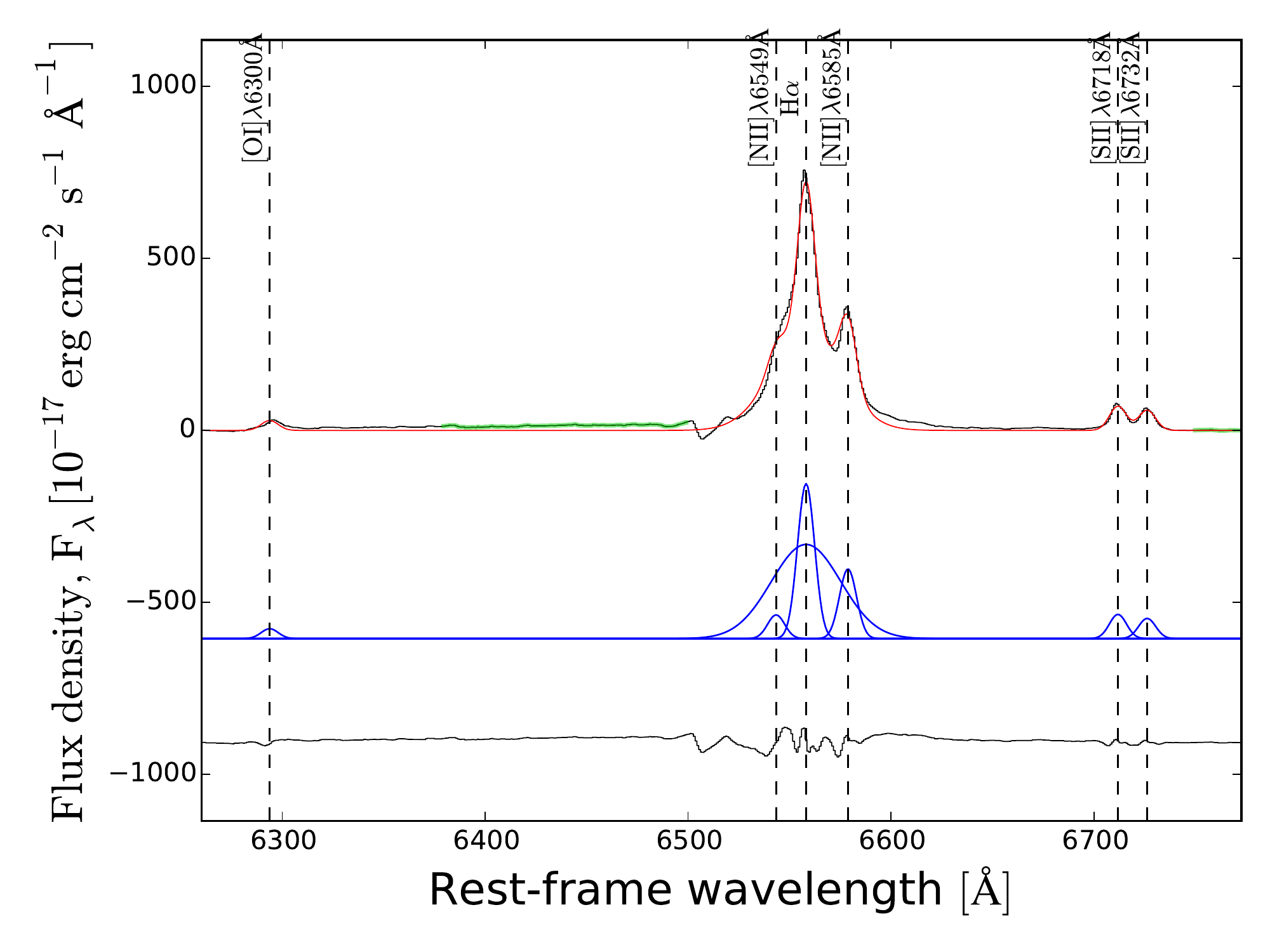}
\caption{Examples of unusual AGN.  \textit{Top Row}: Optical spectrum of ``naked'' candidate 2MASX J19263018+4133053 showing the \Halpha\ region (left) and \swiftxrt (red) and \swiftbat (black) spectrum (right).  We see no evidence of a broad \Halpha\ component in the optical and no evidence of obscuration in the X-rays. \textit{Middle Row} Optical spectrum of double broad line sources showing the \Halpha\ region of ESO 359-G 019 (left) and IGR J22292+6647 (right). The fit is composed of narrow Gaussian components added to two broad \Halpha\ components (in blue). The residual is shown below in grey. \textit{Bottom Row}:  Optical spectrum of the type 1 optically elusive Mrk 507 showing the \Hbeta\ region (left) and the \Halpha\ region (right). Both \Halpha\ and \Hbeta\ show a clear broad component, but weak narrow emission lines characteristic of an HII region. The residual is shown below in grey.}
\label{intobfig}
\end{figure*}

As we studied the properties of our sample in the optical and X-rays, a number of objects showed interesting and unusual characteristics. Some examples of these objects are presented in Figure~\ref{intobfig}. 
We discuss here four different types:  AGN with very low X-ray column density, but lacking broad emission lines, or vice versa; double broad line AGN; and weak line AGN.  We note that because the X-ray and optical spectroscopy are not simultaneous the contradictory optical and X-ray classification could be caused by variability.  \\

\subsubsection{AGN with contradictory optical and X-ray classification}
\label{subsec:naked_agn}

%'naked AGN' have been reported by \cite{hawk} who studied 800 quasars and emission line galaxies during a period of 25 years. Around 10\% of the type-II AGNs of his sample show strong variability in optical luminosity. His claim is that their variation in the continuum implies that we see the active nucleus directly. So if we do not see any broad emission, it means that there is no high velocity gas clouds in the BLR.

So-called ``naked'' AGN candidates \citep{Hawkins:2004:519, Panessa:2006:173} are objects showing an optical spectrum with no detectable broad emission lines in the optical (Seyfert 2) and no obscuration in the X-rays (\NH$<10^{20.5}$
\nhunit). Therefore, they are intriguing because they contradict the basic expectation from the geometrical unification scheme of AGN. Six AGN in our sample ($\approx1\%$) satisfy the ``naked" AGN candidate criteria 
(2MASX J01302127-4601448, SDSS J155334.73+261441.4,
LCRS B232242.2-384320, 2MASX J11271632+1909198, 2MASX J19263018+4133053, and PKS 2331-240): their optical spectra classify them as Seyfert 2, but we observe little obscuration in the X-rays (\NH$<10^{20.5}$ \nhunit) with the 90\% error bars below (\NH$<10^{21}$ \nhunit).  \\

%\subsection{
Another interesting class of AGN are objects which have broad Balmer emission lines (Seyfert 1, 1.2, and 1.5 AGN), but very high column densities of \NH$>10^{23}$\,\nhunit with \NH$>10^{22.5}$\,\nhunit for all 90\% error bars.  The five AGN in our sample that satisfy these criteria are Mrk 975, CGCG 031-072, WISE J144850.99-400845.6, 3C 445, and 2MASXJ19301380+3410495.  3C 445 was already known to be a peculiar broad line radio galaxy with an X-ray absorbed spectrum that has multiple X-ray absorption components consistent with our findings \citep[][]{Grandi:2007:L21,Reeves:2010:803}.  We note however that we do not find any Compton-thick Seyfert 1, 1.2, or 1.5, with the maximum column density of these sources never exceeding \NH$= 10^{23.7}$\,\nhunit.
%

%\begin{figure}[h!]
%\hfill
%\subfigure{\includegraphics[width=0.49\textwidth]{spec-2091-53447-0584Ha_very_BROAD.png}}
%\hfill
%\subfigure{\includegraphics[width=0.49\textwidth]{MGC+09-21-096_n12_deredHa_very_BROAD.png}}
%\hfill
%\caption{Optical spectrum of the \Halpha\ region of FBQS  J110340.2+372925 (left) and MGC +09-21-096 (right). The fit is composed of 6 narrow gaussian components added to 2 broad \Halpha\ components (in blue). The residual is shown below in grey. The spectrum was previously redshift and Miky-Way extinction corrected, and also continuum subtracted.}
%\label{broad_spectrum}
%\end{figure}

%\subsubsection{XBONGs and optically elusive AGNs}
%\begin{figure}[h!]
%\hfill
%\subfigure{\includegraphics[width=0.49\textwidth]{Mrk_507hb.png}}
%\hfill
%\subfigure{\includegraphics[width=0.49\textwidth]{Mrk_507ha.png}}
%\hfill
%\caption{Optical spectrum of the \Hbeta\ region (left) and the \Halpha\ region (right) of the type-I optically elusive Mrk 507. Both \Halpha\ and \Hbeta\ show a clear broad component. The residual is shown below in grey. The spectrum was previously redshift and Miky-Way extinction corrected, and also continuum subtracted.}
%\label{lia}
%\end{figure}
%%

\subsubsection{Double Broad Line AGN}
\label{subsec:double_broad_agn}

This sub-class of broad line AGN show two broad and well-separated (in velocity space) \Halpha\ emission profiles. 
Previous studies have suggested several possible explanations for the origin of there double broad lines, including: the relativistic accretion disk; a binary BLR in a binary BH system; bipolar outflows, or a spherically symmetric BLR illuminated by an anisotropic ionizing radiation source \citep[see, e.g.,][]{Eracleous:1994:1,Eracleous:2009:133}. 
A close visual inspection of our sample reveals only seven sources with such features (FBQS J110340.2, 2MASX J08032736, 3C 332, NGC 4235, MCG +09-21-096, 2MASX J21320220, ESO 359-G019).\\

\subsubsection{Weak Line AGN}
\label{subsec:elusive_agn}

The last category of peculiar objects we consider are AGN which lack some or all of the narrow line emission typical of AGN and cannot be studied using emission line diagnostics. This category comprises \NBPTniiweaklineper\ of the sample (\NBPTniiweakline/\Ntot) using the \NII/\Halpha\ emission line diagnostic.  Only three of the weak line AGN lack \emph{any} detectable emission lines despite having high quality spectra. These sources are consistent with X-ray bright optically normal galaxies \citep[e.g. XBONGS,][]{Comastri:2002:771}. The XBONGS are 2MASX J04595677+3502536, 2MASX J13553383+3520573, and ESO 436-G034.  We have verified that the association of the BAT X-ray sources to other counterparts was not erroneous, by verifying that their soft X-ray counterparts are the brightest counterpart in the field-of-view; that these AGN are not associated with known (background) blazars or beamed AGN; and that the optical spectra of these sources have high signal to noise in the continuum ($S/N>10$).  Our results are consistent with the idea that XBONGs are exceedingly rare ($<0.5$\% at most) and confirms the idea that the large fractions found in distant X-ray surveys is likely because of host galaxy dilution and the difficulty detecting emission lines in dusty galaxies \citep[e.g.,][]{Moran:2002:L71}.    \\

%Optically elusive AGNs refer to active galaxies with emission line ratios that classify them in the HII or Composite region.  Among our sample MCG +04-22-042, Mrk 1044, and SWIFT J1143 satisfy this criteria.

% Number of AGN from Hbeta measurements

\section{Summary, Conclusions and Future Work}
\label{sec:summary}

We present the first catalog and data release of the \emph{Swift-BAT Spectroscopic Survey} (``BASS''). 
Starting from an all-sky catalog of AGN detected in the 14--195\,keV band, we analyze a total of \Ntot\ AGN, and host galaxies, using a compilation of optical spectra from public surveys and dedicated campaigns. 
This spectroscopic data set allows us to measure strong, narrow and broad emission lines, stellar velocity dispersions, and to derive estimates of black hole masses (\mbh) and accretion rates (\lledd).
Our main findings are:

\begin{enumerate}
\renewcommand{\theenumi}{(\roman{enumi})}
\renewcommand{\labelenumi}{\theenumi}

\item 
There is a continuous increase in the fraction of broad line (type 1) AGN, both with broad \Hbeta\ and/or \Halpha, with increasing 14--195\,keV and 2--10\,keV X-ray luminosities. 
Also, the classification of obscured and unobscured sources based on the FWHM of the Balmer emission lines shows broad agreement with those based on the X-ray obscuration, with about 94\% of AGN being consistently classified, for the threshold set at \NH$\simeq 10^{21.9}$\,\nhunit. The sources classified as Seyfert 1.9 show a range of column densities, however.

\item 
Compared to narrow line AGN in the SDSS, the X-ray selected AGN in our sample that have emission lines include a much larger fraction of dustier galaxies ($\ha/\hb>5$). We find that the X-ray selected AGN show a relatively constant type 1 to type 2 fraction of about 60\%, over a broad range of redshift, while the same fraction among SDSS AGN is much lower and shows a strong dependence on redshift (2\%--30\%). The average \OIII\ luminosity and velocity dispersion of BAT AGN are higher than SDSS AGN, consistent with their brighter X-ray emission, and the smaller number of BAT AGN per sky area.

\item Using the \NII/\Halpha\ emission line diagnostic, about half (\NBPTniiseyper, \NBPTniisey/\Ntot) of the BAT AGN are classified as Seyferts, with a few percent classified in each of the sub-classes of LINERs, Composite galaxies, or HII regions. Another 15\% reside in dusty galaxies, where the upper limits on \Hbeta\ imply either a Seyfert or LINER (\NBPTniiagnper, \NBPTniiagn/\Ntot). Finally, about 20\% reside in galaxies with weak or no emission lines or are associated with known blazars or beamed AGN. The weaker lines involved in other diagnostics (\SII/\Halpha, \OI/\Halpha, \OIII/\OII, and \heii/\Hbeta) have a lower detection fraction, but overall the sample is dominated by Seyfert AGN.  

\item 
We find that the accretion rates of Seyfert 1 AGN (in terms of \lledd) are higher than those of Seyfert 2, mainly because Seyfert 1 AGN have higher bolometric luminosities. With increasing redshift the survey tends to find higher \lledd\ systems. Finally, using Eddington ratios is highly efficient at separating beamed AGN from unbeamed sources because typical beamed AGN are above the Eddington limit.

%\item Among ten sources where the BAT SC is well measured ($SC_{BAT \ error}<0.08$), we find that the SC of the majority of sources 7/10 vary significantly when compared to the single \nustar observation.  This suggests that the majority of AGN have significant variability above 10~keV.

\end{enumerate}

The present work provides a broad overview of the optical spectra of the BAT hard X-ray selected AGN ($>10$\,keV). 
In future studies of this sample we will address in detail specific aspects of AGN physics and SMBH growth. 
Among the many follow up opportunities, we note our recently published study of correlations between X-ray continuum emission and narrow line emission \cite[e.g., \oiii;][]{Berney:2015:3622};
a large study of the NIR spectra of over 100 of the BAT AGN \citep{Lamperti17}; %Additionally, many of the spectra of merger companions were taken with the primary AGN, so 
a study of trends in accretion rate with merger stage in interacting AGN hosts (Koss et al., in prep).
Another study investigates the role of accretion rate in emission line ratios \citep{Oh:2017:1466} and X-ray properties such as $\Gamma$ (Trakhtenbrot et al., submitted), and obscuration (Ricci et al., in prep). 
Finally, future optical spectroscopy studies will use deeper BAT maps that are now available to study fainter sources (Oh et al., in prep).
We therefore expect that the BASS sample will enable a wide variety of AGN studies in the local universe, and will serve as an important benchmark for high-redshift AGN detected in deep, small-area surveys.

%%%%%% BT GOT HERE %%%%%%%%%%%%%%%%%%%%%%%%%%%%%%%%%%%%%%%%%%%%%%%%
%{\color{red}\todo\todo\todo\ -- {\Large BT got up to here} \todo}\\
%%%%%%%%%%%%%%%%%%%%%%%%%%%%%%%%%%%%%%%%%%%%%%%%%%%%%%%%%%%%%%%%%

%%%% ACK  %%%%%%%%%%%%%%%%%%%%%%%%%%%%%%%%%%%%%%%%%%%%%%%%%%%%%%%%%%%%%%%%%%%%%%%%%%%%%%%%%%%%%%%%%%%
\acknowledgements
M.\,K. acknowledges support from the Swiss National Science Foundation (SNSF) through the Ambizione fellowship grant PZ00P2\textunderscore154799/1 and SNSF grant PP00P2 138979/1, K.\,O. and K.\,S. acknowledge support from the SNSF through Project grant 200021\textunderscore157021.    M.\,B. acknowledges support from NASA Headquarters under the NASA Earth and Space Science Fellowship Program, grant NNX14AQ07H.  Support for the work of E.\,T. was provided by the Center of Excellence in Astrophysics and Associated Technologies (PFB 06), by the FONDECYT regular grant 1120061 and by the CONICYT Anillo project ACT1101.  M.\,K. would like to thank Di Harmer at the NOAO for teaching him how use the Goldcam spectrograph on his first optical spectroscopy run.  \\

This paper used archival optical spectroscopic data from several telescopes.  Kitt Peak National Observatory, National Optical Astronomy Observatory, is operated by the Association of Universities for Research in Astronomy (AURA), Inc., under cooperative agreement with the National Science Foundation.The Kitt Peak National Observatory observations were obtained using MD-TAC time as part of the thesis of M.K. (2008A-0393,2009B-0295) and L.W. at the University of Maryland. We also acknowledge the following people who assisted in the Palomar observations presented herein: Kristen Boydstun, Clarke Esmerian, Carla Fuentes, David Girou, Ana Glidden, Hyunsung Jun, George Lansbury, Ting-Ni Lu, Alejandra Melo, Eric Mukherjee, Becky Tang, and Dominika Wylezalek.  This paper uses observations made at the South African Astronomical Observatory (SAAO).\\

Data in this paper were acquired through the Gemini Science Archive and processed using the Gemini IRAF package and Gemini python.  Data from Gemini programs GN-2009B-Q-114, GN-2010A-Q-35, GN-2011A-Q-81, GN-2011B-Q-96, GN-2012A-Q-28, GN-2012B-Q-25, GS-2010A-Q-54, and GS-2011B-Q80 were used in this publication and included NOAO-granted community-access time for 2011B-0559 (PI Koss).  Based on observations obtained at the Gemini Observatory and processed using the Gemini IRAF package, which is operated by the Association of Universities for Research in Astronomy, Inc., under a cooperative agreement with the NSF on behalf of the Gemini partnership: the National Science Foundation (United States), the National Research Council (Canada), CONICYT (Chile), Ministerio de Ciencia, Tecnolog\'{i}a e Innovaci\'{o}n Productiva (Argentina), and Minist\'{e}rio da Ci\^{e}ncia, Tecnologia e Inova\c{c}\~{a}o (Brazil). The authors wish to recognize and acknowledge the very significant cultural role and reverence that the summit of Mauna Kea has always had within the indigenous Hawaiian community. We acknowledge the efforts of the staff of the Australian Astronomical Observatory (AAO), who developed the 6dF instrument and carried out the observations for the survey.  We are most fortunate to have the opportunity to conduct observations from this mountain.\\

Funding for SDSS-III has been provided by the Alfred P. Sloan Foundation, the Participating Institutions, the National Science Foundation, and the U.S. Department of Energy Office of Science. The SDSS-III web site is http://www.sdss3.org/.  SDSS-III is managed by the Astrophysical Research Consortium for the Participating Institutions of the SDSS-III Collaboration including the University of Arizona, the Brazilian Participation Group, Brookhaven National Laboratory, University of Cambridge, Carnegie Mellon University, University of Florida, the French Participation Group, the German Participation Group, Harvard University, the Instituto de Astrofisica de Canarias, the Michigan State/Notre Dame/JINA Participation Group, Johns Hopkins University, Lawrence Berkeley National Laboratory, Max Planck Institute for Astrophysics, Max Planck Institute for Extraterrestrial Physics, New Mexico State University, New York University, Ohio State University, Pennsylvania State University, University of Portsmouth, Princeton University, the Spanish Participation Group, University of Tokyo, University of Utah, Vanderbilt University, University of Virginia, University of Washington, and Yale University.\\

Finally, we wish to acknowledge several community software resources and websites.
IRAF is distributed by the National Optical Astronomy Observatory, which is operated by the Association of Universities for Research in Astronomy (AURA) under a cooperative agreement with the National Science Foundation.             This research has made use of the NASA/IPAC Extragalactic Database (NED) which is operated by the Jet Propulsion Laboratory, California Institute of Technology, under contract with the National Aeronautics and Space Administration.   This research made use of Astropy, a community-developed core Python package for Astronomy \citep{Collaboration:2013:A33}.  This research made use of APLpy, an open-source plotting package for Python hosted at http://aplpy.github.com.  This research has made use of the SIMBAD database, operated at CDS, Strasbourg, France.\\

%%%%%%%%%%%%%%%%%%%%%%%%%%%%%%%%%%%%%%%%%%%%%%%%%%%%%%%%%%%%%%%%%%%%%%%%%%%%%%%%%%%%%%%%%%%%%%%

%%%%%%%%%%%%%%%%%%%%%%%%%%%%%%%%%%%%%%%%%%%%%%%%%%%%%%%%%%%%%%%%%%%%%%%%%%%%%%%%%%%%%%%%%%%%%%%

%%%%%%%%%%%%%%%%%%%%%%%%%%%%%%%%%%%%%%%%%%%%%%%%%%%%%%%%%%%%%%%%%%%%%%%%%%%%%%%%%%%%%%%%%%%%%%%

%L_bol from X-ray vs L_bol from optical

% Spectra of high Eddington ratio sources
%\begin{figure*}
%\centering
%\subfigure{\includegraphics[width=8.2cm]{Mrk348_single_3900-5500.pdf}}
%\subfigure{\includegraphics[width=8.2cm]{Mrk348_single_3900-5500_zoom.pdf}}
%\vfill
%\subfigure{\includegraphics[width=8.2cm]{2MASXJ18241083+1846088_single_3900-5500.pdf}}
%\subfigure{\includegraphics[width=8.2cm]{2MASXJ18241083+1846088_single_3900-5500_zoom.pdf}}
%\vfill
%\caption{Fit of the spectra of Mrk348 (upper panel) and 2MASXJ18241083+1846088 (bottom panels) taken with the Perkins telescope. The left panels show the fit of whole wavelength range (3900-7000\,\AA), the right panel the fit of the CaH+K and MgIb regions. The quality flag is 2 for  and the absorption features are fitted quite well by \ppxf.}
%\label{2sources}
%\end{figure*}

%\begin{figure*}
%\centering
%\subfigure{\includegraphics[width=8.2cm]{spec-2142-54208-0436_3900-7000.pdf}}
%\subfigure{\includegraphics[width=8.2cm]{spec-2142-54208-0436_3900-7000_zoom.pdf}}
%\vfill
%\caption{Fit of the spectrum of CGCG 164-019 taken with SDSS. The left panel shows the fit of whole wavelength range (3900-7000\,\AA), the right panel the fit of the CaH+K and MgIb regions. The quality flag is 1 and the absorption features are well fitted by \ppxf.}
%\label{CGCG}
%\end{figure*}

\begin{turnpage}
\begin{center}
\begin{table*}
\begin{minipage}{\textwidth}
%\tabletypesize{\footnotesize}
\caption{Optical spectra} 
\begin{tabular}{llccccccccccc}
\tableline
\tableline
\multirow{2}{*}{ID\tablenotemark{a}}
&\multirow{2}{*}{Counterpart Name}
&\multirow{2}{*}{Source}
&\multirow{2}{*}{$z_{\rm \big[OIII\big]}$\tablenotemark{b}}
&Distance
&log $L_{\rm 14-195}$\tablenotemark{c}
&log $L_{\rm bol}$\tablenotemark{d}
&Date
&Exp. 
&Slit width 
%&\multirow{2}{*}{\ha/\hb\tablenotemark{d}}
&\multirow{2}{*}{Type\tablenotemark{e}}
%&\multirow{2}{*}{pPXF\tablenotemark{f}}
&\multirow{2}{*}{Beamed\tablenotemark{f}}
&\multirow{2}{*}{$N_{\rm H}$\tablenotemark{g}} \\
&
&
&
& (Mpc)
& (\ergs)
& (\ergs)
& dd/mm/yyyy
& (s)
& (kpc) \\
\tableline
   1&        2MASX J00004876-0709117&                           SDSS&                         0.037&                        165.14&                         43.63&                         44.53&                    25/10/2013&                          5401&                          1.54&                 1.9&                   0&               Obs.\\
2&                   Fairall 1203&                            6DF&                         0.058&                        261.64&                         43.92&                         44.82&                    01/09/2005&                          1200&                          8.02&                 1.9&                   0&  $<10^{20}cm^{-2}$\\
4&        2MASX J00032742+2739173&                           SDSS&                         0.040&                        175.13&                         43.68&                         44.58&                    09/09/2013&                          4500&                          1.63&                 2.0&                   0&               Obs.\\
5&        2MASX J00040192+7019185&                        Masetti&                         0.096&                        442.73&                         44.47&                         45.38&                    27/11/2006&                          1800&             $\cdot\cdot\cdot$&                 1.9&                   0&               Obs.\\
6&                        Mrk 335&                        Perkins&                         0.026&                        113.28&                         43.45&                         44.36&                    01/04/2011&                          1800&                          1.07&                 1.2&                   0&             Unobs.\\
7&        2MASX J00091156-0036551&                           SDSS&                         0.073&                        331.27&                         44.09&                         44.99&                    06/09/2000&                          2700&                          4.49&                 2.0&                   0&               Obs.\\
8&                       Mrk 1501&                         Gemini&                         0.089&                        408.45&                         44.80&                         45.70&                    17/08/2012&                           595&                          1.82&                 1.5&                   1&             Unobs.\\
10&        2MASX J00210753-1910056&                            6DF&                         0.096&                        439.38&                         44.60&                         45.50&                    27/08/2003&                          1200&                         13.01&                 1.9&                   0&             Unobs.\\
13&        2MASX J00253292+6821442&                        Palomar&                         0.012&                         53.99&                         42.80&                         43.71&                    23/12/2014&                           150&                          0.50&                 2.0&                   0&               Obs.\\
14&        2MASX J00264073-5309479&                            6DF&                         0.063&                        283.63&                         44.13&                         45.04&                    03/07/2005&                          1200&                          8.62&                 1.9&                   0&  $<10^{20}cm^{-2}$\\
\tableline
\tablenotetext{a}{\textit{Swift}-BAT 70-month hard X-ray survey ID (http://swift.gsfc.nasa.gov/results/bs70mon/).}
\tablenotetext{b}{Redshift measured from \OIII. If this line was not available due to instrumental features, the \Hbeta\ or \Halpha\ emission line redshift was measured. For high redshift sources ($z>0.3$), the \MgII\ or \CIV\ emission line was measured.}
\tablenotetext{c}{\textit{Swift}-BAT X-ray luminosity (14-195 keV).}
\tablenotetext{d}{Bolometric luminosity estimated from the \textit{Swift}-BAT X-ray luminosity (14-195 keV).}
\tablenotetext{e}{AGN classification following \citet{Osterbrock:1981:462}.}
\tablenotetext{f}{Flag presenting beamed AGN (`1').}
\tablenotetext{g}{Obscuration flag distinguished by hydrogen column density: `Obs.' for $N_{\rm H}>10^{22} {\rm cm}^{-2}$ and `Unobs.' for $N_{\rm H}< 10^{22} {\rm cm}^{-2}$. Further details on the column density can be found in Ricci et al. (submitted).  }
\end{tabular}
\tablecomments{(This table is available in its entirety in a machine-readable form in the online journal. A portion is shown here for guidance regarding its form and content.)}
\label{tab:general_info}
\end{minipage}
\end{table*}
\end{center}
\end{turnpage}

\begin{table}
%\begin{minipage}{\textwidth}
%\tabletypesize{\scriptsize}
\begin{center}
\caption{Emission lines masked in \ppxf\ host galaxy fitting} 
\begin{tabular}{ll}
\tableline
\tableline
Emission line
&Wavelength [\AA]  \\
\tableline
[OII] & 3726.03 \\ 
 & 3728.82 \\ 
CaH & 3968.47 \\ 
H$\delta$ & 4101.76 \\ 
H$\gamma$ & 4340.47 \\ 
HeII & 4686.00 \\
H$\beta$ & 4861.33 \\ 

 [OIII] & 4958.92 \\ 
 & 5006.84 \\ 
 
[NI] & 5200.00 \\ 

[FeVII] & 5721.00 \\ 
NaD & 5890.00 \\ 
NaD & 5896.00\\
 
[OI] & 6300.3 \\ 

[NII] & 6548.03 \\ 
 & 6583.41 \\ 
H$\alpha$ & 6562.8 \\ 

[SII] & 6716.47 \\ 
 & 6730.85 \\ 
Sky & 5577.00 \\ 
Sky & 6300.00 \\ 
Sky & 6363.00 \\ 
Sky & 6863.00 \\ 
\tableline
\end{tabular}
\label{tab:emlinesmask}
\end{center}
%\end{minipage}
\end{table}

\begin{table*}
\begin{minipage}{\textwidth}
%\tabletypesize{\scriptsize}
\begin{center}
\caption{Stellar velocity dispersion measurements} 
\begin{tabular}{lcccccccccccc}
\tableline
\tableline
\multirow{2}{*}{ID\tablenotemark{a}}
& \multirow{2}{*}{Source}
& \multirow{2}{*}{Redshift\tablenotemark{b}}
& $\sigma$
& \multirow{2}{*}{log $M_{\rm BH}/M_{\odot}$}
& \multirow{2}{*}{${\rm flag}_{\sigma}$\tablenotemark{c}}
& \multirow{2}{*}{Ca H+K\tablenotemark{d}}
& \multirow{2}{*}{Mgb\tablenotemark{e}}
& $\sigma_{\rm CaT}$\tablenotemark{f} 
& \multirow{2}{*}{CaT\tablenotemark{g}}
& $\sigma_{\rm lit.}$
& \multirow{2}{*}{$\log (M_{\rm BH}/M_{\odot})_{\rm lit.}$\tablenotemark{h}}
& \multirow{2}{*}{Ref.\tablenotemark{i}}\\
&
&
& (\kms)
&
&
&
&
& (\kms)
&
& (\kms) 
&
&
\\
\tableline
1&      SDSS&              0.03767&            $152\pm6$&        $7.97\pm0.30$&    1&     1&   1&  $\cdot \cdot \cdot$&  $\cdot \cdot \cdot$&    $\cdot\cdot\cdot$&       $\cdot \cdot \cdot$&      $\cdot \cdot \cdot$\\
2&       6DF&              0.05846&           $180\pm38$&  $\cdot \cdot \cdot$&    7&     1&   1&  $\cdot \cdot \cdot$&  $\cdot \cdot \cdot$&    $\cdot\cdot\cdot$&       $\cdot \cdot \cdot$&      $\cdot \cdot \cdot$\\
4&      SDSS&              0.03970&            $142\pm7$&        $7.85\pm0.30$&    1&     1&   1&  $\cdot \cdot \cdot$&  $\cdot \cdot \cdot$&    $\cdot\cdot\cdot$&       $\cdot \cdot \cdot$&      $\cdot \cdot \cdot$\\
5&   Masetti&  $\cdot \cdot \cdot$&  $\cdot \cdot \cdot$&  $\cdot \cdot \cdot$&    9&     1&   1&  $\cdot \cdot \cdot$&  $\cdot \cdot \cdot$&    $\cdot\cdot\cdot$&       $\cdot \cdot \cdot$&      $\cdot \cdot \cdot$\\
6&   Perkins&  $\cdot \cdot \cdot$&  $\cdot \cdot \cdot$&  $\cdot \cdot \cdot$&    9&     1&   1&  $\cdot \cdot \cdot$&  $\cdot \cdot \cdot$&    $\cdot\cdot\cdot$&       $\cdot \cdot \cdot$&      $\cdot \cdot \cdot$\\
7&      SDSS&              0.07334&           $250\pm16$&        $8.91\pm0.32$&    1&     1&   1&  $\cdot \cdot \cdot$&  $\cdot \cdot \cdot$&    $\cdot\cdot\cdot$&       $\cdot \cdot \cdot$&      $\cdot \cdot \cdot$\\
8&    Gemini&  $\cdot \cdot \cdot$&  $\cdot \cdot \cdot$&  $\cdot \cdot \cdot$&    9&     0&   1&  $\cdot \cdot \cdot$&  $\cdot \cdot \cdot$&    $\cdot\cdot\cdot$&       $\cdot \cdot \cdot$&      $\cdot \cdot \cdot$\\
10&       6DF&              0.09561&           $284\pm21$&        $9.16\pm0.32$&    2&     1&   1&  $\cdot \cdot \cdot$&  $\cdot \cdot \cdot$&    $\cdot\cdot\cdot$&       $\cdot \cdot \cdot$&      $\cdot \cdot \cdot$\\
13&   Palomar& $\cdot \cdot \cdot$& $\cdot \cdot \cdot$&  $\cdot \cdot \cdot$&    9&     1&   1&  $\cdot \cdot \cdot$&                    9&    $\cdot\cdot\cdot$&       $\cdot \cdot \cdot$&      $\cdot \cdot \cdot$\\
14&       6DF&              0.06286&           $326\pm49$&  $\cdot \cdot \cdot$&    7&     1&   1&  $\cdot \cdot \cdot$&  $\cdot \cdot \cdot$&    $\cdot\cdot\cdot$&       $\cdot \cdot \cdot$&      $\cdot \cdot \cdot$\\

\tableline
\tablenotetext{a}{\swiftbat 70-month hard X-ray survey ID (http://swift.gsfc.nasa.gov/results/bs70mon/).}
\tablenotetext{b}{Redshift measured from the stellar template.}
\tablenotetext{c}{Quality flag: 1= excellent fit with small error ($\langle \sigs \rangle$ error=11\,\kmpssh, $\sigs<34$\,\kmpssh), 2= larger errors than flag 1 ($\langle \sigs \rangle$ error=23\,\kmpssh, $\sigs<60$\,\kmpssh), but acceptable fit, 3= bad fit with high S/N, 7= presence of broad component at \hb\ or \ha, 8= very weak absorption features, 9=bad fit.}
\tablenotetext{d}{flag= 1 when \Cahk\ is fitted.}
\tablenotetext{e}{flag= 1 when \mgb\ is fitted.}
\tablenotetext{f}{\sigs\ measured from \caii\ triplet.}
\tablenotetext{g}{flag= 1 when \caii\ triplet is fitted.}
\tablenotetext{h}{Black hole mass from literature: C05 \citep{Capetti:2005:465}; C09 \citep{Cappellari09}; D03 \citep{Devereux03}; H05 \citep{Herrnstein05};  K08 \citep{Kondratko:2008:87};  K11 \citep{Kuo:2011:20}; L03 \citep{Lodato03};  M11 \citep{Medling:2011:32};  O14 \citep{Onken:2014:37}; RJ06 \citep{Rothberg2006}; T03 \citep{Tadhunter:2003:861}; TYK05 \citep{Trotter98, Yamauchi04, Kondratko05}; W06 \citep{Wold06}; W12 \citep{Walsh12}}
\tablenotetext{i}{Reference for $\sigma_{\rm lit.}$: C04 \citep{CidFernandes:2004:273}; F00 \citep{Ferrarese:2000:L9}; G05 \citep{Garcia-Rissmann:2005:765}; G13 \citep{Grier:2013:90}; Hy (http://leda.univ-lyon1.fr); H09 \citep{Ho:2009:1}; L17 \citep{Lamperti17}; M13 \citep{McConnell:2013:184}; NW95 \citep{Nelson:1995:67}; N04 \citep{Nelson:2004:652}; RJ06 \citep{Rothberg2006}; V15 \citep{vandenBosch:2015:10}} 
\end{tabular}
\tablecomments{(This table is available in its entirety in a machine-readable form in the online journal. A portion is shown here for guidance regarding its form and content.)}
\label{tab:ppxf}
\end{center}
\end{minipage}
\end{table*}

\begin{table*}
\begin{minipage}{\textwidth}
\begin{center}
\caption{Emission line Measurements - \OII\ Spectral Region}
\begin{tabular}{lccccccc}
\tableline
\tableline
\multirow{2}{*}{ID\tablenotemark{a}}
&FWHM\tablenotemark{b}
&\NeVa\tablenotemark{c}
&\NeVb\tablenotemark{c}
&\OII\tablenotemark{c}
&\NeIIIa\tablenotemark{c}
&\NeIIIb\tablenotemark{c}
&\multirow{2}{*}{Flag\tablenotemark{d}} \\
& (\kms)
& (\ergcms)
& (\ergcms)
& (\ergcms)
& (\ergcms)
& (\ergcms) \\
\tableline
   1&                     $382\pm4$&             $\cdot\cdot\cdot$&             $\cdot\cdot\cdot$&                   $1.5\pm0.0$&                   $0.7\pm0.0$&                        $<1.2$&    1\\
   2&                    $632\pm55$&             $\cdot\cdot\cdot$&             $\cdot\cdot\cdot$&             $\cdot\cdot\cdot$&                   $5.5\pm0.2$&                        $<4.7$&    2f\\
   4&                     $493\pm2$&             $\cdot\cdot\cdot$&             $\cdot\cdot\cdot$&                   $5.5\pm0.1$&                   $1.4\pm0.0$&                        $<1.1$&    1\\
   5&           $\cdot \cdot \cdot$&                      $<104.0$&                       $<72.9$&                       $<24.0$&                        $<6.5$&                       $<21.2$&    9\\
   6&                    $801\pm16$&             $\cdot\cdot\cdot$&             $\cdot\cdot\cdot$&             $\cdot\cdot\cdot$&                       $<38.0$&                  $28.9\pm0.8$&    2\\
   7&                     $671\pm8$&             $\cdot\cdot\cdot$&             $\cdot\cdot\cdot$&                  $13.1\pm0.1$&                   $2.8\pm0.0$&                        $<1.9$&    1\\
   8&           $\cdot \cdot \cdot$&             $\cdot\cdot\cdot$&             $\cdot\cdot\cdot$&             $\cdot\cdot\cdot$&             $\cdot\cdot\cdot$&             $\cdot\cdot\cdot$&    9\\
  10&           $\cdot \cdot \cdot$&             $\cdot\cdot\cdot$&             $\cdot\cdot\cdot$&             $\cdot\cdot\cdot$&             $\cdot\cdot\cdot$&                        $<1.5$&    2f\\
  13&           $\cdot \cdot \cdot$&                      $<261.4$&                      $<150.9$&                       $<45.3$&                       $<41.5$&                       $<49.5$&    9\\
   14&           $\cdot \cdot \cdot$&             $\cdot\cdot\cdot$&             $\cdot\cdot\cdot$&                        $<4.3$&                        $<7.4$&                        $<5.5$&   2f\\

\tableline
\tablenotetext{a}{\textit{Swift}-BAT 70-month hard X-ray survey ID (http://swift.gsfc.nasa.gov/results/bs70mon/).}
\tablenotetext{b}{FWHM measured from \OII.}
\tablenotetext{c}{Emission line flux ($\times10^{-15}$). Symbols `$\cdot \cdot \cdot$' and `$<$' indicate lack of spectral coverage and 3$\sigma$ upper limit estimation, respectively.}
\tablenotetext{d}{Spectral fitting quality flag: 1= a good fit with small error, 2= acceptable fit, 3= bad fit for high S/N source due to either the presence of broad line component or offset in emission lines, 9= lack of spectral coverage or no emission line is detected, f= poor calibration because a single flux calibration was applied to optical spectra taken over several different nights as in the 6DF spectra.}
\end{tabular}
\tablecomments{(This table is available in its entirety in a machine-readable form in the online journal. A portion is shown here for guidance regarding its form and content.)}
\label{tab:OII_complex}
\end{center}
\end{minipage}
\end{table*}

\begin{table*}
\begin{minipage}{\textwidth}
\begin{center}
\caption{Emission line Measurements - narrow \hb\ Spectral Region}
\begin{tabular}{lcccccc}
\tableline
\tableline
\multirow{2}{*}{ID\tablenotemark{a}}
&FWHM\tablenotemark{b}
&\HeII\tablenotemark{c}
&\hb\tablenotemark{c}
&\multirow{2}{*}{${\rm Flag}_{\rm bH\beta}$\tablenotemark{d}}
&\OIII\tablenotemark{c}
& \multirow{2}{*}{Flag\tablenotemark{e}} \\
& (\kms)
& (\ergcms)
& (\ergcms)
&
& (\ergcms) \\
\tableline
  
 1&                     $249\pm1$&                   $0.3\pm0.0$&                   $1.1\pm0.0$&                             n&                  $16.1\pm0.0$&    1\\
 2&                    $560\pm3$&                        $<5.4$&                   $7.5\pm0.4$&                             n&                  $44.6\pm0.3$&    1f\\
 4&                     $382\pm1$&                   $0.3\pm0.0$&                   $1.2\pm0.0$&                             n&                  $49.4\pm0.0$&    1\\
 5&                    $1336\pm3$&                        $<7.7$&                  $12.4\pm0.1$&                             n&                 $139.6\pm0.2$&    2\\
 6&                     $609\pm27$&             $\cdot\cdot\cdot$&                  $82.0\pm15.7$&                             b&                 $276.2\pm6.5$&    1\\
 7&                     $637\pm2$&                        $<0.9$&                   $4.5\pm0.1$&                             n&                  $38.9\pm0.1$&    1\\
 8&                     $416\pm6$&             $\cdot\cdot\cdot$&                  $14.2\pm2.7$&                             b&                 $181.9\pm2.6$&    1\\
10&                   $463\pm98$&                        $<1.5$&                        $<2.6$&                             n&                  $11.0\pm1.8$&    2f\\
13&                     $330\pm3$&                       $<22.5$&                       $<17.6$&                             n&                  $43.3\pm1.3$&    1\\
 14&                    $506\pm7$&                        $<5.1$&                        $<6.2$&                             n&                  $39.8\pm0.3$&    1f\\

\tableline
\tablenotetext{a}{\textit{Swift}-BAT 70-month hard X-ray survey ID (http://swift.gsfc.nasa.gov/results/bs70mon/).}
\tablenotetext{b}{FWHM measured from narrow \hb.}
\tablenotetext{c}{Emission line flux ($\times10^{-15}$). Symbols `$\cdot \cdot \cdot$' and `$<$' indicate lack of spectral coverage and 3$\sigma$ upper limit estimation, respectively.}
\tablenotetext{d}{Flag discriminating narrow \ha\ (`n') and broad \ha\ (`b').  `h' denotes high-redshift source (see Table~\ref{tab:high_z}).}
\tablenotetext{e}{Spectral fitting quality flag: 1= a good fit with small error, 2= acceptable fit, 3= bad fit for high S/N source due to either the presence of broad line component or offset in emission lines, 9=  lack of spectral coverage or no emission line is detected, f= poor calibration because a single flux calibration was applied to optical spectra taken over several different nights as in the 6DF spectra.}
\end{tabular}
\tablecomments{(This table is available in its entirety in a machine-readable form in the online journal. A portion is shown here for guidance regarding its form and content.)}
\label{tab:Hb_complex}
\end{center}
\end{minipage}
\end{table*}

\begin{table*}
\begin{minipage}{\textwidth}
\begin{center}
\caption{Emission Line Measurements - narrow \ha\ Spectral region} 
\begin{tabular}{lcccccccc}
\tableline
\tableline
\multirow{2}{*}{ID\tablenotemark{a}}
&FWHM\tablenotemark{b}
%&${\rm FWHM}_{\rm br}$\tablenotemark{b}
&\OI\tablenotemark{c}
&\ha\tablenotemark{c}
& \multirow{2}{*}{${\rm Flag}_{\rm bH\alpha}$\tablenotemark{d}}
%&H$\alpha_{\rm br}$\tablenotemark{c}
&\NII\tablenotemark{c}
&\SIIa\tablenotemark{c}
&\SIIb\tablenotemark{c}
&\multirow{2}{*}{Flag\tablenotemark{e}} \\
& (\kms)
& (\ergcms)
& (\ergcms)
& 
& (\ergcms)
& (\ergcms)
& (\ergcms) \\
\tableline
  
  1&                     $221\pm2$&                   $0.6\pm0.0$&                   $7.6\pm0.0$&   b&                   $1.9\pm0.0$&                   $1.4\pm0.0$&                   $1.4\pm0.0$&    1\\
  2&                    $510\pm5$&                   $5.7\pm0.3$&                  $24.4\pm0.2$&   b&                  $22.1\pm0.3$&                   $7.8\pm0.3$&                   $6.9\pm0.2$&    2f\\
  4&                     $426\pm1$&                   $1.1\pm0.0$&                  $10.5\pm0.0$&   n&                   $5.8\pm0.0$&                   $3.4\pm0.0$&                   $2.5\pm0.0$&    1\\
  5&                     $269\pm0$&                   $1.3\pm0.0$&                   $1.7\pm0.1$&   b&                   $1.9\pm0.0$&                   $5.8\pm0.0$&                   $3.9\pm0.1$&    3\\
  6&                     $597\pm53$&                       $<17.5$&                 $243.4\pm6.9$&   b&                  $22.9\pm2.6$&                       $<50.8$&                       $<50.8$&    2\\
  7&                     $624\pm1$&                   $6.8\pm0.1$&                  $24.5\pm0.1$&   n&                  $22.6\pm0.1$&                  $10.6\pm0.1$&                   $9.7\pm0.1$&    2\\
  8&                     $411\pm0$&                  $10.9\pm0.9$&                  $58.9\pm0.8$&   b&                  $24.3\pm0.6$&                  $17.0\pm0.7$&                  $14.9\pm0.8$&    2\\
 10&                   $558\pm70$&                        $<1.9$&                   $4.6\pm1.5$&   n&                   $3.3\pm0.6$&                        $<2.4$&                        $<2.4$&    2f\\
 13&                     $449\pm3$&                        $<5.7$&                  $24.0\pm0.4$&   n&                  $25.0\pm0.2$&                   $9.2\pm0.1$&                   $8.4\pm0.4$&    1\\
  14&                     $506\pm52$&                      $<5.7$&                  $12.2\pm1.4$&   b&                  $13.8\pm2.5$&                   $4.0\pm1.4$&                   $2.0\pm2.2$&    2f\\
  
\tableline
\tablenotetext{a}{\textit{Swift}-BAT 70-month hard X-ray survey ID (http://swift.gsfc.nasa.gov/results/bs70mon/).}
\tablenotetext{b}{FWHM measured from narrow \ha.}
%\tablenotetext{b}{FWHM measured from broad \ha.}
\tablenotetext{c}{Emission line flux ($\times10^{-15}$). Symbols `$\cdot \cdot \cdot$' and `$<$' indicate lack of spectral coverage and 3$\sigma$ upper limit estimation, respectively.}
\tablenotetext{d}{Flag discriminating narrow \ha\ (`n') and broad \ha\ (`b').  `h' denotes high-redshift source (see Table~\ref{tab:high_z}).}
\tablenotetext{e}{Spectral fitting quality flag: 1= a good fit with small error, 2= acceptable fit, 3= bad fit for high S/N source due to either the presence of broad line component or offset in emission lines, 9= lack of spectral coverage or no emission line is detected, f= poor calibration because a single flux calibration was applied to optical spectra taken over several different nights as in the 6DF spectra.}
\end{tabular}
\tablecomments{(This table is available in its entirety in a machine-readable form in the online journal. A portion is shown here for guidance regarding its form and content.)}
\label{tab:Ha_complex}
\end{center}
\end{minipage}
\end{table*}

\begin{table*}
\begin{minipage}{\textwidth}
\begin{center}
\caption{Strong emission line classification}
\begin{tabular}{llccccc}
\tableline
\tableline
ID\tablenotemark{a}
& Counterpart Name
&[NII]/H$\alpha$
&[SII]/H$\alpha$
&[OI]/$\Halpha$
&HeII
&[OIII]/[OII] \\
\tableline 
  
  1&     2MASXJ00004876-0709117 (B)\tablenotemark{b}&              Seyfert &              Seyfert &              Seyfert &              Seyfert &              Seyfert\\
  2&                Fairall1203 (B)&              Seyfert &              Seyfert &              Seyfert &    No HeII detection &  $\cdot \cdot \cdot$\tablenotemark{c}\\
  4&         2MASXJ00032742+2739173&              Seyfert &              Seyfert &              Seyfert &              Seyfert &              Seyfert\\
  5&     2MASXJ00040192+7019185 (B)&    Optically elusive &    Optically elusive &    Optically elusive &    Optically elusive &    Optically elusive\\
  6&                     Mrk335 (B)&                  HII &    Optically elusive &    Optically elusive &              Seyfert &    Optically elusive\\
  7&         2MASXJ00091156-0036551&              Seyfert &              Seyfert &              Seyfert &    No HeII detection &              Seyfert\\
  8&                    Mrk1501 (B)&              Seyfert &              Seyfert &              Seyfert &              Seyfert &  $\cdot \cdot \cdot$\\
 10&         2MASXJ00210753-1910056&            AGN Limit\tablenotemark{d} &    Optically elusive &    Optically elusive &    No HeII detection &    Optically elusive\\
 13&         2MASXJ00253292+6821442&            AGN Limit &            AGN Limit &    Optically elusive &    No HeII detection &  $\cdot \cdot \cdot$\\
 14&     2MASXJ00264073-5309479 (B)&            AGN Limit &            AGN Limit &    Optically elusive &    No HeII detection &    Optically elusive\\  
  
\tableline
\tablenotetext{a}{\textit{Swift}-BAT 70-month hard X-ray survey ID (http://swift.gsfc.nasa.gov/results/bs70mon/).}
\tablenotetext{b}{The symbol (B) indicates broad-line source.}
\tablenotetext{c}{The symbol $\cdot \cdot \cdot$ indicates lack of wavelength coverage referring 'no wave' classification which listed in Fig.~\ref{BPTnormal} and Fig.~\ref{BPTblue}.}
\tablenotetext{d}{AGN Limit refers to objects which have an \hb\ upper limit either in the Seyfert or in the LINER region.}
%\tablenotetext{e}{Low SNR refers to objects which have lack enough emission line detections to be placed on the diagnostic diagram.}
\end{tabular}
\tablecomments{(This table is available in its entirety in a machine-readable form in the online journal. A portion is shown here for guidance regarding its form and content.)}
\label{tab:BPT}
\end{center}
\end{minipage}
\end{table*}

\begin{turnpage}
\begin{table*}
\begin{minipage}{1.2\textwidth}
\begin{center}
\caption{Properties derived from spectral decomposition of broad \hb\ and \ha\ sources}
\begin{tabular}{lcccccccccccccc}
\tableline
\tableline
\multirow{2}{*}{ID\tablenotemark{a}}
%&\multirow{2}{*}{Source\tablenotemark{b}}
&$\log L_{5100}$
&$\log L_{\rm bol}$\tablenotemark{b}
&${\rm FWHM}_{\rm bH\beta}$
&${\rm EW}_{\rm bH\beta}$
&\multirow{2}{*}{$\log M_{\rm BH}/M_{\odot}$\tablenotemark{c}}
&$\log L/L_{\rm Edd}$\tablenotemark{c}
&\multirow{2}{*}{${\rm Flag}_{b\hb}$\tablenotemark{d}}
& b\ha\tablenotemark{d}
&${\rm FWHM}_{\rm bH\alpha}$
&${\rm EW}_{\rm bH\alpha}$
&\multirow{2}{*}{$\log M_{\rm BH}/M_{\odot}$\tablenotemark{e}}
&\multirow{2}{*}{${\rm Flag}_{b\ha}$\tablenotemark{f}}
&\multirow{2}{*}{$\log L/L_{\rm Edd}$\tablenotemark{g}}
&\multirow{2}{*}{$\log (M_{\rm BH}/M_{\odot})_{\rm lit.}$\tablenotemark{h}} \\
& (\ergs)
& (\ergs)
& (\kms)
& (\AA)
&
& 5100\AA
&
& (\ergcms)
& (\kms)
& (\AA)
&
& 
& \\
\tableline
 
 1&  $\cdot \cdot \cdot$&  $\cdot \cdot \cdot$&  $\cdot \cdot \cdot$&  $\cdot \cdot \cdot$&  $\cdot \cdot \cdot$&  $\cdot \cdot \cdot$&  $\cdot \cdot \cdot$&                    $9.5\pm0.1$&                    $3680\pm25$&                          14.79&                           6.42&                    1&                -0.00&            $\cdot \cdot \cdot$\\
6&                43.90&                44.80&                 2065&                99.42&                 7.29&                -0.67&                    1&                 $1929.3\pm8.7$&                   $1706\pm669$&                         297.48&                           6.86&                    2&                -0.61&          $7.23^{+0.04}_{0.04}$\\
8&                44.43&                45.28&                 5295&               100.66&                 8.45&                -1.34&                    1&                 $1076.6\pm2.0$&                    $4704\pm33$&                         492.98&                           8.26&                    2&                -0.67&          $8.07^{+0.17}_{0.12}$\\
16&                44.66&                45.50&                 2451&                69.27&                 7.93&                -0.61&                    1&            $\cdot \cdot \cdot$&                               &            $\cdot \cdot \cdot$&            $\cdot \cdot \cdot$&  $\cdot \cdot \cdot$&  $\cdot \cdot \cdot$&          $8.49^{+0.12}_{0.10}$\\
18&  $\cdot \cdot \cdot$&  $\cdot \cdot \cdot$&  $\cdot \cdot \cdot$&  $\cdot \cdot \cdot$&  $\cdot \cdot \cdot$&  $\cdot \cdot \cdot$&  $\cdot \cdot \cdot$&                  $319.3\pm1.9$&                   $11519\pm66$&                         421.97&                           8.84&                    2&                -1.46&            $\cdot \cdot \cdot$\\
28&  $\cdot \cdot \cdot$&  $\cdot \cdot \cdot$&  $\cdot \cdot \cdot$&  $\cdot \cdot \cdot$&  $\cdot \cdot \cdot$&  $\cdot \cdot \cdot$&  $\cdot \cdot \cdot$&                   $53.7\pm0.3$&                   $5469\pm110$&                          18.83&                           6.93&                    2&                -0.42&            $\cdot \cdot \cdot$\\
33&  $\cdot \cdot \cdot$&  $\cdot \cdot \cdot$&  $\cdot \cdot \cdot$&  $\cdot \cdot \cdot$&  $\cdot \cdot \cdot$&  $\cdot \cdot \cdot$&  $\cdot \cdot \cdot$&                  $110.2\pm1.2$&                    $2010\pm26$&                          35.70&                           6.02&                    1&                 0.66&            $\cdot \cdot \cdot$\\
36&                43.43&                44.37&                 5446&               133.99&                 7.82&                -1.63&                    1&            $\cdot \cdot \cdot$&                               &            $\cdot \cdot \cdot$&            $\cdot \cdot \cdot$&  $\cdot \cdot \cdot$&  $\cdot \cdot \cdot$&            $\cdot \cdot \cdot$\\
39&                44.74&                45.58&                 5205&               126.00&                 8.64&                -1.24&                    1&            $\cdot \cdot \cdot$&                               &            $\cdot \cdot \cdot$&            $\cdot \cdot \cdot$&  $\cdot \cdot \cdot$&  $\cdot \cdot \cdot$&          $8.46^{+0.09}_{0.08}$\\
43&                42.83&                43.85&                 4214&                57.64&                 7.21&                -1.54&                    2&                  $491.4\pm4.2$&                  $3536\pm1881$&                         225.14&                           6.90&                    2&                -0.93&            $\cdot \cdot \cdot$   \\
\tableline
\tablenotetext{a}{\textit{Swift}-BAT 70-month hard X-ray survey ID (http://swift.gsfc.nasa.gov/results/bs70mon/).}
%\tablenotetext{b}{Source for spectral decomposition of broad \hb.} 
\tablenotetext{b}{$L_{\rm bol}$ is estimated from $L_{5100}$, following  \citet{Trakhtenbrot:2012:3081}.}
%\tablenotetext{c}{Narrow emission-line flux ($\times10^{-15}$).}
\tablenotetext{c}{Following \citet{Trakhtenbrot:2012:3081}, with $L_{\rm bol}$ estimated from $L_{5100}$.}
\tablenotetext{d}{Spectral fitting quality flag for broad \hb: 1= a good fit with small error, 2= acceptable fit.}
\tablenotetext{e}{Black hole mass derived from broad \ha\ following \citet{Greene:2005:122}.}
\tablenotetext{f}{Spectral fitting quality flag for broad \ha: 1= a good fit with small error, 2= acceptable fit.}
%1= a good fit with small error, 2= acceptable fit, 3= bad fit for high S/N source due to either the presence of broad line component or offset in emission lines, 9=  lack of spectral coverage or no emission line is detected.}
\tablenotetext{g}{Eddington ratio derived from the \textit{Swift} BAT survey (14-195 keV) and $M_{\rm BH}$ following \citet{Greene:2005:122}.}
\tablenotetext{h}{Black hole mass from \citet[(http://www.astro.gsu.edu/AGNmass/)]{Bentz:2015:67}}.
\end{tabular}
\tablecomments{(This table is available in its entirety in a machine-readable form in the online journal. A portion is shown here for guidance regarding its form and content.)}
\label{tab:broad}
\end{center}
\end{minipage}
\end{table*}
\end{turnpage}

\begin{table*}
\begin{minipage}{\textwidth}
\begin{center}
\caption{Properties derived from spectral decomposition of high redshift sources}
\begin{tabular}{llccccccccc}
\tableline
\tableline
\multirow{2}{*}{ID\tablenotemark{a}}
& \multirow{2}{*}{Counterpart Name}
& \multirow{2}{*}{redshift}
& \multirow{2}{*}{line\tablenotemark{b}}
& $\log L_{\rm cont.}$\tablenotemark{c}
& $F_{\rm line}$\tablenotemark{d}
& ${\rm EW}_{\rm line}$
& ${\rm FWHM}_{\rm line}$
& \multirow{2}{*}{$\log {M_{\rm BH}/M_{\odot}}$}
& \multirow{2}{*}{$\log {L/L_{\rm Edd}}$}
& \multirow{2}{*}{Flag\tablenotemark{e}}\\
&
&
&
& (\ergs)
& (\ergcms)
& (\AA)
& (\kms)
& 
& 
& \\
\tableline
 120&                 [HB89]0212+735&        2.367&  CIV&    46.26&            6.02&           21.13&  8207&     9.69&    -0.89& 2\\
  188&                       4C+32.14&        1.258& MgII&    46.26&           12.19&           13.56&  2790&     9.04&    -0.46& 1\\
  311&                 [HB89]0552+398&        2.365&  CIV&    46.04&            2.46&           15.11&  3134&     8.74&    -0.16& 1\\
  387&                     B2 0743+25&        2.994&  CIV&    46.24&            3.66&           26.81&  5689&     9.36&    -0.58& 2\\
  428&                 [HB89]0836+710&        2.172&  CIV&    46.75&           15.36&           16.18&  7754&     9.90&    -0.61& 1\\
  445&          1RXS J090915.6+035453&        3.288&  CIV&    46.30&            3.36&           25.64&  6000&     9.43&    -0.60& 1\\
  545&                     PKS1127-14&        1.184& MgII&    45.65&            3.74&           18.26&  3208&     8.78&    -0.79& 1\\
  601&                     B2 1210+33&        2.504&  CIV&    46.30&            9.44&           38.44&  4299&     9.15&    -0.31& 2\\
  645&                          3C279&        0.536&  CIV&    45.11&           20.74&           27.27&  6196&     8.83&    -1.19& 2\\
  693&                 LBQS 1344+0233&        1.310& MgII&    45.89&            6.55&           23.24&  2326&     8.65&    -0.43& 1\\
  752&                        3C309.1&        0.905& MgII&    45.27&            9.15&           52.32&  3374&     8.59&    -0.93& 2\\

\tableline
\tablenotetext{a}{\textit{Swift}-BAT 70-month hard X-ray survey ID (http://swift.gsfc.nasa.gov/results/bs70mon/).}
\tablenotetext{b}{Broad emission line (\CIV\ or \MgII) fitted to measure spectral quantities listed in this table.}
\tablenotetext{c}{Monochromatic luminosity at rest-frame 1450 \AA\ (\CIV) or 3000 \AA\ (\MgII).} 
\tablenotetext{d}{Emission line flux ($\times10^{-15}$).}
\tablenotetext{e}{Spectral fitting quality flag: 1= a good fit with small error, 2= acceptable fit.}
\end{tabular}
\tablecomments{(This table is available in its entirety in a machine-readable form in the online journal. A portion is shown here for guidance regarding its form and content.)}
\label{tab:high_z}
\end{center}
\end{minipage}
\end{table*}

\section{Appendix}
We performed a variety of checks on our galaxy template fitting, velocity dispersion, emission line fitting, black hole mass, and bolometric luminosity measurements which we describe here.

\subsection{Galaxy Template Fitting and Velocity Dispersion Measurements} 
\label{sec:template}

\begin{figure*}
\centering
\subfigure{\includegraphics[width=8.2cm]{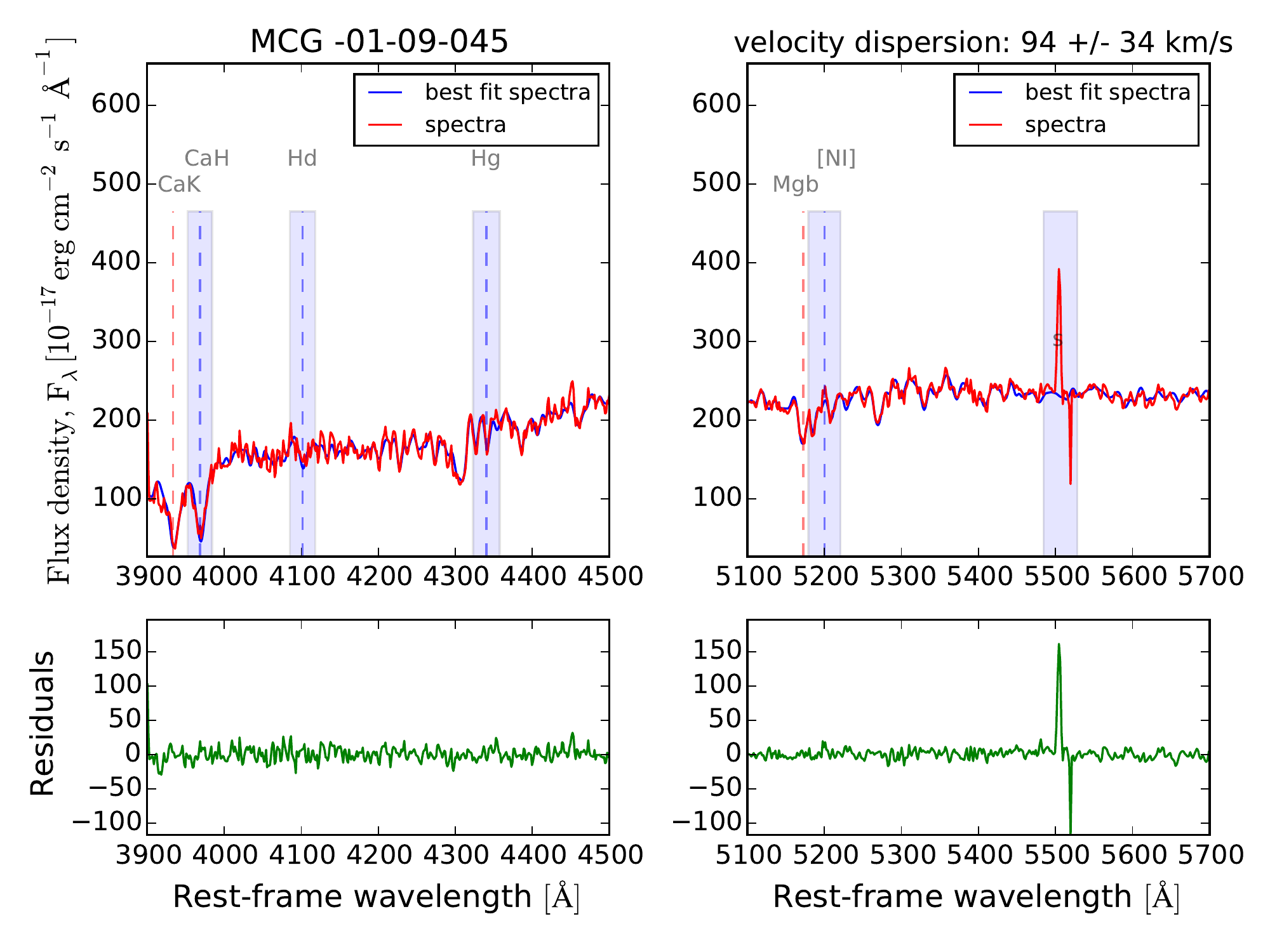}}
\subfigure{\includegraphics[width=8.2cm]{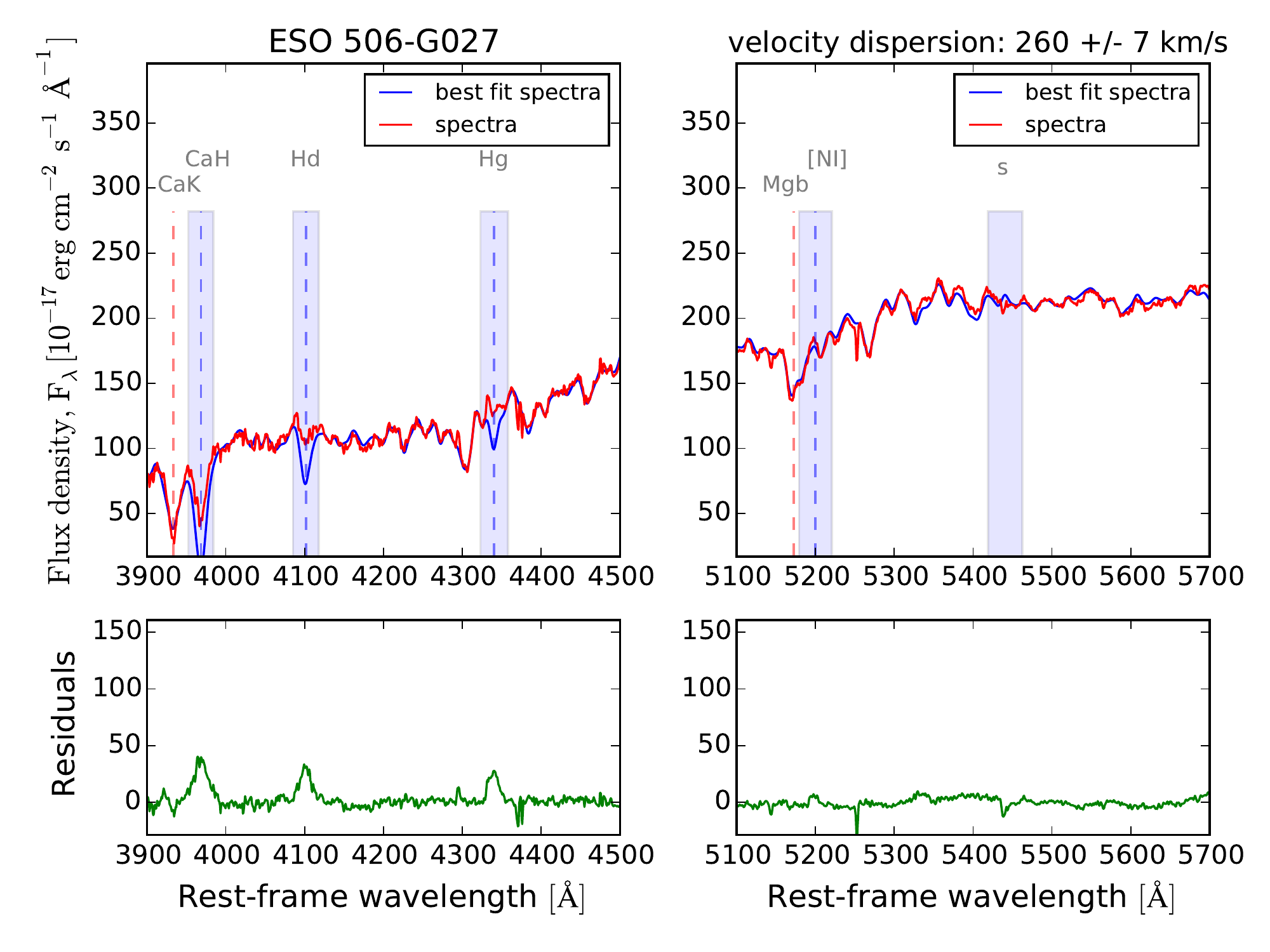}}
\vfill
\subfigure{\includegraphics[width=8.2cm]{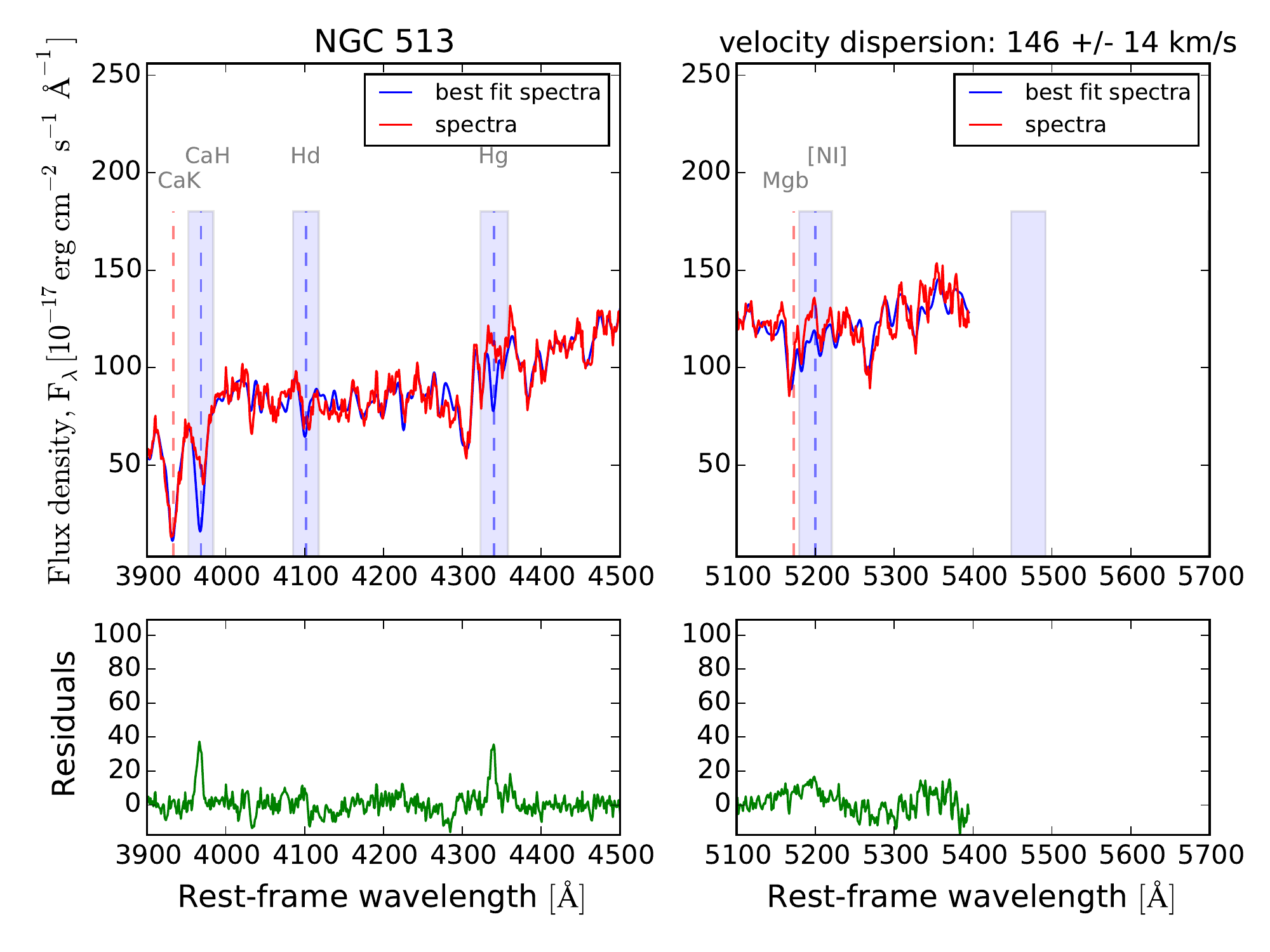}}
\subfigure{\includegraphics[width=8.2cm]{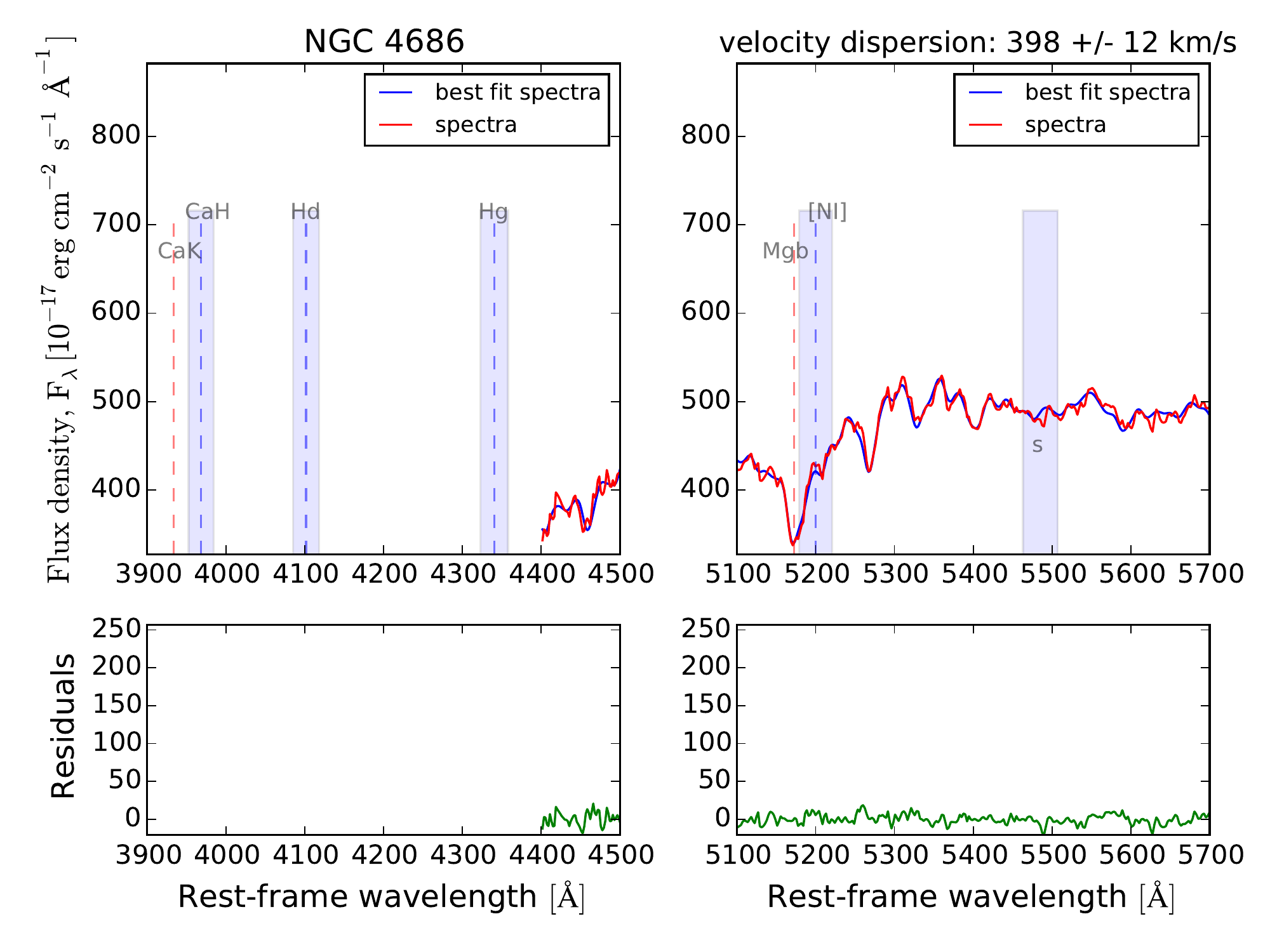} }
\caption{Example \ppxf fits for galaxy MCG -01-09-045 (upper left) taken with UK Schmidt,  ESO 506-G027 (upper right) Gemini, NGC 513 (bottom left) CTIO 1.5m and NGC 4686 (bottom right) Kitt Peak 2.1m. The green bottom portion shows the residuals which include the masked emission lines. The fits of these galaxies have flag 1, meaning that the residuals are small and the error in the value of the velocity dispersion is small. The blue regions are the wavelength regions that we mask in order to exclude the contamination on the stellar continuum due to emission lines and sky features. For NGC 4686 we fit only from 4400\,\AA\ because the wavelength range of this spectrum do not cover the CaH+K region.}
\label{flag1}
\end{figure*}

\begin{figure*}
\centering
\subfigure{\includegraphics[width=8.2cm]{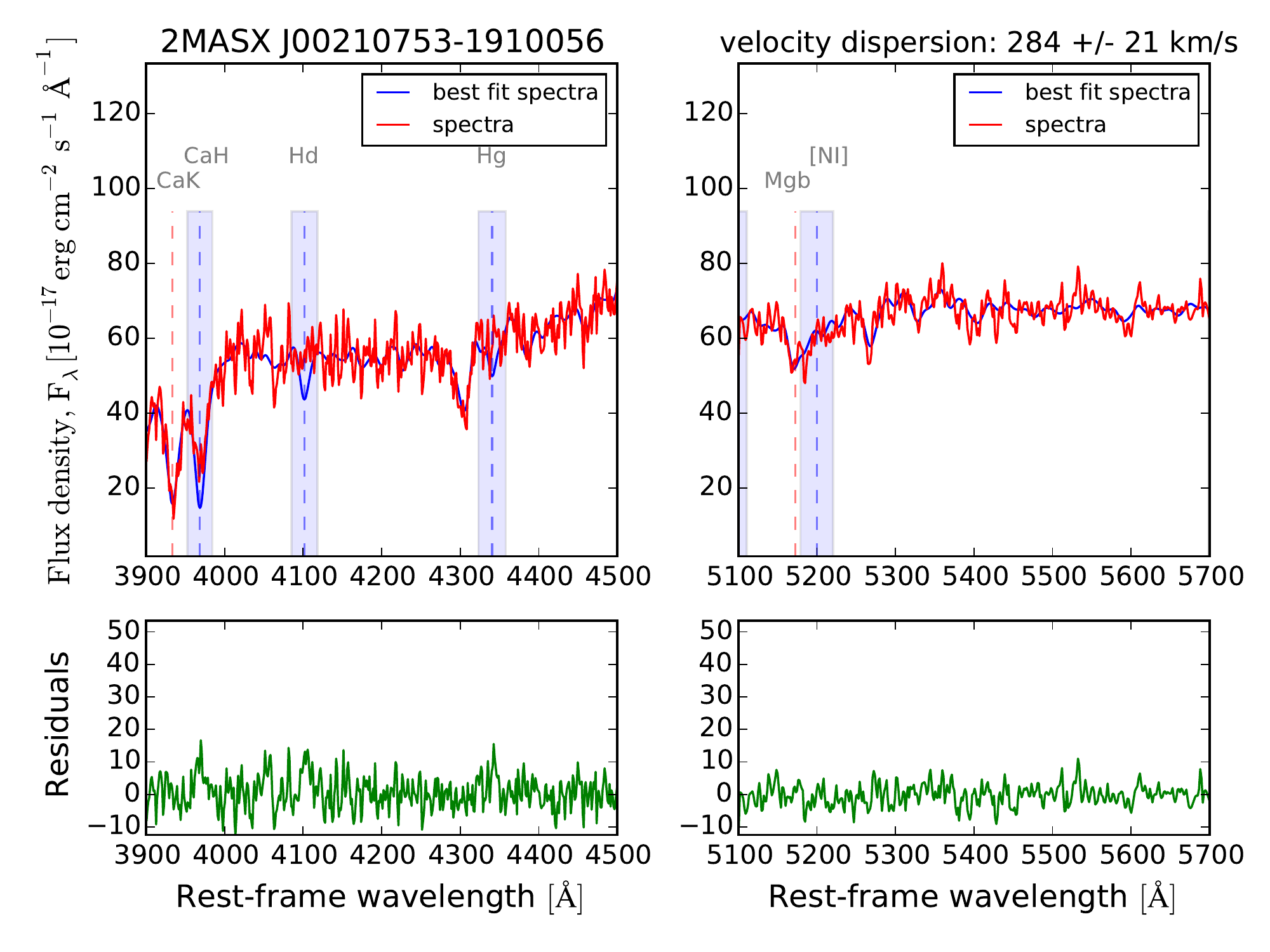}}
\subfigure{\includegraphics[width=8.2cm]{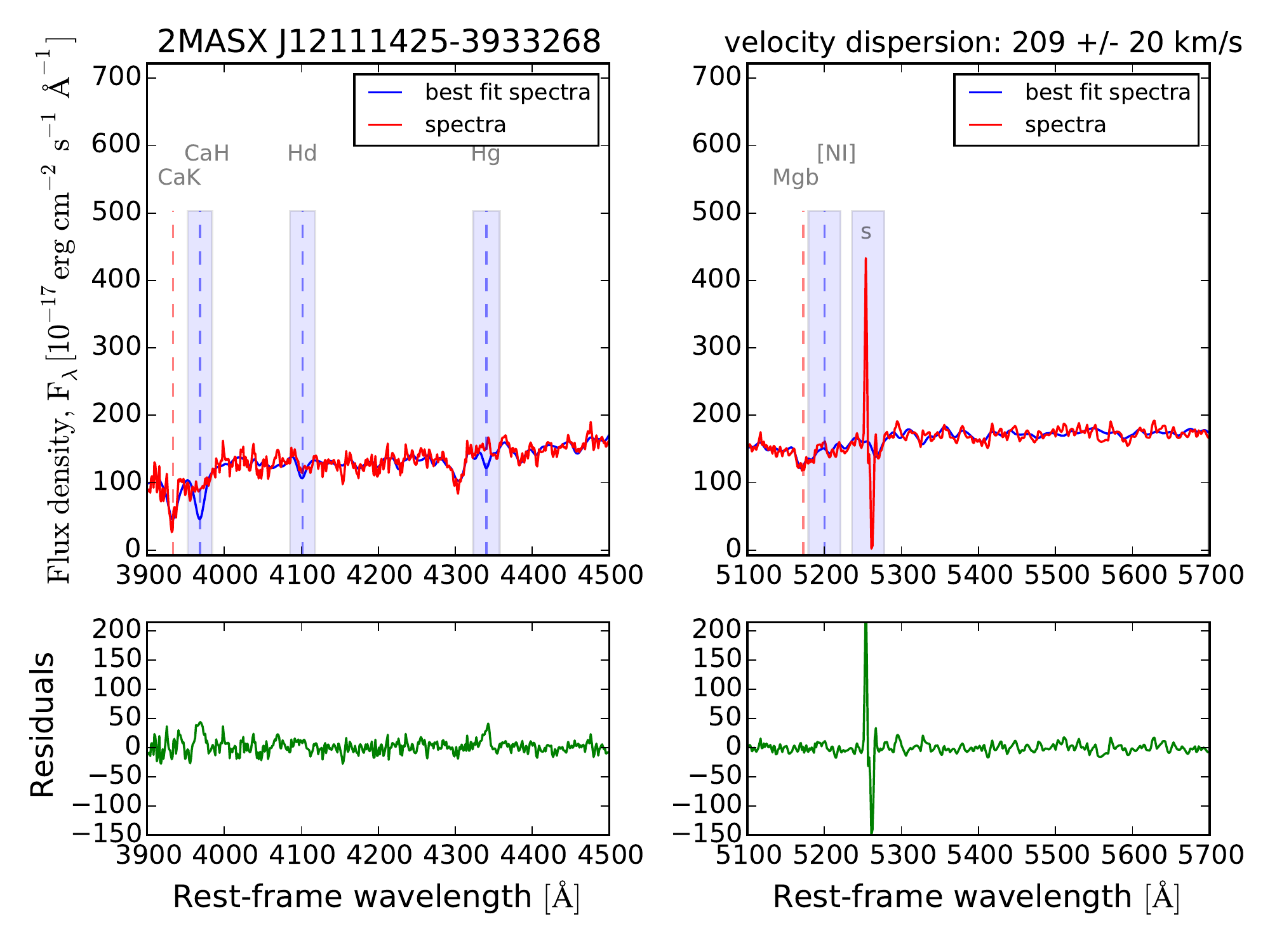}}
\vfill
\subfigure{\includegraphics[width=8.2cm]{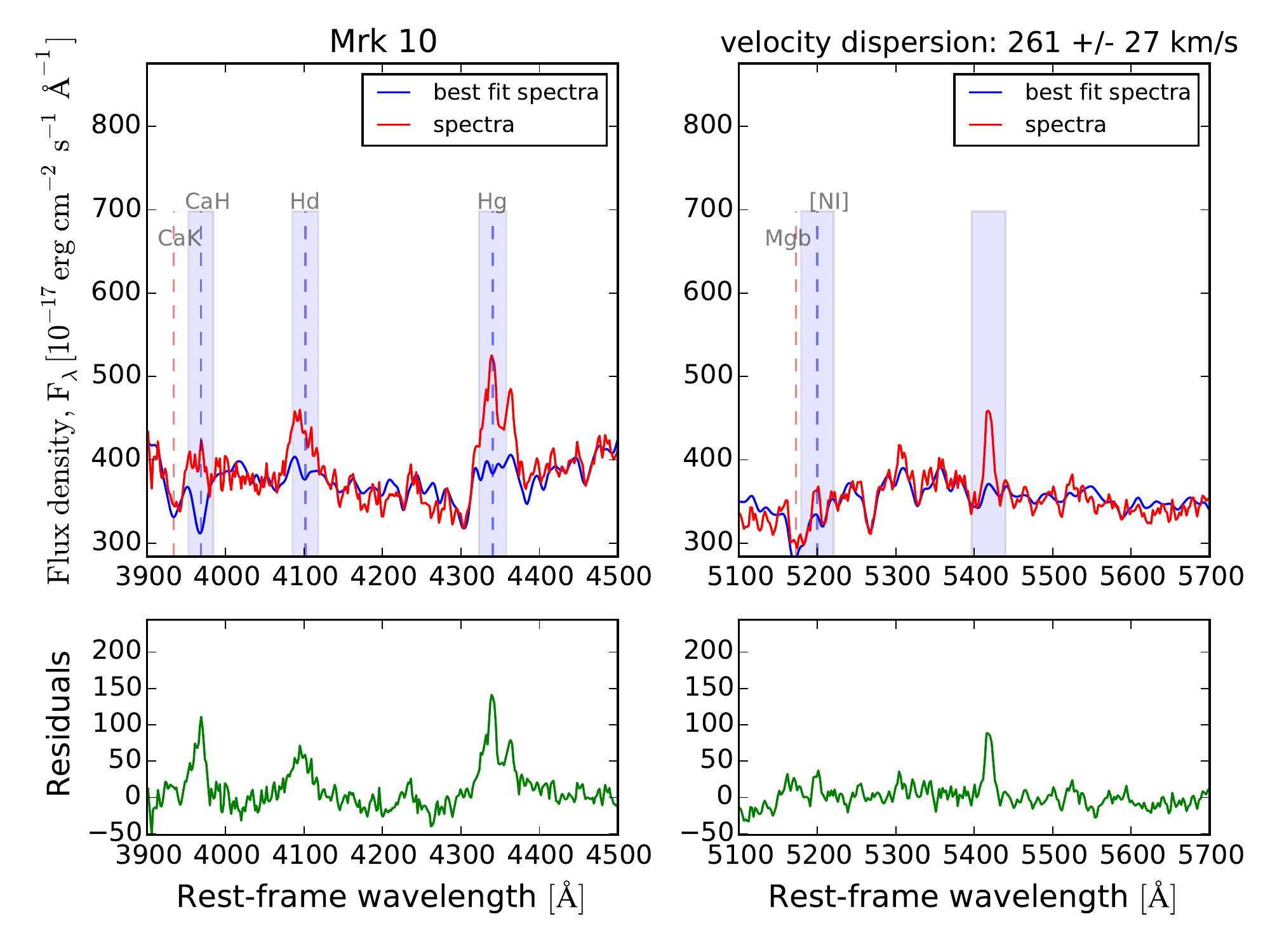}}
\subfigure{\includegraphics[width=8.2cm]{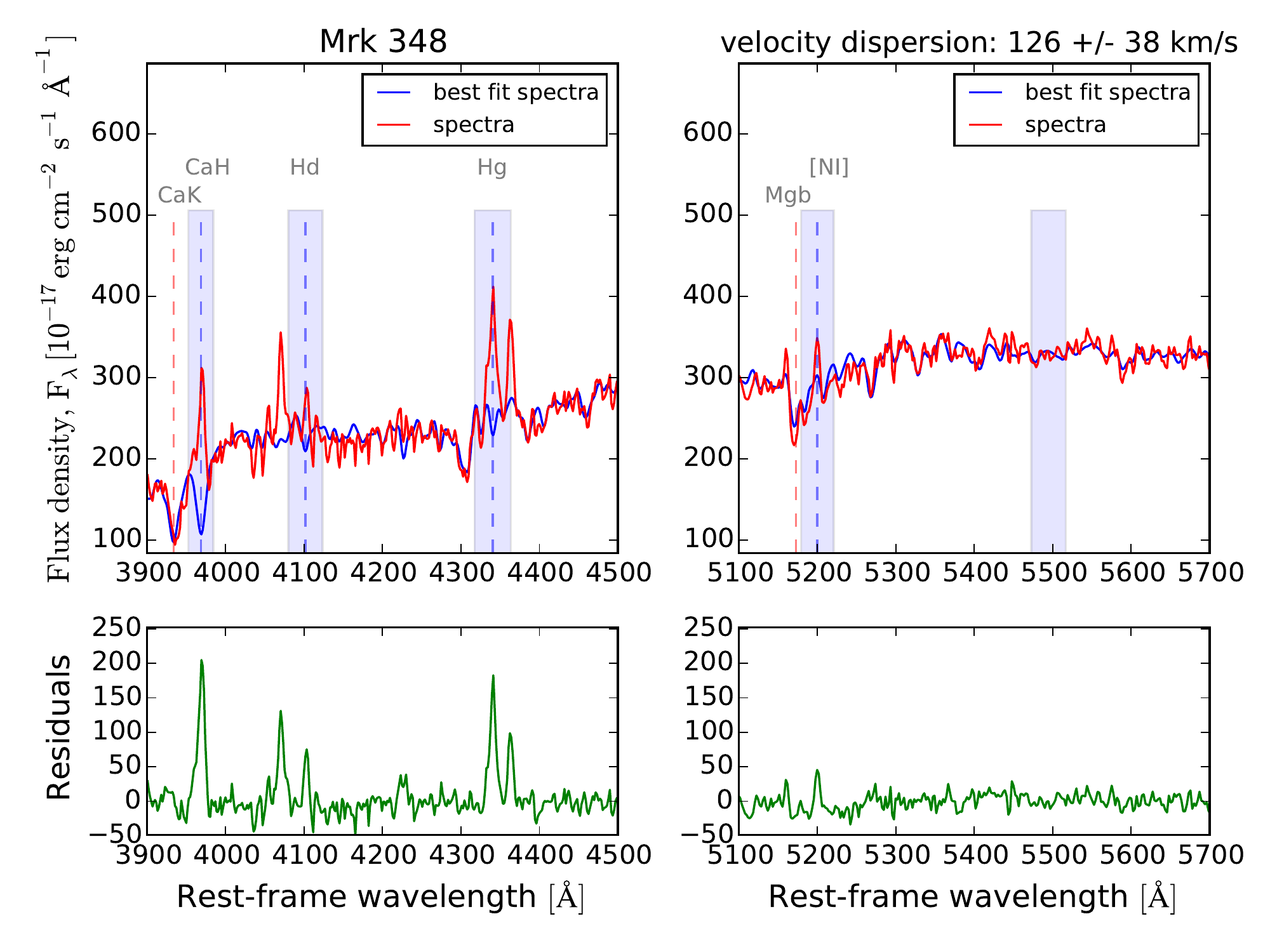}}

\caption{Example \ppxf fits for galaxy 2MASX J00210753-1910056 (upper left) taken with UK Schmidt,  2MASX J12111425-3933268 (upper right) UK Schmidt, Mrk 10 (bottom left) CTIO 1.5m and Mrk 348 (bottom right) CTIO 1.5m. The fits of these galaxies have flag 2, meaning that the fit of the stellar continuum is not very precise, but the absorption lines of Ca H+K and \mgb\ are well fitted and the value of the velocity dispersion is sufficiently precise to rely on it. The blue regions are the wavelength regions that we mask in order to exclude the contamination on the stellar continuum due to emission lines and sky features.}
\label{flag2}
\end{figure*}

\begin{figure}
\includegraphics[scale=0.45]{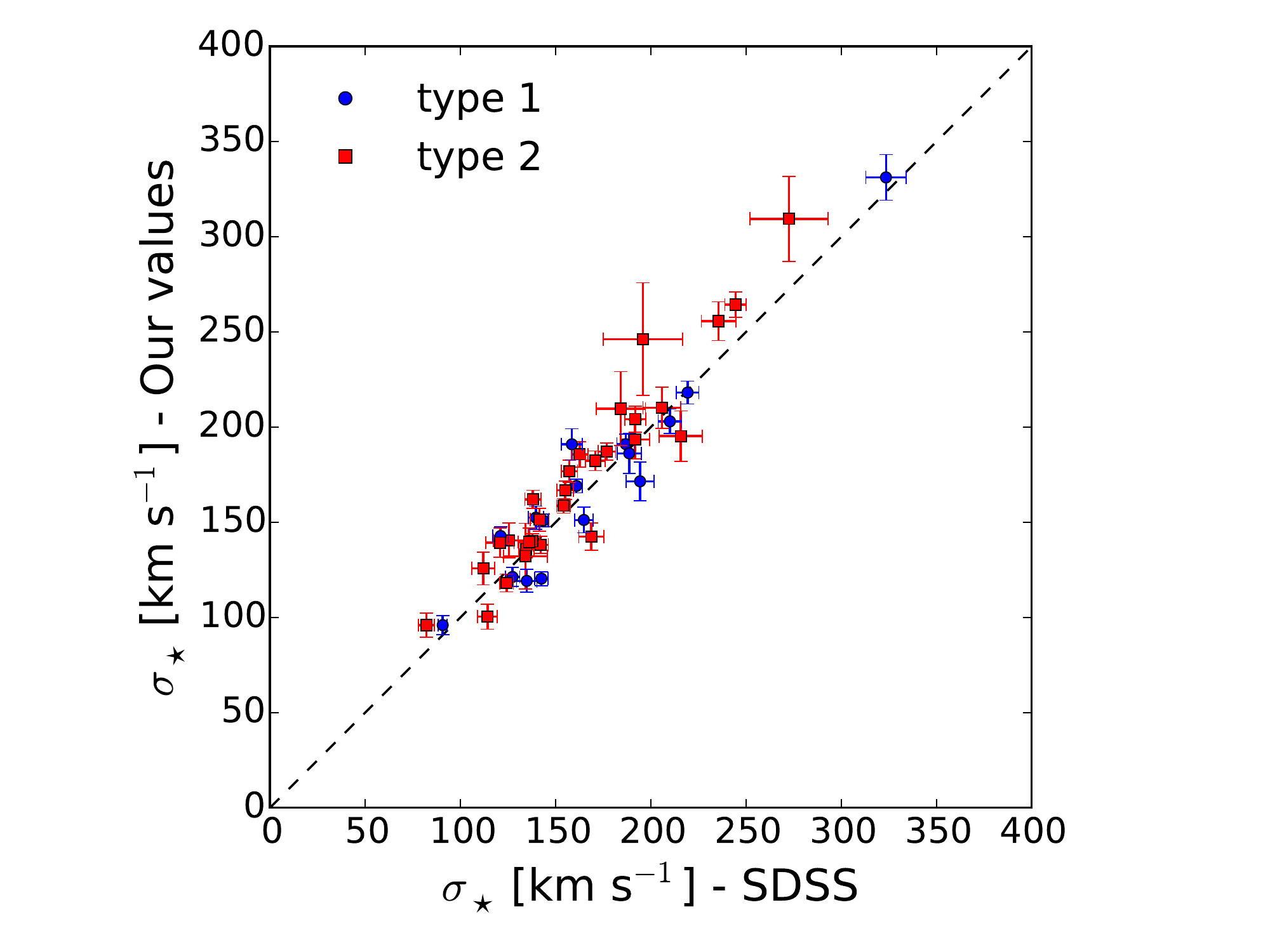}
\caption{Comparison plot between the values of the velocity dispersion from the SDSS 12th data release, measured with a direct fitting method and our values obtained using \ppxf.}
\label{SDSScompfig}
\end{figure}

Examples of the velocity dispersion fits from different telescopes can be found in Figure \ref{flag1} and Figure \ref{flag2}.  We first compared our results for the spectra from SDSS with the values given from the SDSS 12th data release, measured using a direct fitting method rather than \ppxf\ (Fig. \ref{SDSScompfig}). The values calculated using \ppxf\ are in good agreement, with maximal differences always less than 50\,\kmpssh. \\

\begin{figure}
\centering
\includegraphics[scale=0.45]{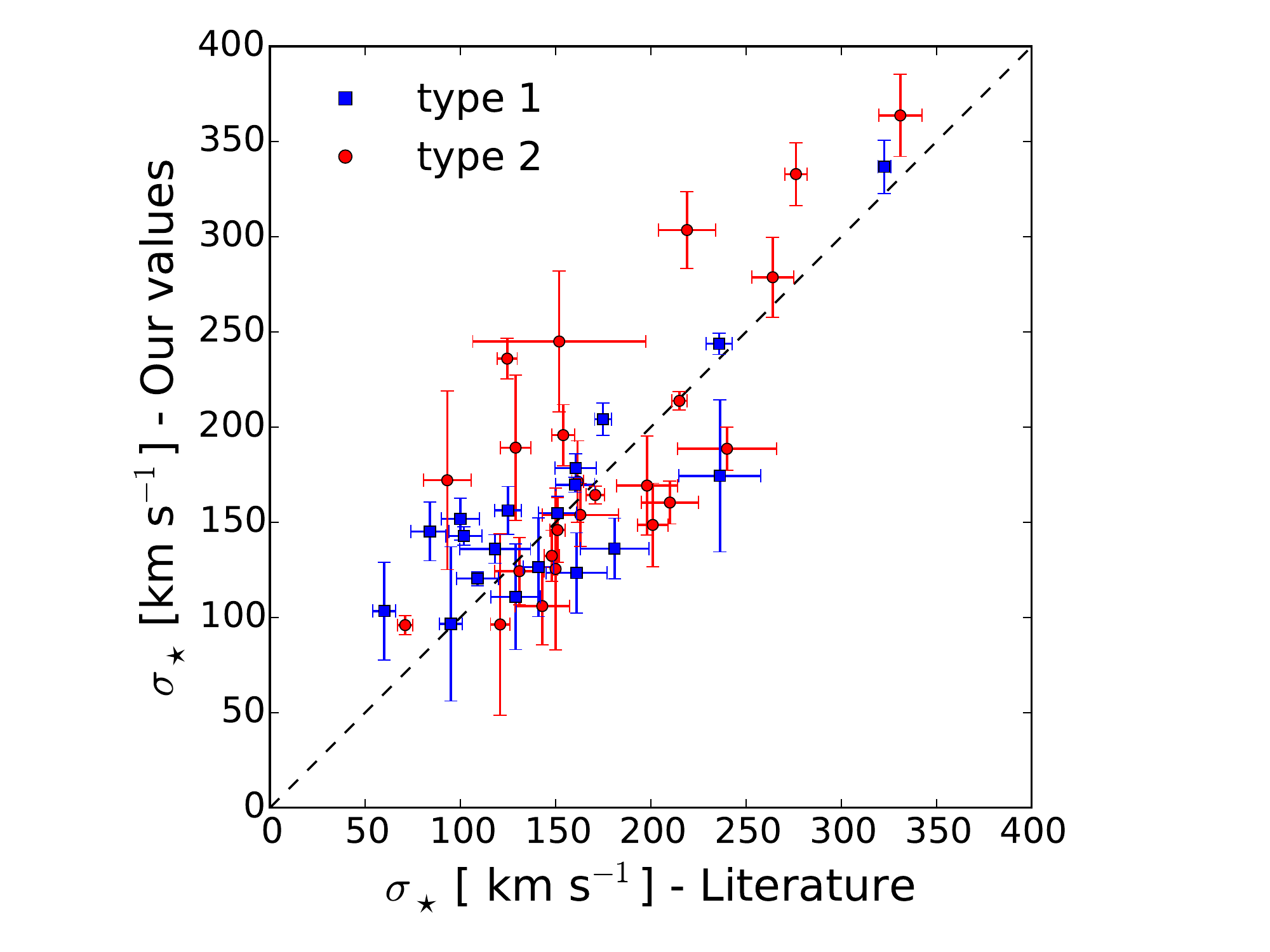} 
\caption{Comparison plot between the values of the velocity dispersion from the literature and our values obtained using \ppxf.}
\label{complit}
\end{figure}

We also compared velocity dispersion \sigs\ using \ppxf\ to literature values obtained with other telescopes. We used the data from the Hyperleda catalogue \citep{Paturel:2003:45} and literature references \citep{CidFernandes:2004:273,Garcia-Rissmann:2005:765, Ho:2009:1,vandenBosch:2015:10}. For 41 galaxies in our sample we found a value of \sigs\ in the literature. Figure  \ref{complit} shows the comparison plot.  Most (31/41, 76\%) of the obtained values are in agreement with the literature values, with the differences between our values and the literature at $<$50\,\kmpssh. For 10/41 (24\%) of sources, the difference is between 50\,\kmps and 100\,\kmps and none above this value.\\

%Flag 2

%Comparison plots

\begin{figure}
\centering
\subfigure{\includegraphics[width=8cm]{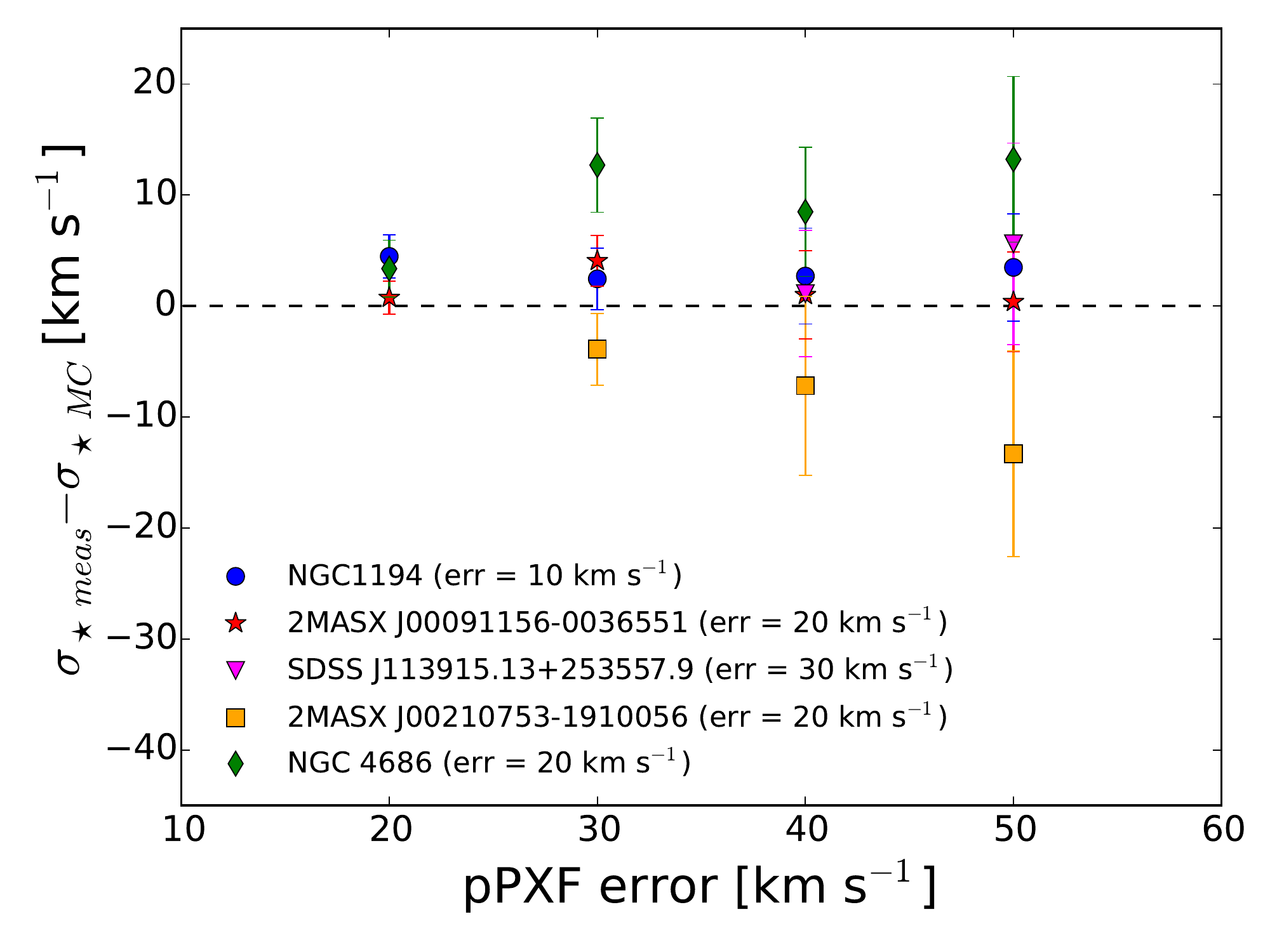}}
%\subfigure{\includegraphics[width=8cm]{Comp_Lum_14-195keV_vs_Lum_opt.pdf}}
\caption{Difference between the value of the velocity dispersion measured by \ppxf\ $\sigma_{*, meas}$ and the mean value obtained from the MC simulations $\sigma_{*, MC}$  as a function of the level of the error on \sigs\ measured by \ppxf\  for 5 different spectra. The spectra are taken from SDSS, KPNO and 6dF.  The values of the error on \sigs\ measured by \ppxf\ (before adding noise) for each spectrum are listed in bracket in the legend.}
\label{MC}
\end{figure}

We finally used Monte Carlo (MC) simulations to check the  velocity dispersion \sigs\ measured by \ppxf  and any possible systematics with larger noise.  We considered five spectra from different telescopes (SDSS, KPNO, 6dF) with different values of the \ppxf  errors (10\,\kmpssh, 20\,\kmpssh, 30\,\kmpssh). Then we added random noise to the spectrum  to increase the values of the \ppxf error until an established level (20\,\kmpssh, 30\,\kmpssh, 40\,\kmpssh, 50\,\kmpssh). For every level of noise, we ran 100 MC simulations and we calculated the mean of the results obtained for \sigs.  The difference between the original value $\sigma_{*, ppxf}$ and the mean value from the MC simulations $\sigma_{*, MC}$ is always smaller than 20\,\kmpssh (Figure \ref{MC}).

\subsection{Emission Line Measurements}

\begin{figure*}
\centering
%\subfigure{\includegraphics[width=8.2cm]{cgcg164m019_bfall.png}}
\subfigure{\includegraphics[width=5.5cm]{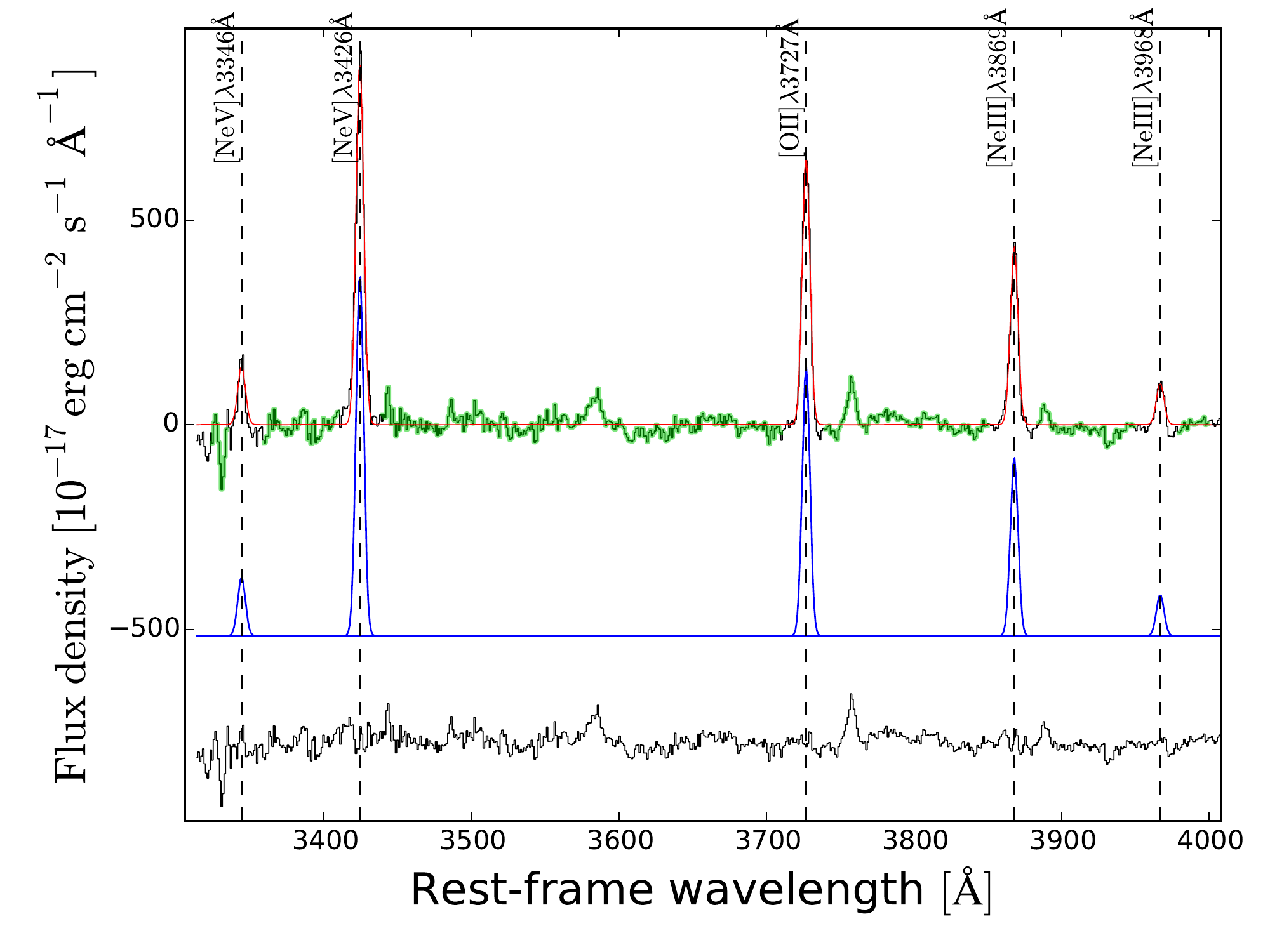}}
%\vfill
\subfigure{\includegraphics[width=5.5cm]{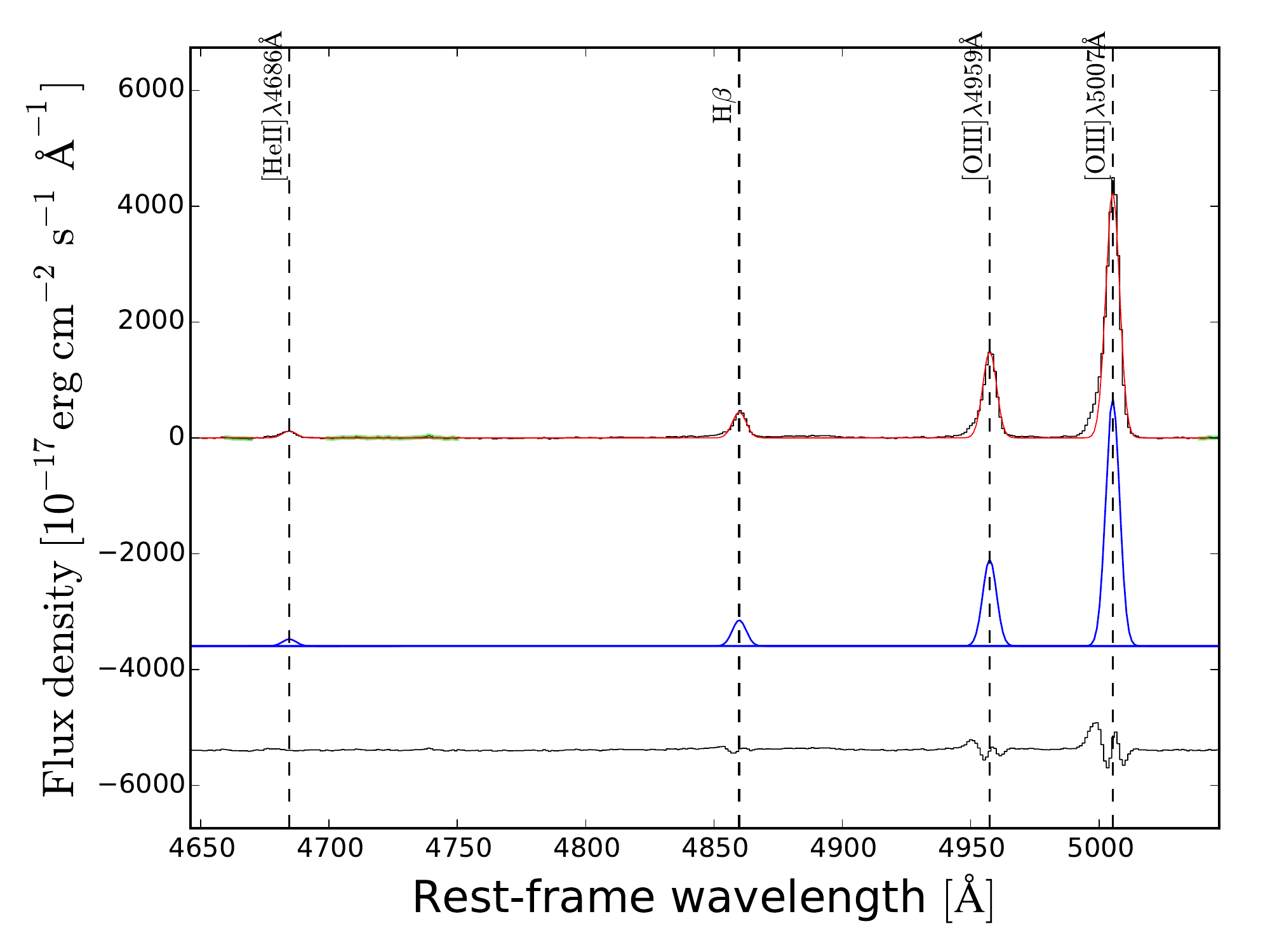}}
\subfigure{\includegraphics[width=5.5cm]{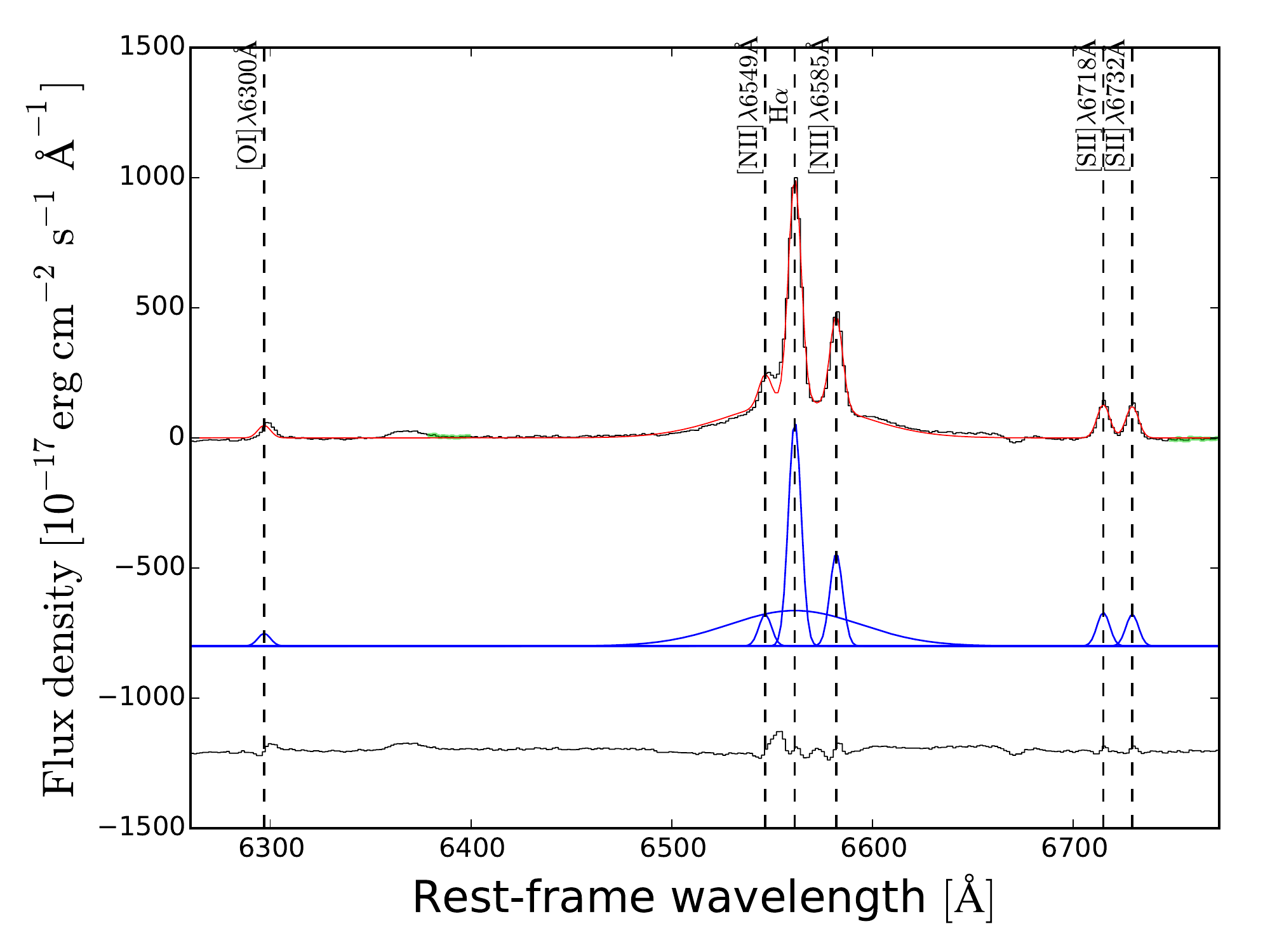}}
\vfill
%\subfigure{\includegraphics[width=5.5cm]%{Mrk975aNeV.png}}
\subfigure{\includegraphics[width=5.5cm]{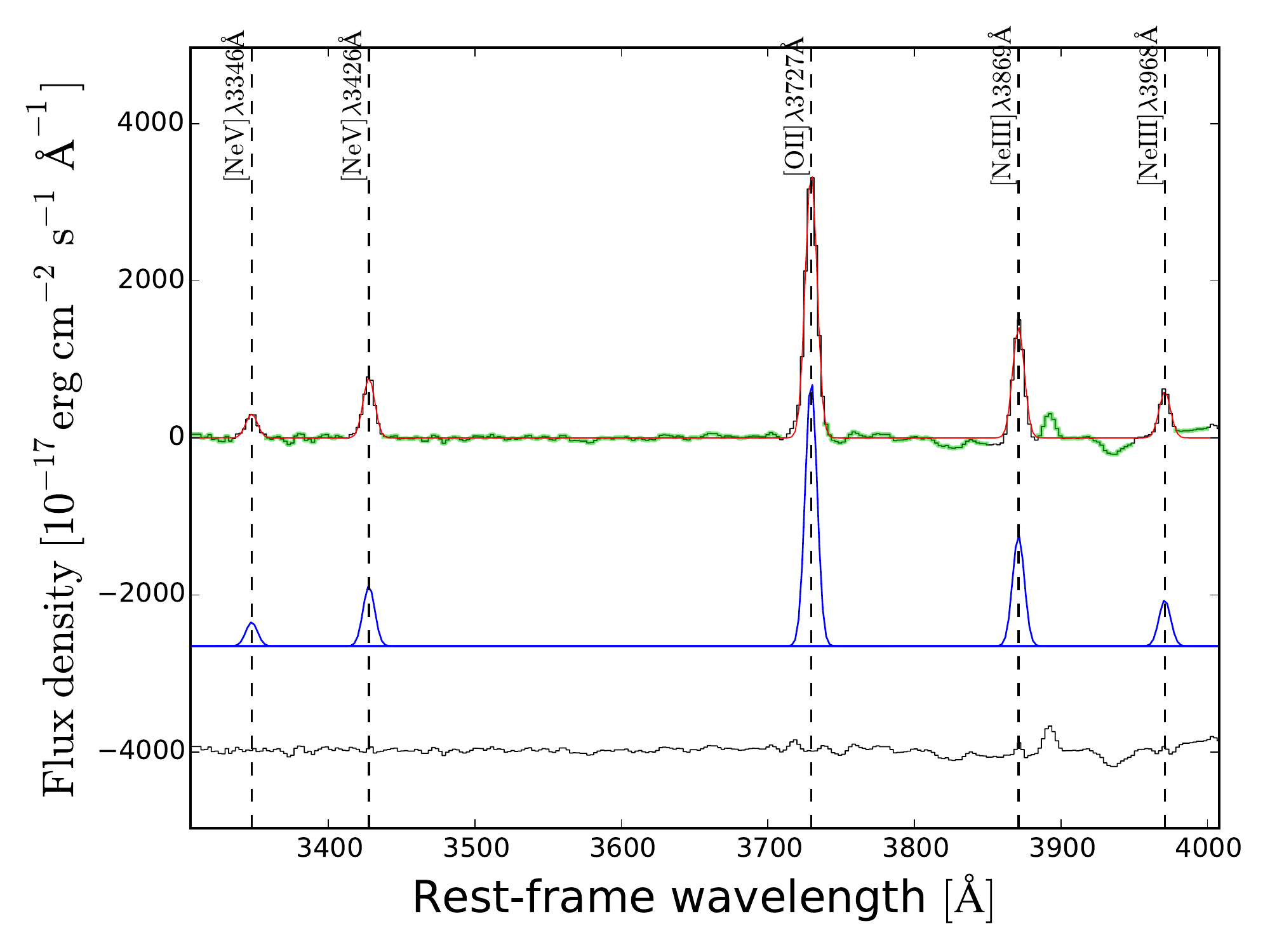}}
\subfigure{\includegraphics[width=5.5cm]{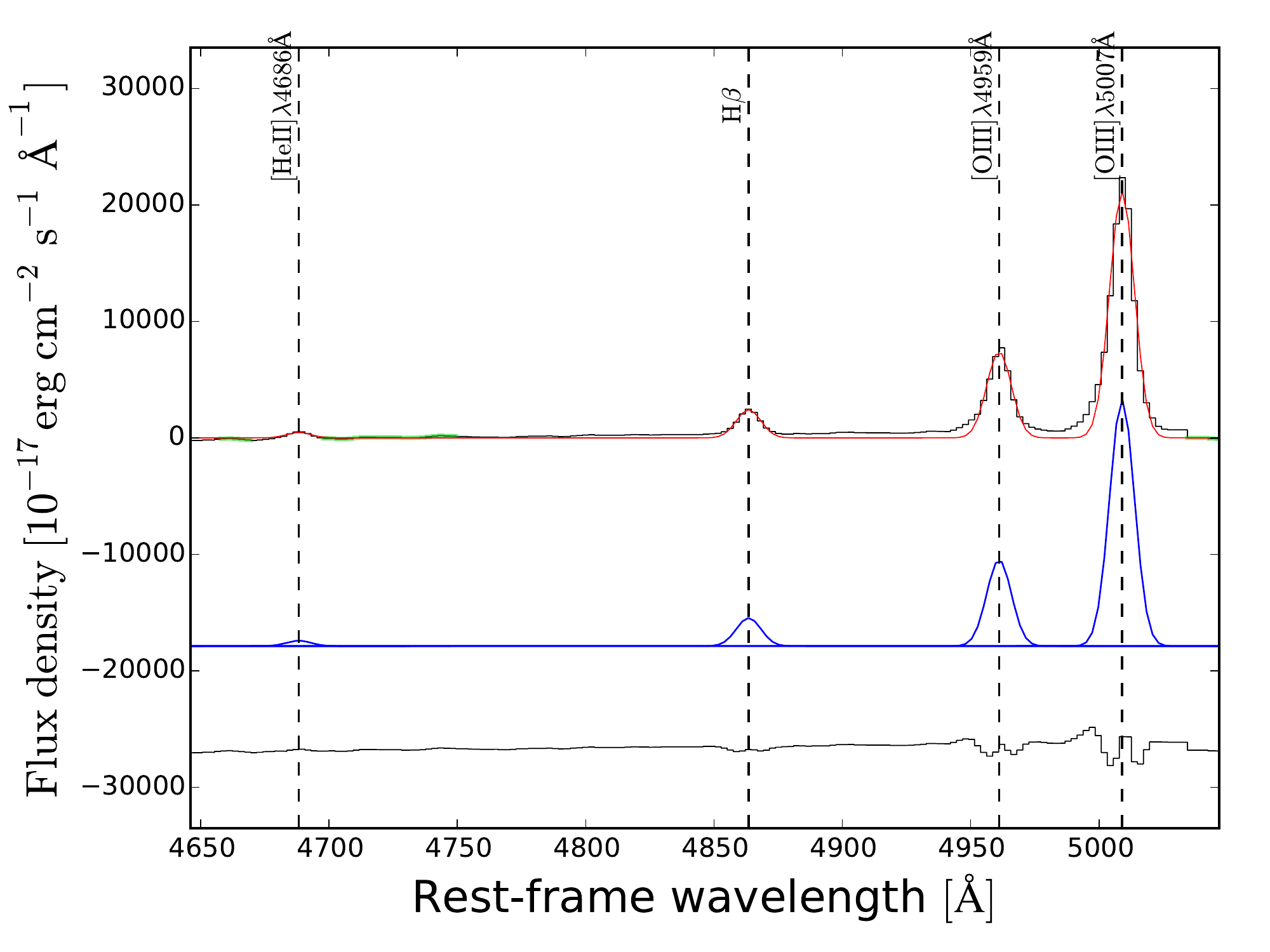}}
\subfigure{\includegraphics[width=5.5cm]{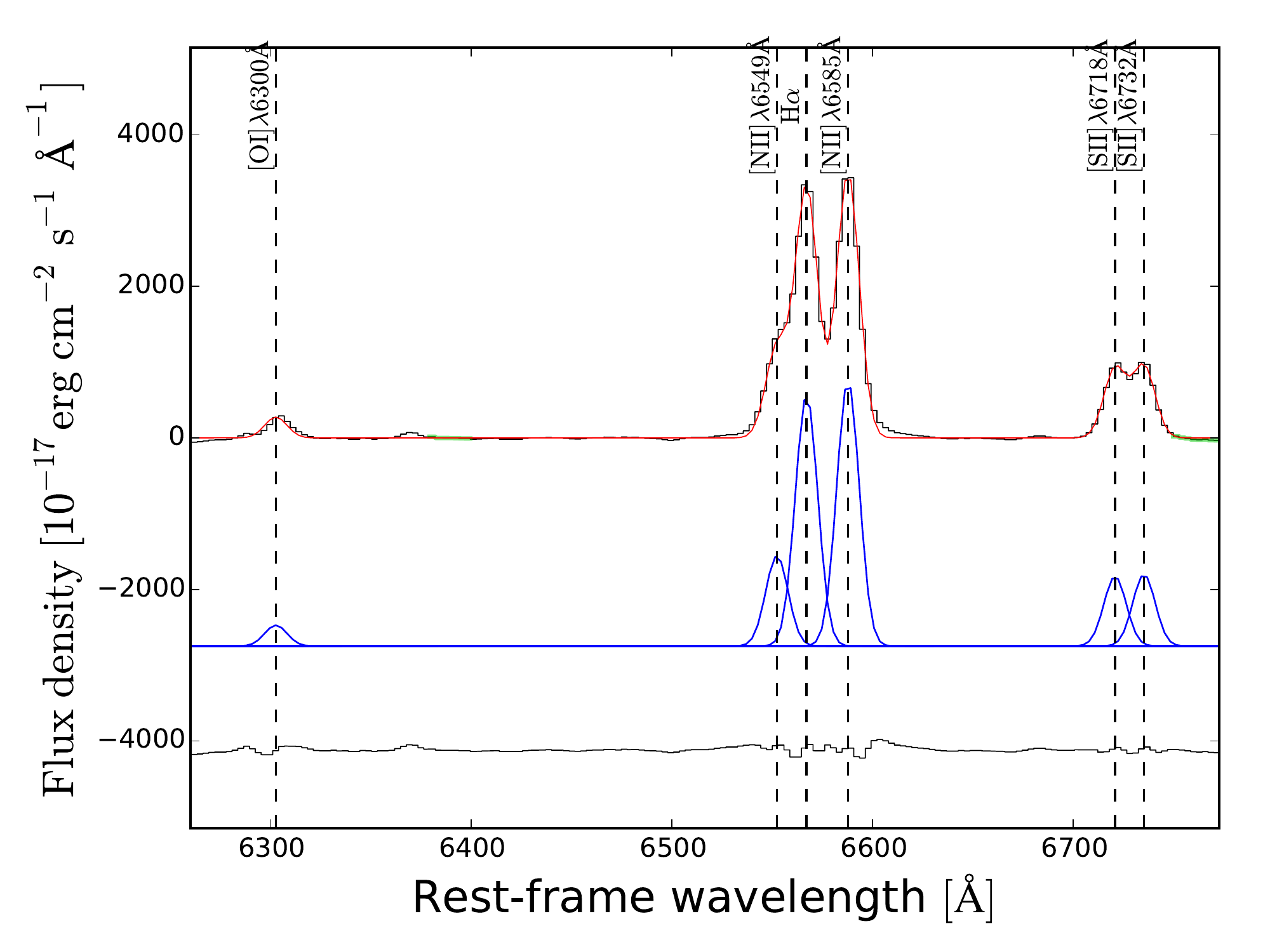}}

\caption{Example of emission line fits for galaxy CGCG 164-019 (upper row) taken with Palomar and  NGC 3393 (lower row) taken with UH 2.2m. The left panels show the fits of the blue \NeV, \OII, and \NeIII\ region, the middle panels the fits of the \Hbeta\ region, and the right panels the fits of the \Halpha\ region.}
%The fits of these galaxies have flag 2, meaning that the fit of the stellar continuum is not very precise, but the absorption lines of CaK and Mgb are well fitted and the value of the velocity dispersion is sufficiently precise to rely on it. The blue regions are the wavelength regions that we mask in order to exclude the contamination on the stellar continuum due to emission lines and sky features.}
\label{em_fit_3_regions}
\end{figure*}

\begin{figure*}[h!] 
\centering

\subfigure{\includegraphics[width=0.49\textwidth]{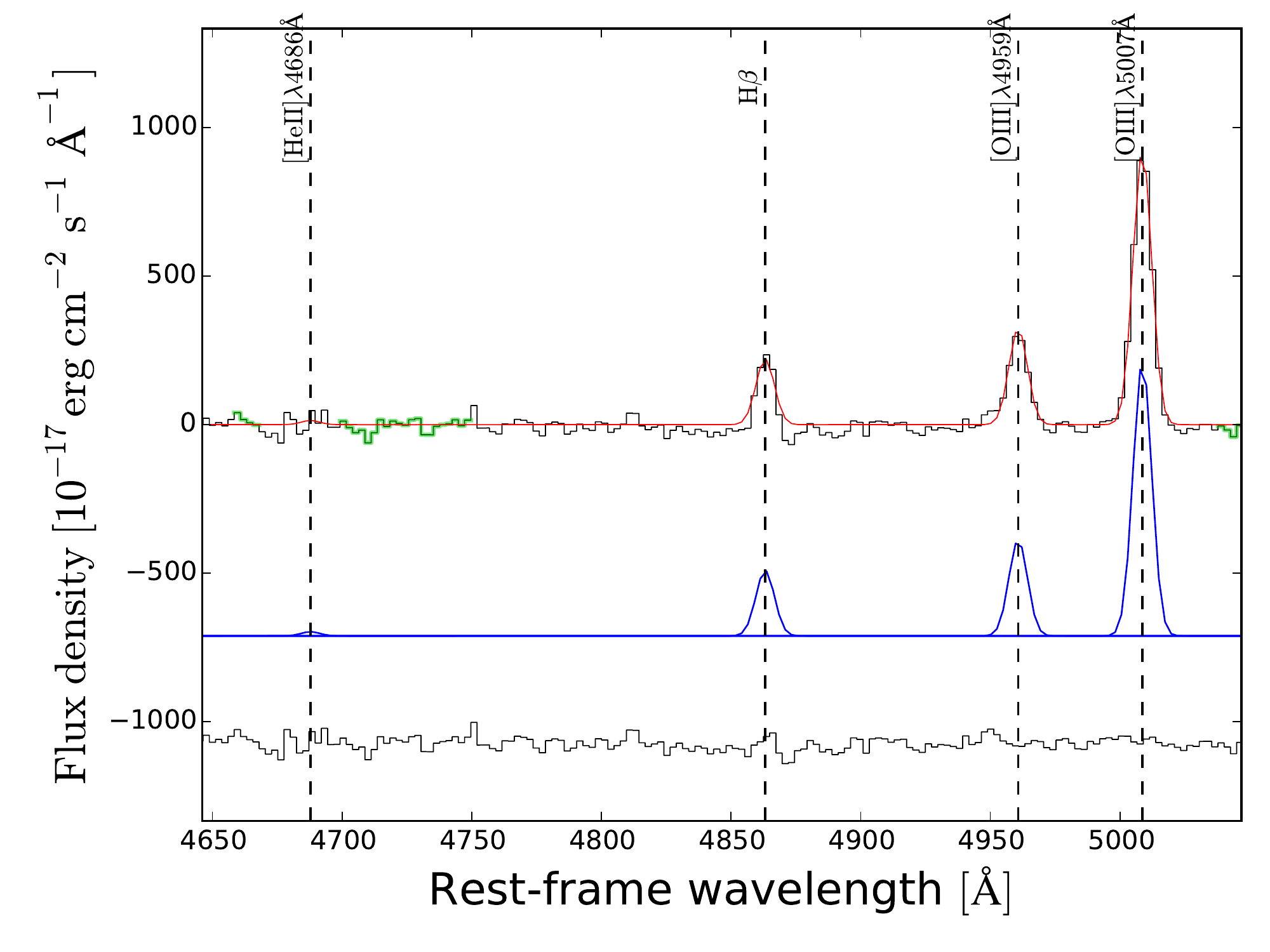}}
\subfigure{\includegraphics[width=0.49\textwidth]{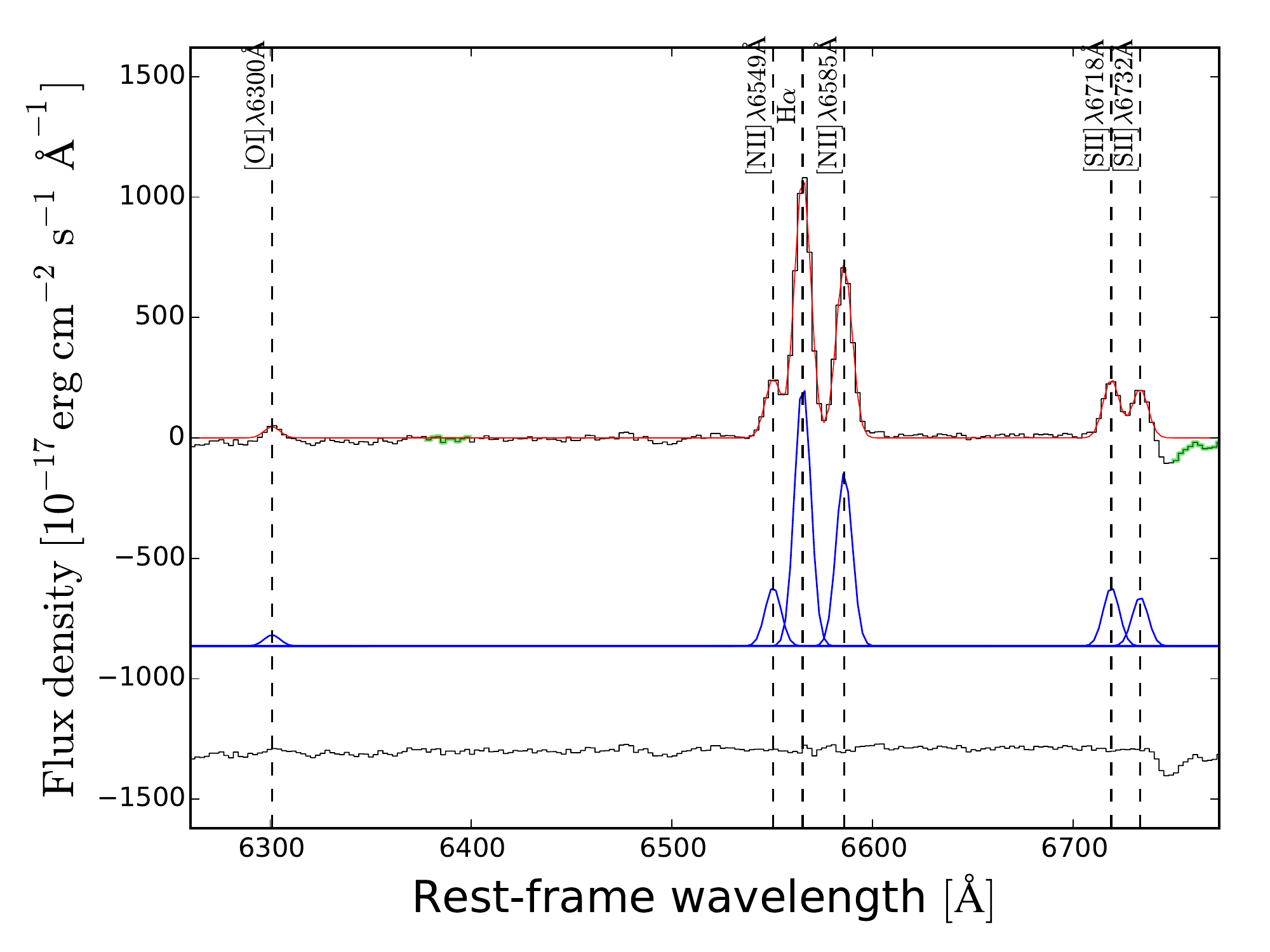}}
\vfill

\subfigure{\includegraphics[width=0.49\textwidth]{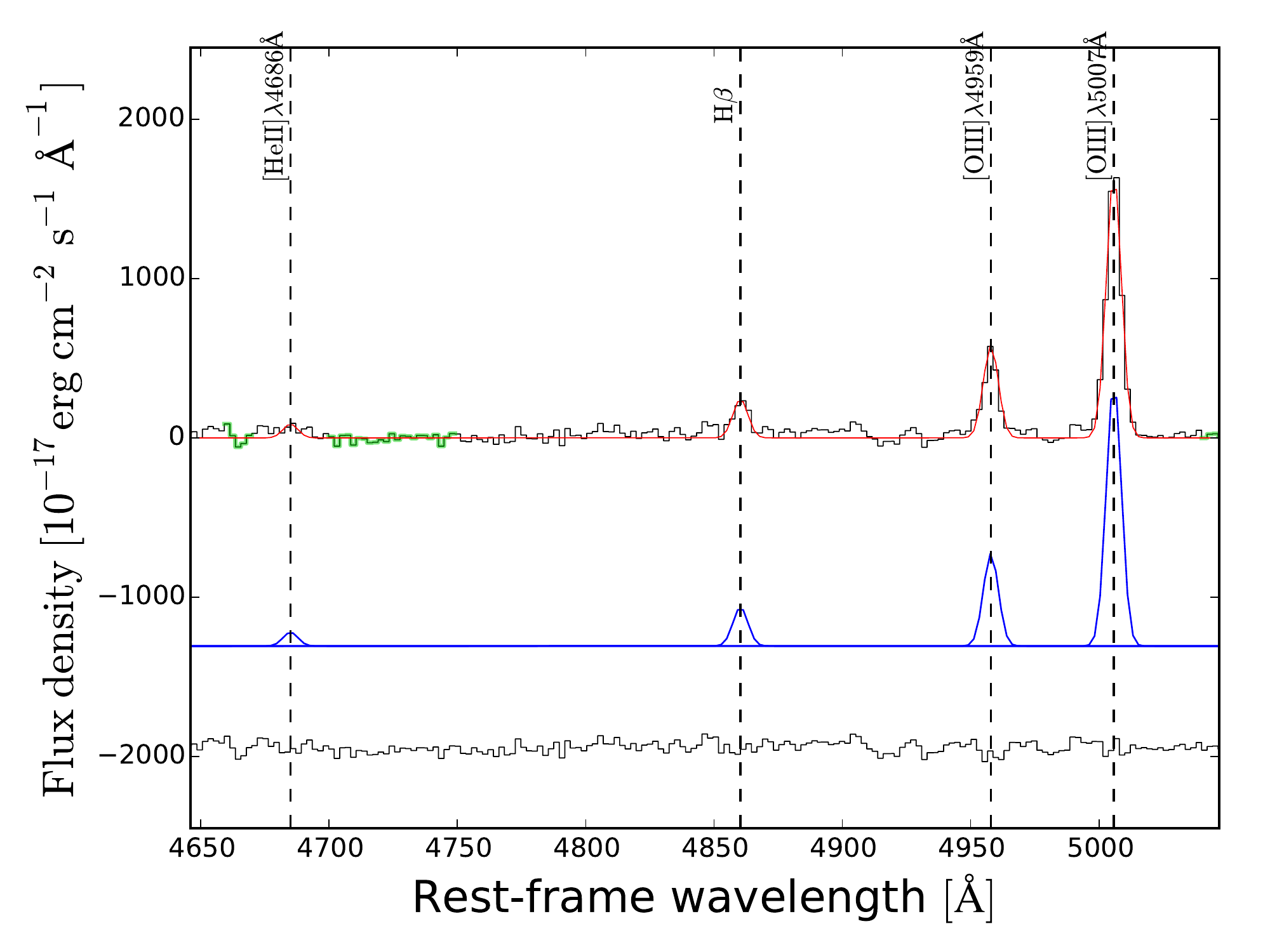}}
\subfigure{\includegraphics[width=0.49\textwidth]{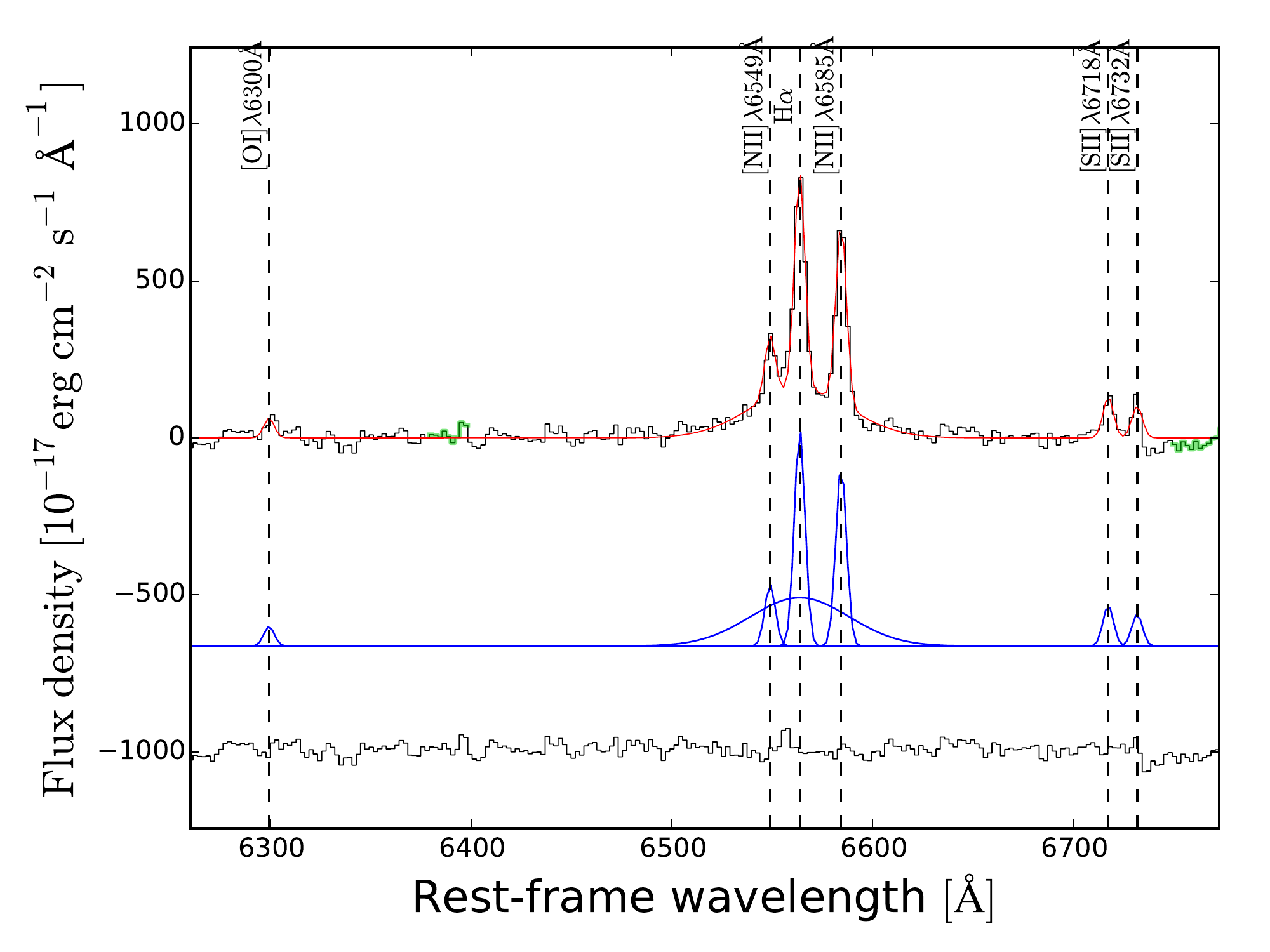}}

\caption{Example of emission line fits for galaxy 2MASX J03534246+3714077 (upper row) taken with KPNO and  2MASX J05353211+4011152 (lower row) taken with the Perkins 1.8m. The left panels show the fits of the H$\beta$ region, and the right panels show the fits of the H$\alpha$ region.}
\label{em_fit_2_regions}
\end{figure*}

\begin{figure}[h!] 
\centering
\includegraphics[width=0.6\textwidth]{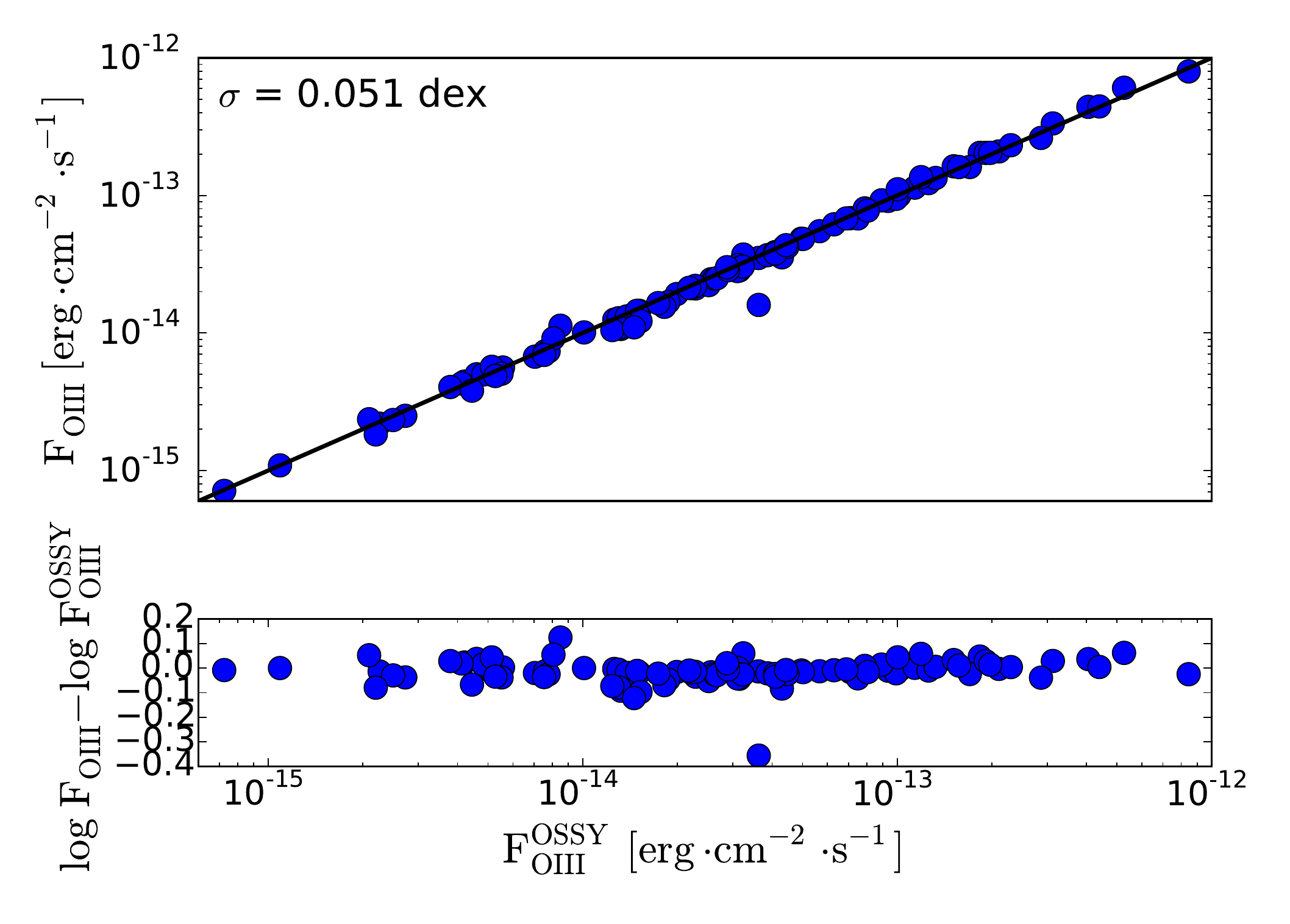}
\caption{Log-log plot of the \OIII\ flux values from 96 objects in the OSSY catalog \citep{Oh:2011:13} on the x-axis, and our measured values on the y-axis. The standard deviation and the median of the offset are written in the top left corner. The bottom panel shows the offset in log space between the OSSY values and our values.}
\label{ossy}
\end{figure}

We provide several examples of our emission line fits using different telescopes. 
A separate BASS study found that the physical slit size, telescope sample, or X-ray obscuration level was not a significant contributor to the scatter in emission line measurements compared to the X-ray emission \citep{Berney:2015:3622} for sources with $0.01<z<0.4$ where the size of the slit is typically kpc scales.      With some setups we were able to cover the  blue \NeV, \OII, and \NeIII\ region (e.g., Fig. \ref{em_fit_3_regions}) in addition to the \Hbeta\ and \Halpha\ regions (e.g., Fig. \ref{em_fit_2_regions}).  We tested our emission line measurements with a number of literature values to confirm their accuracy.  Figure \ref{ossy} shows the comparison between values from the OSSY catalog \citep{Oh:2011:13} and our measured values for the \oiii\ emission line flux derived from the same SDSS spectra.  We obtain a standard deviation of $\sigma$=0.051 dex and a median offset of 0.014 dex for flux values over three orders of magnitude.\\

%Example of emission lines fit of the three regions (Halpha, Hbeta, NeV)

%plot with an example Blazar, Blended, Low SNR, and No wave (four panels in one figure)
%\begin{figure*}[h!] 
%\centering
% blazar/beamed AGN
%\subfigure{\includegraphics[width=0.49\textwidth]{g0325235-563545all_blazar.pdf}}
% blended
%\subfigure{\includegraphics[width=0.49\textwidth]{2MAS2014all_lowSNR_KPNO.pdf}}
% No wave
%\subfigure{\includegraphics[width=0.49\textwidth]{3C062all_nowave_Gemini.pdf}}

%\caption{Example of spectra classified as ``beamed AGN" (2MASX J03252346-5635443 taken from 6dF, upper left) and ``No wave" (3C 062 taken with Gemini, lower right).}
%emission lines fits for galaxy 2MASX J03534246+3714077 (upper row) taken with KPNO and  2MASX J05353211+4011152 (lower row) taken with Perkins 1.8m. The left panels show the fits of the H$\beta$ region, and the right panels show the fits of the H$\alpha$ region.}
%\label{example_classification}
%\end{figure*}

\begin{figure}[h!] 
\centering
\includegraphics[width=0.4\textwidth]{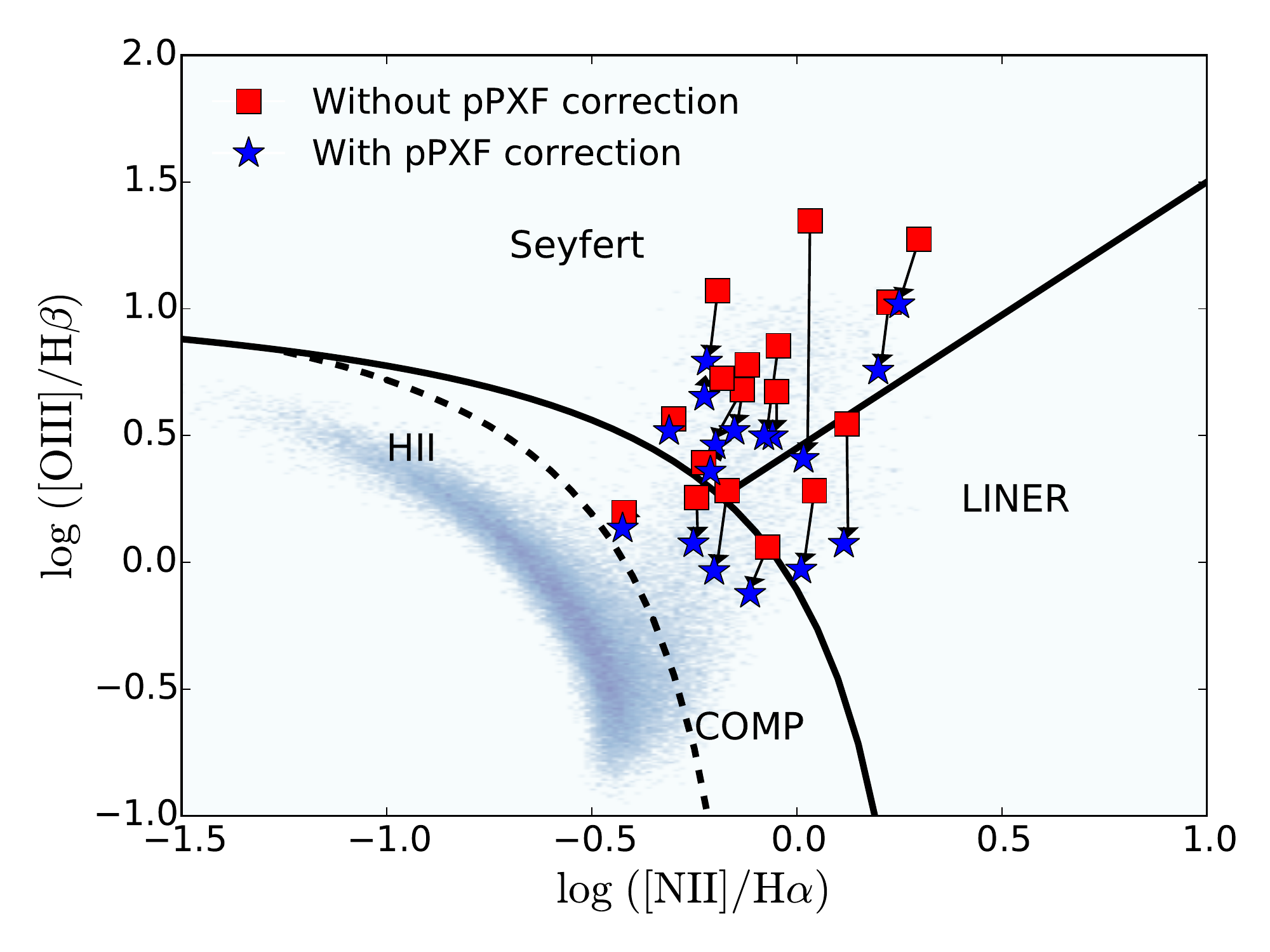}
\caption{Effect of stellar absorption on the line diagnostic classification \citep{Kewley:2006:961} for 16 objects.  The red squares show the classification without \ppxf\ absorption line fitting, and the blue stars after the absorption line fitting. The arrows represent the movement of each point on the diagram. The gray area represents the SDSS sample (SNR$>$2.5 for lines). The solid line that separates Seyferts from LINERs is from \citet{Schawinski:2007:1415}.}
\label{ppxf}
\end{figure}

We also tested how the inclusion of the effect of fitting empirical models with absorption lines which we applied to all narrow line sources changes the emission line measurements. Due to the absorption features on \Halpha\ and \Hbeta, the obtained values for the two Balmer lines are often under-evaluated. The effect is more pronounced for the \Hbeta\ emission line. The result on the line diagnostic diagram is a strong shift to the bottom of the diagram and a relatively small shift to left, which means towards the HII and Composite regions (See Figure \ref{ppxf}). We note that in this study this fitting correction for absorption has only been done for narrow line sources, where the empirical model fits are not biased by AGN light.

\subsection{Black Hole Mass}

\begin{figure}[hbtp]
\centering
\subfigure{\includegraphics[width=0.4\textwidth]{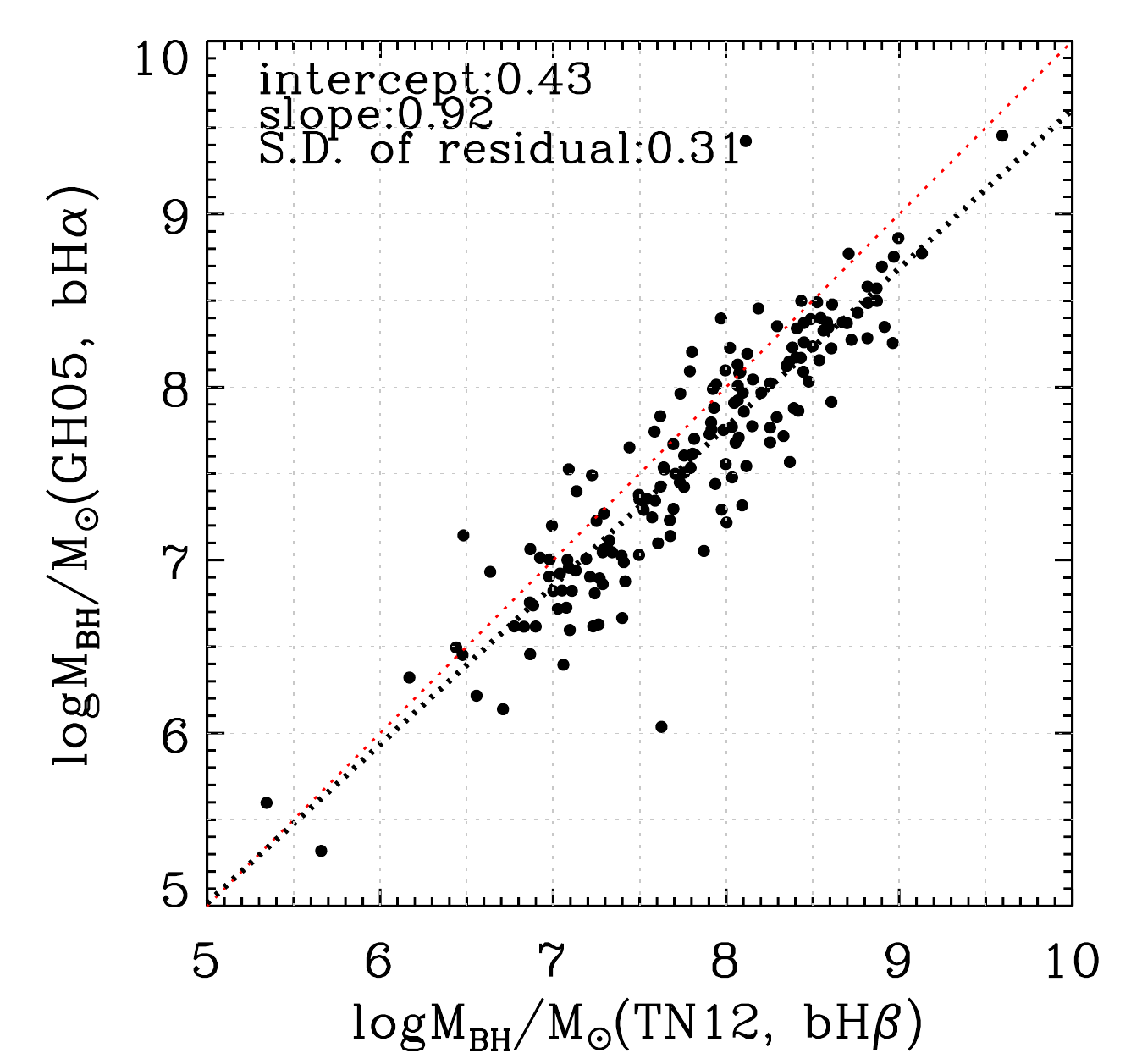}}
\subfigure{\includegraphics[width=0.4\textwidth]{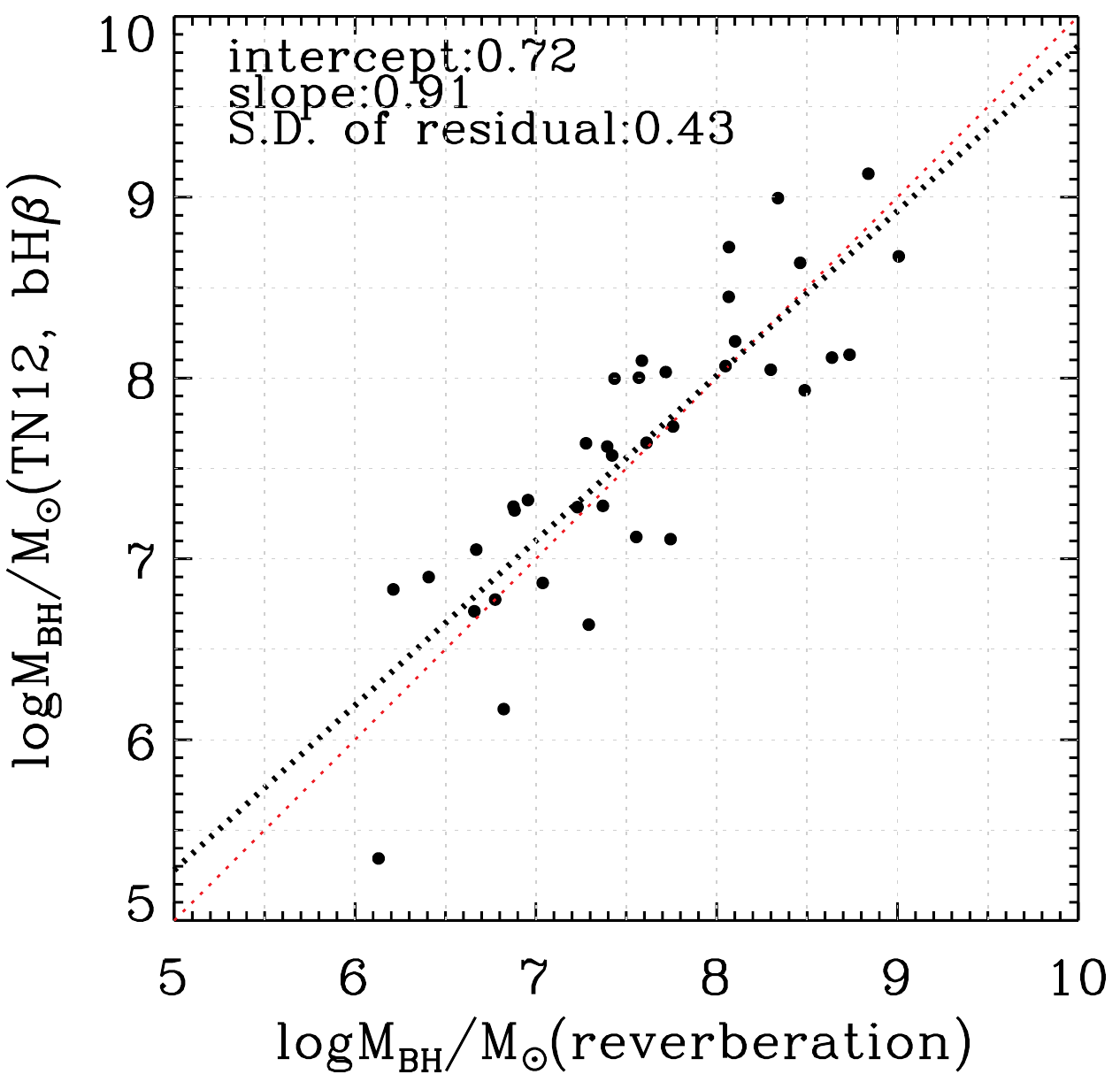}}
\caption{{\em Top}: Comparison of black hole masses from broad \hb\ \citep{Trakhtenbrot:2012:3081} and broad \ha\ \citep{Greene:2005:122}.  {\em Bottom}: Comparison of single epoch broad \hb\ measurements to reverberation mapping \citep{Bentz:2015:67}. One-to-one fiducial line and linear regression fit are shown with red and black dotted lines, respectively, for both figures.}
\label{Comp_Mbh}
\end{figure}

\begin{figure*}[h!] 
\centering
\includegraphics[width=0.49\textwidth]{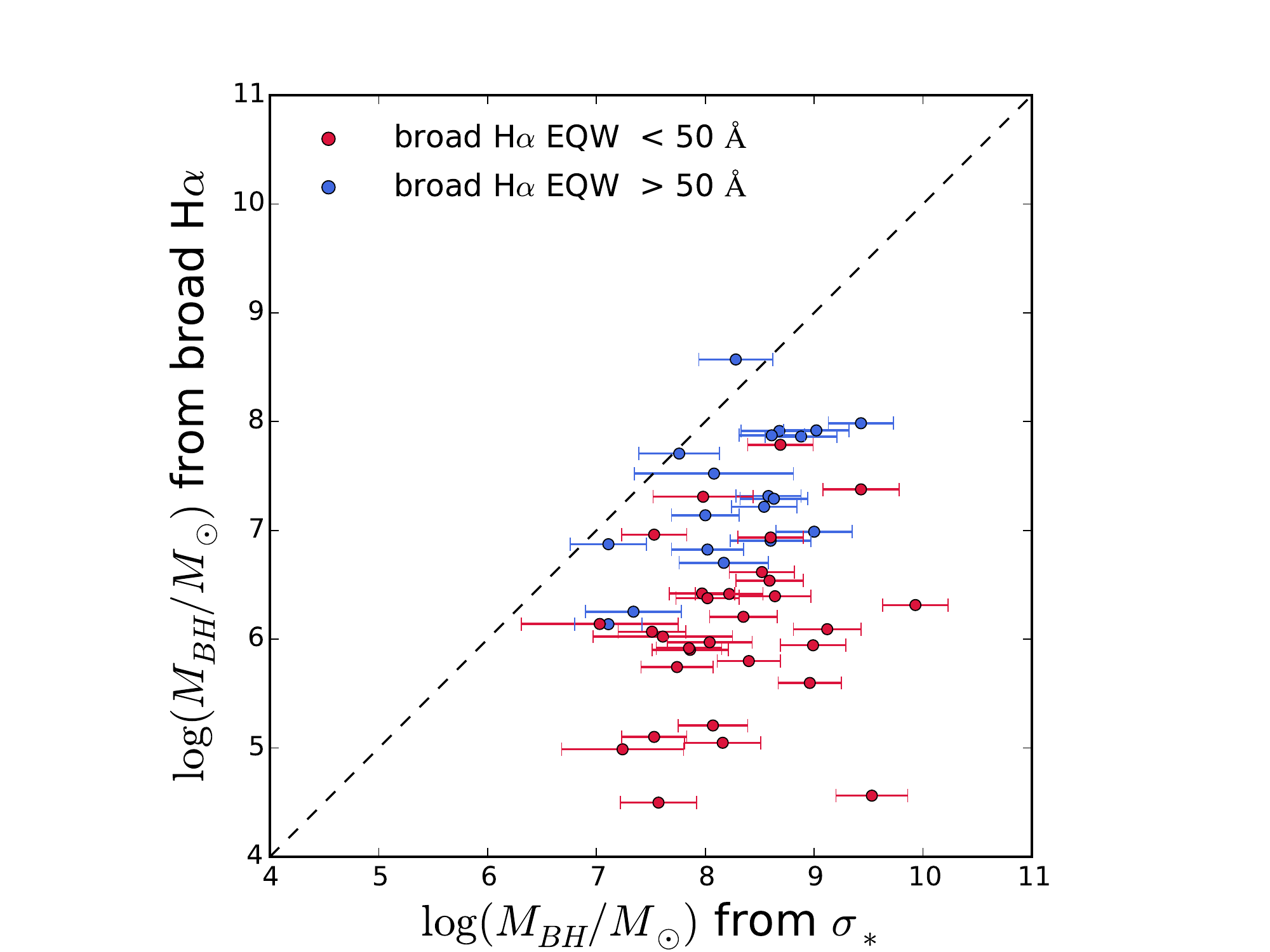}
\includegraphics[width=0.49\textwidth]{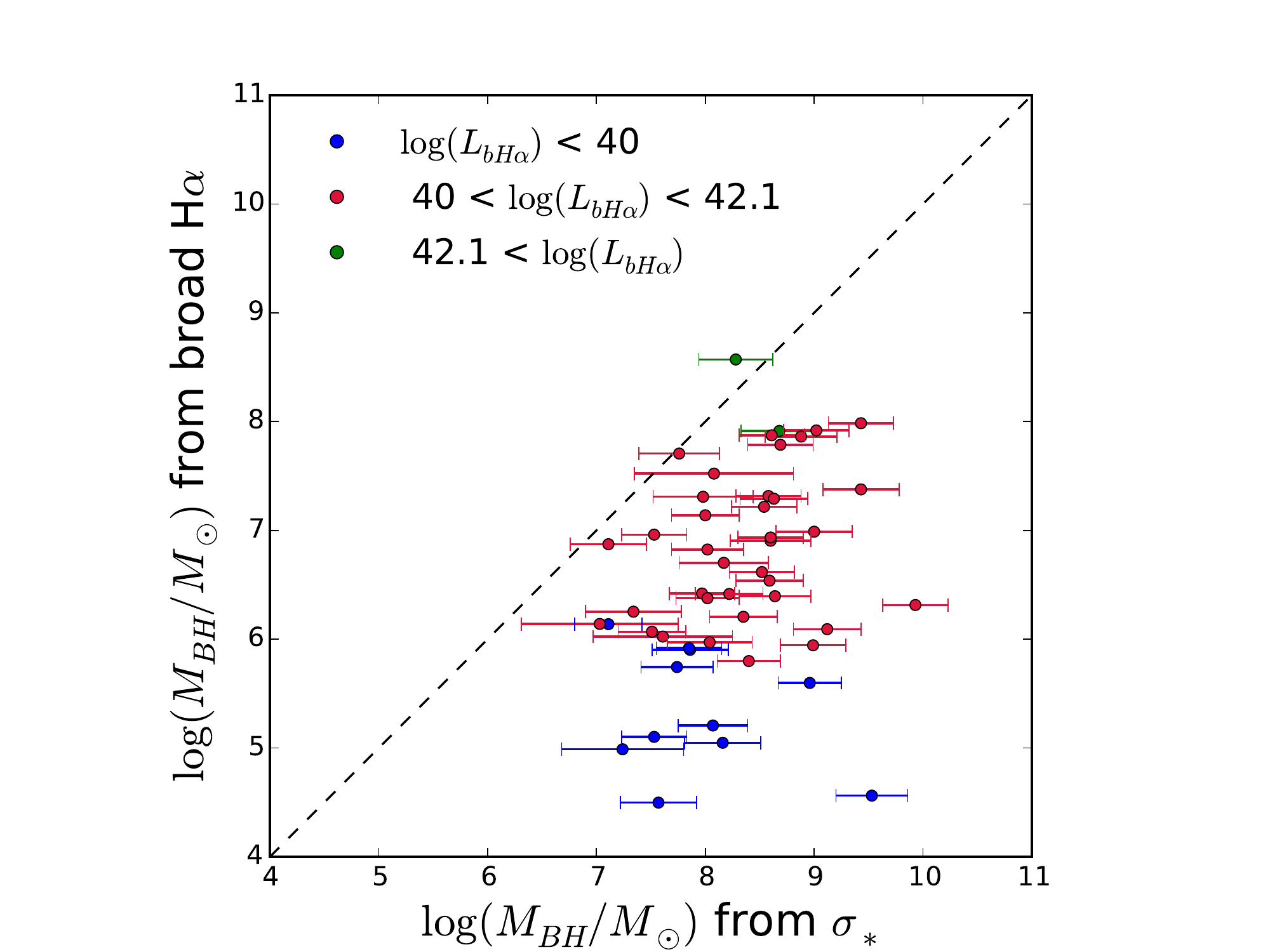}
\caption{Comparison of black hole masses for Seyfert 1.9 galaxies derived from the velocity dispersion and from the broad H$\alpha$.  We find that at low EW ($<$50\,\AA) there is a significant difference between measurements of \sigs and broad \Halpha.
}
\label{MHa_vs_Msigma}
\end{figure*}

We compared our black hole mass measurements (Fig. \ref{Comp_Mbh}) using broad line measurements of the \Hbeta\ or \Halpha\ regions in single epoch observations.  We also compare the mass measurements to higher quality reverberation mapping measurements for the small sample where these are available.  We find that the slope is consistent with unity with fairly large scatter ($\approx$0.4-0.5 dex).

We also made a comparison of Seyfert 1.9 black hole masses measured with very weak broad \Halpha\ lines with no broad \Hbeta\ detected (Fig. \ref{MHa_vs_Msigma}).  We find that below equivalent widths (EW) of 50\,\AA\ in \Halpha, the velocity dispersion differs significantly suggesting these broad features could be spurious  and thus we have used this limit for flag 2 sources in our sample.  The value of this limit is consistent with many past quasar studies \citep[e.g. EW$<$45\,\AA;][]{Shen:2011:45} though some studies of nearby AGN have used much weaker broad lines \citep[e.g. EW$<$15\,\AA;][]{Greene:2007:131}.  

The offset in black hole mass measurements between Seyfert 1.9 using velocity dispersion measurements and broad line $\Halpha$ is concerning though offsets have been found in AGN samples  \citep[e.g.,][]{Shankar:2017:4029}.  We note, however, that these two methods are tied to reproduce similar masses for systems where both are applicable \citep[e.g.,][]{Graham:2011:2211,Woo:2013:49}, so it is likely that the subsample of Seyfert 1.9 explored here is not representative.  One particular concern is that the weak broad line $\Halpha$ may suffer high extinction and is underestimated, which we are exploring in a current VLT/XSHOOTER program using the NIR broad Paschen emission lines (Oh et al., in prep.).

\subsection{Bolometric Luminosity}
\begin{figure*}[hbtp]
\centering
\includegraphics[width=6.0cm]{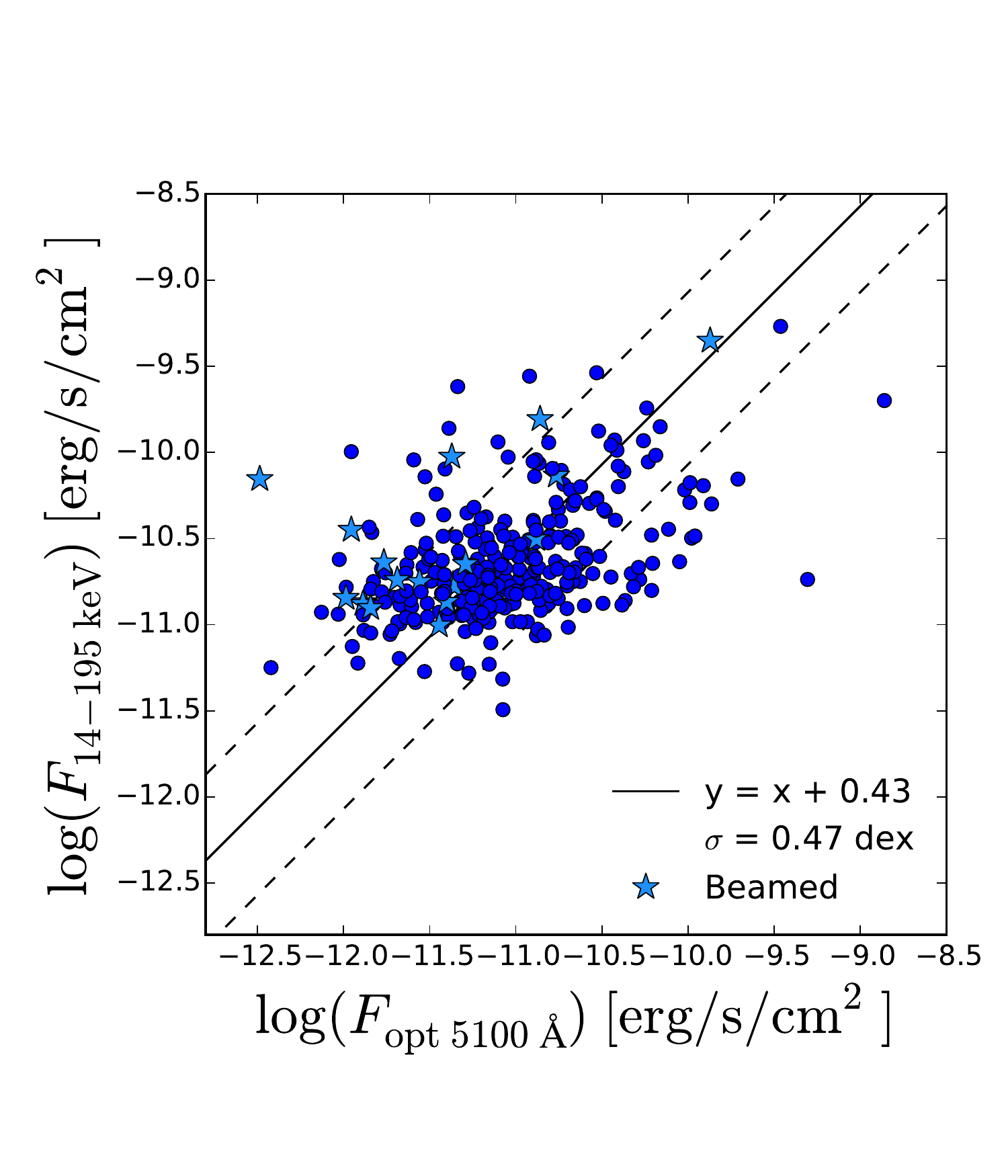}
\includegraphics[width=5.85cm]{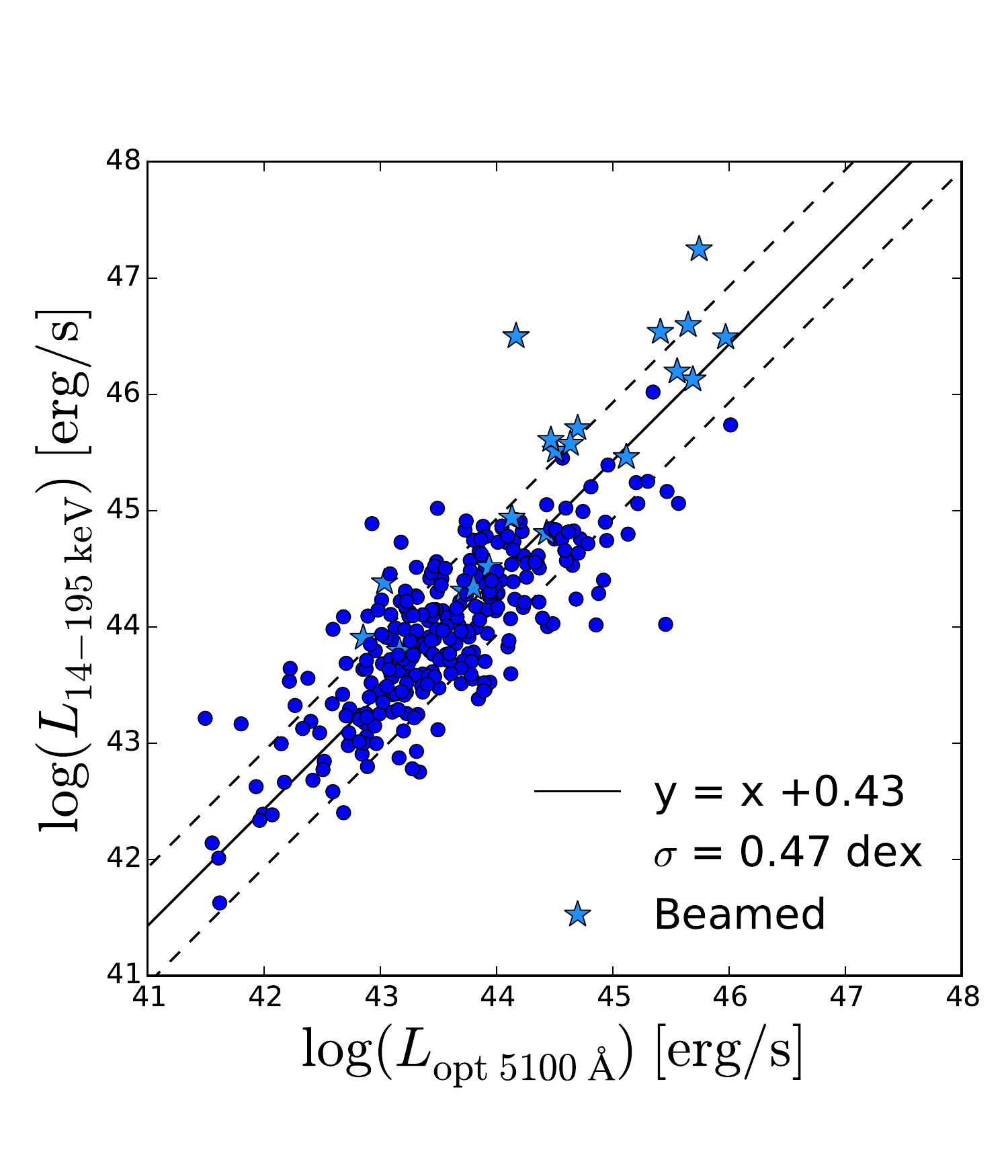}
\includegraphics[width=6.0cm]{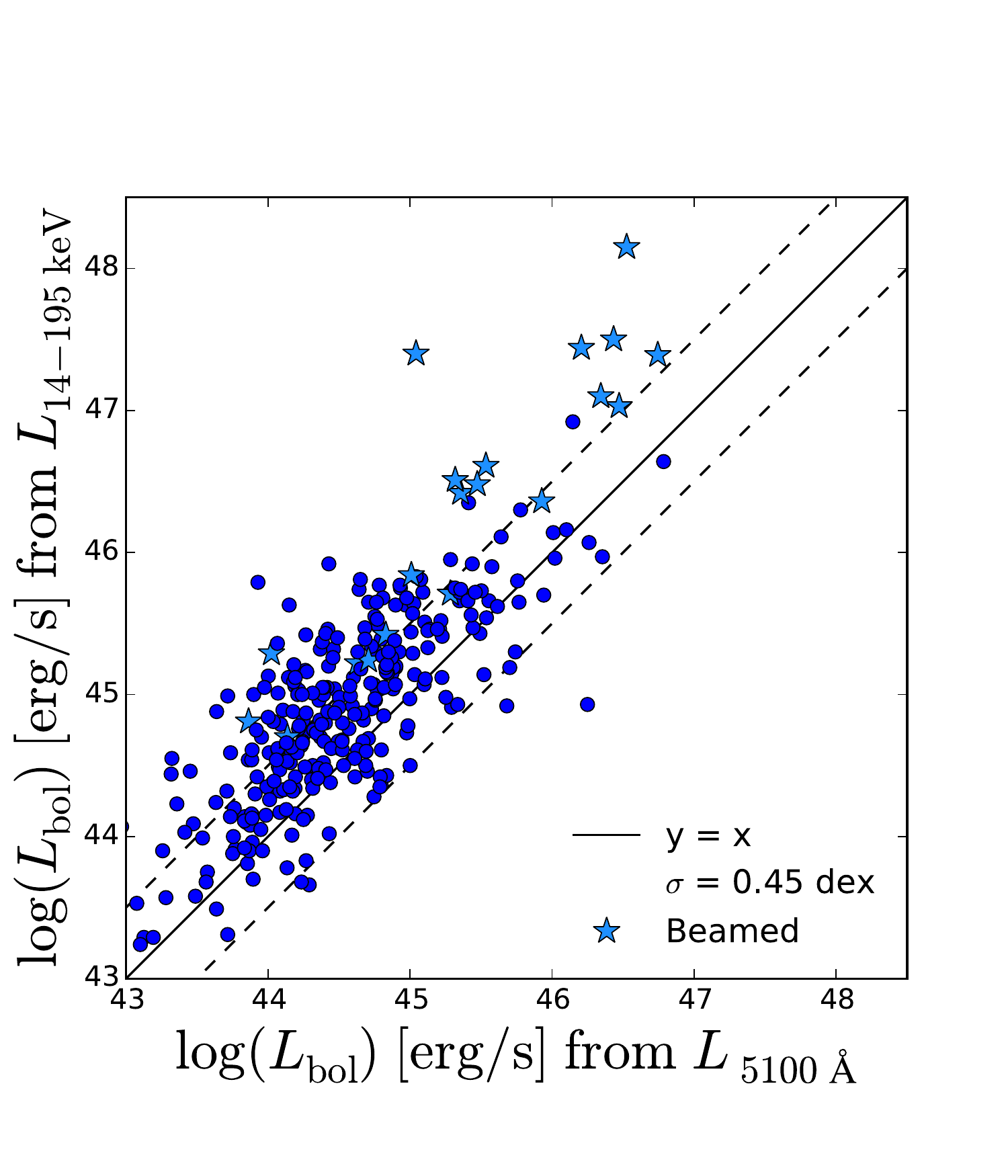}
\caption{Relation between the X-ray emission and optical 5100\,\AA\ emission shown in flux (left panel) and luminosity (middle panel). The right panel shows the comparison between the bolometric luminosity derived from the X-ray luminosity and the bolometric luminosity derived from the optical luminosity for broad line Seyfert 1 galaxies.}
\label{Comp_Lbol}
\end{figure*}

For consistency, we compare our values of the bolometric luminosity $L_{bol}$, derived from the X-ray luminosity, with the values of \Lbol\  obtained from the optical luminosity at 5100\,\AA\ \citep[e.g.,][]{Wandel:1999:579} for Seyfert 1 AGN.  In  Figure \ref{Comp_Lbol} the comparison between $L_{bol}$  obtained from the X-ray bolometric corrections and $L_{bol}$  derived from the optical luminosity are shown.   The two methods to infer $L_{bol}$ show fairly larger scatter (0.46 dex), with more scatter at high luminosities.
%Since our survey did not include UV measurements where the majority of the AGN emission occurs, and the differences do not affect the conclusions

%\begin{figure*}[hbtp]
%\centering
%\subfigure{\includegraphics[width=0.49\textwidth]
%{redshift_offset_vs_Lbol_v3.pdf}}
%\subfigure{\includegraphics[width=0.49\textwidth]
%{redshift_offset_vs_Edd_v3.pdf}}

%\caption{Velocity offset between \oiii\ emission lines and stellar templates as a function of bolometric luminosity (left panel) and Eddington ratio (right panel).}
%\label{Comp_Lbol}
%\end{figure*}

%%%% BIB  %%%%%%%%%%%%%%%%%%%%%%%%%%%%%%%%%%%%%%%%%%%%%%%%%%%%%%%%%%%%%%%%%%%%%%%%%%%%%%%%%%%%%%%%%%%
{\it Facilities:} \facility{Swift}, \facility{UH:2.2m},  \facility{Sloan}, \facility{KPNO:2.1m}, \facility{FLWO:1.5m (FAST)}, \facility{Shane (Kast Double spectrograph)}, \facility{CTIO:1.5m}, \facility{Hale}, \facility{Gemini:South}, \facility{Gemini:North}, \facility{Radcliffe}, \facility{Perkins}

\end{document}